\documentclass{aa}
\usepackage{graphicx}
\usepackage{txfonts}
\usepackage{natbib}
\usepackage{hyperref}
\hypersetup{colorlinks=true, citecolor=blue}
\usepackage{graphicx}
\usepackage{txfonts}
\usepackage{rotating}
\usepackage{tabularx}
\bibpunct{(}{)}{;}{a}{}{,}

\begin{document} 

\title{Towards universal hybrid star formation rate estimators}
\author{M. Boquien\inst{1,2} \and R. Kennicutt\inst{1} \and D. Calzetti\inst{3} \and D. Dale\inst{4} \and M. Galametz\inst{5} \and M. Sauvage\inst{6} \and K. Croxall\inst{7} \and B. Draine\inst{8} \and A. Kirkpatrick\inst{3} \and N. Kumari\inst{1} \and L. Hunt\inst{9} \and I. De Looze\inst{1,10} \and E. Pellegrini\inst{11} \and M. Rela\~no\inst{12} \and J.--D. Smith\inst{13} \and F. Tabatabaei\inst{14}}
\institute{
  Institute of Astronomy, University of Cambridge, Madingley Road, Cambridge CB3 0HA, UK
  \and Unidad de Astronomía, Fac. Cs. Básicas, Universidad de Antofagasta, Avda. U. de Antofagasta 02800, Antofagasta, Chile \email{mederic.boquien@uantof.cl}
  \and Department of Astronomy, University of Massachusetts--Amherst, Amherst, MA 01003, USA
  \and Department of Physics and Astronomy, University of Wyoming, Laramie, WY 82071, USA
  \and European Southern Observatory, Karl-Schwarzschild-Str. 2, D-85748 Garching-bei-München, Germany
  \and Laboratoire AIM, CEA/IRFU/Service d'Astrophysique, CNRS, Université Paris Diderot, Bat. 709, 91191 Gif-sur-Yvette, France
  \and Department of Astronomy, The Ohio State University, 140 W 18th Ave., Columbus, OH 43210, USA
  \and Department of Astrophysical Sciences, Princeton University, Princeton, NJ 08544, USA
  \and INAF-Osservatorio Astrofisico di Arcetri, Largo E. Fermi 5, 50125, Firenze, Italy
  \and Department of Physics \& Astronomy, University College London, Gower Place, London WC1E 6BT, UK
  \and Department of Physics and Astronomy, University of Toledo, 2801 West Bancroft Street, Toledo, OH 43606, USA
  \and Department F\'isica Te\'orica y del Cosmos, Universidad de Granada 18071, Granada, Spain
  \and Ritter Astrophysical Observatory, University of Toledo, Toledo, OH 43606, USA
  \and Instituto de Astrof\'isica de Canarias, V\'ia L\'actea s/n, E-38205 La Laguna, Spain
  }
\date{}
\abstract
  {To compute the star formation rate (SFR) of galaxies from the rest--frame ultraviolet (UV) it is essential to take into account the obscuration by dust. To do so, one of the most popular methods consists in combining the UV with the emission from the dust itself in the infrared (IR). Yet, different studies have derived different estimators, showing that no such hybrid estimator is truly universal.}
  {In this paper we aim at understanding and quantifying what physical processes fundamentally drive the variations between different hybrid estimators. Doing so, we aim at deriving new universal UV+IR hybrid estimators to correct the UV for dust attenuation at local and global scales, taking into account the intrinsic physical properties of galaxies.}
  {We use the CIGALE code to model the spatially--resolved far--UV to far--IR spectral energy distributions of eight nearby star--forming galaxies drawn from the KINGFISH sample. This allows us to determine their local physical properties, and in particular their UV attenuation, average SFR, average specific SFR (sSFR), and their stellar mass. We then examine how hybrid estimators depend on said properties.}
  {We find that hybrid UV+IR estimators strongly depend on the stellar mass surface density (in particular at 70~$\mu$m and 100~$\mu$m) and on the sSFR (in particular at 24~$\mu$m and the TIR). Consequently, the IR scaling coefficients for UV obscuration can vary by almost an order of magnitude: from 1.55 to 13.45 at 24~$\mu$m for instance. This result contrasts with other groups who found relatively constant coefficients with small deviations. We exploit these variations to construct a new class of adaptative hybrid estimators based on observed UV to near--IR colours and near--IR luminosity densities per unit area. We find that they can reliably be extended to entire galaxies.}
  {The new estimators provide better estimates of attenuation--corrected UV emission than classical hybrid estimators published in the literature. Naturally taking into account the variable impact of dust heated by old stellar populations, they constitute an important step towards universal estimators.}

\maketitle

\section{Introduction}

Understanding when, where, and how stars form from the highest redshifts to the local universe is key to understanding the formation and evolution of galaxies. Originating directly from the photosphere of young, massive stars living up to a few 100~Myr, the ultraviolet (UV) emission of galaxies is one of our main windows into star formation. Because rest--frame UV is easily accessible from the ground for distant galaxies, it has even become the star formation tracer of choice for large cosmological surveys. The launch of the Galaxy Evolution Explorer \citep[GALEX,][]{martin2005a} in 2003 has considerably furthered our insight into the UV emission of galaxies by providing us with a large amount of observations of nearby galaxies in the rest--frame far--UV (FUV, 151.6~nm) and near--UV (NUV, 226.7~nm) bands, which are not accessible from the ground.

Yet, as a star formation tracer the UV suffers from a major issue: it is very efficiently attenuated by dust. \cite{burgarella2013a} showed that in the nearby universe $69\pm10$\% of star formation is invisible in the rest--frame FUV because of dust. This obscuration peaks at $89\pm4$\% at $z=1.35$. It means that correcting for the presence of dust is a crucial issue if we want to use the UV as a reliable star formation rate (SFR) estimator.

To correct the UV for dust attenuation, one of the most popular techniques is to translate the UV spectral slope to an attenuation using the IRX--$\beta$ relation. While this method works remarkably well for starburst galaxies \citep{meurer1999a}, it fails, sometimes considerably, for more quiescent galaxies \citep{kong2004a,dale2007a}. This is probably due to a combination of age effects and variations in the attenuation laws \citep[][and many others]{bell2002a,calzetti2005a,boquien2009a,boquien2012a,mao2012a,mao2014a,grasha2013a}.

An alternative approach has seen important developments over the past decade: hybrid (or composite) estimators \citep[e.g.,][]{buat1999a,buat2005a,calzetti2007a,zhu2008a,kennicutt2009a,hao2011a}. Because part of the energy emitted in the mid-- and far--infrared bands is due to the reprocessing by dust of UV photons emitted by massive stars, these hybrid estimators aim to correct the UV for the attenuation using the emission of the dust in a given IR band:
\begin{equation}
 L(UV)_{int}=L(UV)_{obs}+k_i\times L(IR)_i,
\end{equation}
with $L\left(UV\right)_{int}$ the intrinsic UV luminosity (defined as $\nu L_\nu$) in the absence of dust, $L\left(UV\right)_{obs}$ the observed UV luminosity, $L\left(IR\right)_i$ the luminosity in the IR band $i$, and $k_i$ the scaling coefficient for the corresponding IR band. Or equivalently it can be written in terms of SFR:
\begin{equation}
 SFR = c_{UV}\times\left[L(UV)_{obs}+k_i\times L(IR)_i\right],
\end{equation}
with $c_{UV}$ the UV--to--SFR calibration coefficient. The scaling coefficient can then be calibrated observationally by combining the attenuated and attenuation--corrected UV luminosities with the observed luminosities in the given IR band:
\begin{equation}\label{eqn:k-IR-v1}
  k_i=\frac{L\left(UV\right)_{int}-L\left(UV\right)_{obs}}{L\left(IR\right)_i},
\end{equation}
or by assuming $c_{UV}$ and combining the estimated SFR, with the attenuated UV luminosities and the observed luminosities in the given IR band:
\begin{equation}\label{eqn:k-IR-v2}
 k_i=\frac{SFR-c_{UV}\times L(UV)_{obs}}{c_{UV}\times L(IR)_i}.
\end{equation}
While this method can be appealing, different studies have found different $k_i$ coefficients, depending on whether the IR emission unrelated to recent star formation has been subtracted \citep[e.g., the resolved study of][]{liu2011a} or not \citep[e.g., the unresolved study of][]{hao2011a}. This emission is often visible in resolved galaxies as a diffuse component which varies with galaxy type \citep{sauvage1992a}. This perhaps translates differences in the age of stellar populations, or equally in the relative contributions of young and old stellar populations to the dust--heating interstellar radiation field. Even in actively star--forming galaxies, recent results based on idealised hydrodynamical simulations find that up to a third of the total IR luminosity (TIR, which we define here as the integral of dust emission over all wavelengths) can be due to stars older than 100~Myr, and therefore not related to recent star formation \citep{boquien2014a}, in line results assuming a constant SFR \citep{crocker2013a}. This fraction can be even higher in individual IR bands. This is consistent with various \textit{Herschel} \citep{pilbratt2010a} studies, which have shown that the warm dust emission is driven by recent star formation whereas the emission of the colder dust is rather driven by older stellar populations \citep{popescu2000a,bendo2010a,bendo2012a,bendo2015a,boquien2011a,delooze2012a,delooze2012b,lu2014a,natale2015a}.

As dust emission is not entirely related to UV--emitting stars, it is necessary to take this into account when deriving hybrid estimators. One approach to tackle this issue is to systematically subtract the diffuse emission \citep[e.g.,][]{calzetti2007a}, assuming that it is unrelated to recent star formation. This may not be valid if a significant fraction of the diffuse emission is actually due to radiation having escaped from star--forming regions \citep{kirkpatrick2014a}. In any case, this method requires observations at a high enough spatial resolution to enable the subtraction of the diffuse emission on a local basis. This can be achieved for instance by directly processing the images or by estimating the level of the diffuse emission from an ancillary dataset \citep[e.g., from a gas map, assuming the gas--to--dust mass ratio and the radiation field intensity due to old stars,][]{leroy2012a}. An alternative approach is to statistically take the average diffuse dust emission into account by calibrating $k_i$ against a star formation tracer insensitive to the emission of old stars, such as the extinction--corrected H$\alpha$ \citep{kennicutt2009a}. While hybrid SFR estimators provide good results on average, they have so far been limited to constant values in terms of $k_i$. This ignores not only galaxy--to--galaxy variations, which can be large, but also variations within galaxies when applied on local scales. Ultimately, they run the risk of being strongly dependent on the intrinsic properties of the calibration sample. This may induce systematic biases on the SFR if the observed sample has different properties.

The aim of this paper is to derive a parametrisation of $k_i$ that naturally includes the varying degrees of dust heating from old stars at 24~$\mu$m, 70~$\mu$m, 100~$\mu$m and for the TIR. Following this idea, we propose to parametrise $k_i$ on 1. the FUV$-$NIR (J to 3.6~$\mu$m bands) colours, which are sensitive to the SFH, and 2. NIR luminosity densities per unit area, which are sensitive to the stellar mass surface densities. Such an approach would serve as a basis for new hybrid SFR estimators applicable at local and global scales across a wide range of physical conditions. To do so, we first need to quantify and understand how the relation between the UV and the IR varies within and between galaxies. With this in mind, we carry out a spatially resolved, multi--wavelength study of eight star--forming spiral galaxies drawn from the KINGFISH sample (Sect.~\ref{sec:sample}). These data allow us to carry out a detailed Spectral Energy Distribution (SED) modelling including stellar populations of all ages, nebular emission, attenuation by dust in the UV--to--NIR domains, and the re--emission of the absorbed energy at longer wavelengths. In turn, such a modelling allows us to compute their local physical properties such as their UV attenuation, their stellar mass, and their mean intrinsic and specific SFR over 100~Myr (Sect.~\ref{sec:physical-parameters}). Combining these physical properties with the observations, we investigate how $k_i$ depends on the local properties of galaxies in Sect.~\ref{sec:kIR-dependence-parameters}. This allows us to provide new practical methods to correct the FUV for dust attenuation at local scales in Sect.~\ref{sec:hybrid-relations}, which we generalise to entire galaxies in Sect.~\ref{sec:kir-unresolved}. We discuss the limits of this new approach and we provide practical guidelines how to apply it to correct the UV for the attenuation in different cases in Sect.~\ref{sec:limits}. Finally, we summarise our results in Sect.~\ref{sec:conclusion}.

\section{Sample\label{sec:sample}}

\subsection{Selection\label{ssec:selection}}

To carry out this study we need to perform a detailed SED modelling to determine the local physical properties of nearby galaxies and in particular their UV attenuation, which is key as we will see in Sect.~\ref{sec:kIR-dependence-parameters}. To do so we rely on the KINGFISH \citep[Key Insights on Nearby Galaxies: a Far-Infrared Survey with \textit{Herschel},][]{kennicutt2011a} sample of nearby galaxies. It provides us with a broad range of spatially resolved, FUV to far--IR (FIR) data over many galaxy types.

In this paper we restrict our subsample to star--forming spiral galaxies (Sa and later types) that have a minor axis of at least 5\arcmin\ and an inclination no larger than 60$\degr$. This ensures that we have at least several tens of $\sim$1~kpc resolution elements across each galaxy at 24\arcsec, the resolution of \textit{Herschel} SPIRE 350~$\mu$m, which is the coarsest data we consider as we will see later. Such large resolution elements ensure that there is little leakage of energetic photons between neighbouring pixels \citep[leaking photons should be absorbed within a radius of the scale--height of the diffuse ionised gas, 900~pc in the Milky Way,][]{reynolds1989a}, which would affect the energy balance assumption of the modelling (Sect.~\ref{ssec:modelling}). We also only select galaxies without a strong active nucleus (for instance we eliminate galaxies with a Seyfert nucleus but we keep LINER ones) to limit the contamination of the UV and the IR by emission from the nucleus. In addition, we also eliminate galaxies at low Galactic latitude due to the very large number of foreground stars in the field, which makes their removal daunting and uncertain (see Sect.~\ref{ssec:convolution} for the removal procedure). This would lead to a possible contamination of the emission of the galaxies by foreground objects. Finally, we also eliminate galaxies with data of insufficient quality for this study. This is for instance the case for galaxies that have only been observed with the GALEX All--Sky Imaging Survey, which is too shallow to allow for a reliable pixel--by--pixel determination of the UV fluxes.

We have therefore selected eight galaxies in our sample: NGC~628, NGC~925, NGC~1097, NGC~3351, NGC~4321, NGC~4736, NGC~5055, and NGC~7793.

\subsection{Properties}

As we can see in Table~\ref{tab:sample}, the sample is made of fairly regular star--forming galaxies in the nearby universe, spanning the Hubble sequence from Sab to Sd.
\begin{table*}
 \centering
 \begin{tabular}{cccccccc}
  \hline\hline
  Name              &Morphological type&Angular size   &Distance&$12+\log O/H$&$\log sSFR$\\
                    &                  &(arcmin)       &(Mpc)   &\citep{kobulnicky2004a}&(yr$^{-1}$)\\\hline
  NGC \phantom{0}628&SAc               &$10.5\times9.5$&7.2     &9.02&$-$9.73\\
  NGC \phantom{0}925&SABd              &$10.5\times5.9$&9.12    &8.79&$-$9.76\\
  NGC 1097          &SBb               &$9.3\times6.3$ &14.2    &9.09&$-$9.86\\
  NGC 3351          &SBb               &$7.4\times5.0$ &9.33    &9.19&$-$10.48\\
  NGC 4321          &SABbc             &$7.4\times6.3$ &14.3    &9.17&$-$9.88\\
  NGC 4736          &SAab              &$11.2\times9.1$&4.66    &9.01&$-$10.76\\
  NGC 5055          &SAbc              &$12.6\times7.2$&7.94    &9.14&$-$10.53\\
  NGC 7793          &SAd               &$9.3\times6.3$ &3.91    &8.88&$-$9.59\\\hline
 \end{tabular}
 \caption{Main properties of the eight galaxies in the selected sample. Data extracted from Table~1 of \cite{kennicutt2011a}. See references therein.\label{tab:sample}}
\end{table*}
This ensures different levels of star-forming activity (from $\log sSFR=-10.76$~yr$^{-1}$ for NGC~4736 to $\log sSFR=-9.59$~yr$^{-1}$ for NGC~7793) and different levels of diffuse dust emission \citep{sauvage1992a}. This range of star formation activity of over one dex is also expanded by the spatially--resolved nature of this study, each galaxy containing relatively more active and more quiescent regions. In the end, this range of star formation activity represents the sweet spot to calibrate hybrid SFR estimators. More extreme galaxies from the point of view of star formation are either dominated by diffuse emission (such as early type galaxies) or quite the opposite have a negligible diffuse emission compared to dust heating by massive stars (such as young starbursts).

In terms of sizes, all galaxies are well-resolved, with NGC~3351 being the smallest with a size of $7.4\arcmin\times 5.0\arcmin$ and NGC~5055 being the largest with a size of $12.6\arcmin\times 7.2\arcmin$, providing many resolution elements within each object (see Sect.~\ref{sec:data}).

Finally, the metallicities are typically super--solar according to the estimators of \cite{kobulnicky2004a}. This is due to the criteria mentioned in Sect.~\ref{ssec:selection}. They tend to eliminate low metallicity star--forming galaxies as they are often smaller and are not well--resolved with Herschel, if their far--infrared emission is detected at all. While this reduces the volume of the space of physical properties covered by our sample, those galaxies have in general little attenuation and little dust emission. This means that in their case, uncertainties on hybrid estimators will have an especially small impact.

\section{Data\label{sec:data}}

For all of these galaxies we have spatially resolved data in the UV (FUV and NUV bands from GALEX), near--infrared (in the J, H, and Ks bands with 2MASS, and at 3.6~$\mu$m and 4.5~$\mu$m with IRAC\footnote{The calibration of IRAC data has been converted from point--source to extended emission using the aperture corrections described in \url{https://irsa.ipac.caltech.edu/data/SPITZER/docs/irac/iracinstrumenthandbook/29/}.} on--board \textit{Spitzer}), mid--infrared (IRAC 5.8~$\mu$m, 8~$\mu$m, and MIPS 24~$\mu$m from \textit{Spitzer}), and far--infrared (PACS 70~$\mu$m, 100~$\mu$m, 160~$\mu$m, and SPIRE 250~$\mu$m and 350~$\mu$m from \textit{Herschel}\footnote{PACS \citep{poglitsch2010a} and SPIRE \citep{griffin2010a} data were processed to Level 1 using HIPE version 11.1.0 with calibration products PACS\_CAL\_56\_0 and SPIRE\_CAL\_11\_0. Subsequent processing and map making was carried out with Scanamorphos \citep{roussel2013a} version 24. For SPIRE the beam areas at 250~$\mu$m and 350~$\mu$m were assumed to be 465.39\arcsec$^2$ and 822.58\arcsec$^2$.}). This broad set of data enables us to estimate their local physical properties along with their uncertainties from SED modelling (Sect.~\ref{sec:physical-parameters}). We do not consider the SPIRE 500~$\mu$m band because it would further degrade the angular resolution (the beam size at 500~$\mu$m is $\sim35$\arcsec) for no gain in the present context because the TIR luminosity, which is important to constrain the UV attenuation, is already securely determined without the 500~$\mu$m band. Finally, we do not consider optical data either here as our initial investigations on three galaxies in the sample (NGC~628, NGC~1097, and NGC~3351) have shown that their presence does not allow us to obtain significantly better measurements of the physical parameters we are interested in. In addition, they come from heterogeneous sources, contrary to the other bands here.

\subsection{Milky--Way foreground dust extinction correction}

The emission from the FUV to the near--infrared (NIR) is affected by the presence of foreground dust in the Milky Way. We unredden the SED adopting the \cite{cardelli1989a} Milky--Way extinction curve including the \cite{odonnell1994a} update. The $E(B-V)$ normalisation is obtained from NASA/IPAC's Galactic Dust Reddening and Extinction service\footnote{\url{https://irsa.ipac.caltech.edu/applications/DUST/}} as the mean $E(B-V)$ computed from the \cite{schlafly2011a} extinction maps in a 5\arcmin\ radius circle, similar to the typical size of the galaxies in the sample.

\subsection{Convolution\label{ssec:convolution}}

To carry out a spatially resolved modelling and analysis, we need to ensure that for each galaxy all the maps have the same angular resolution and pixel size. To do so, we need to convolve all images to the same point spread function and project them on a common grid of pixels.

In a first step, to ensure there is no contamination from prominent foreground stars, we edit them out from the images using \textsc{iraf}'s \textsc{imedit} procedure. It replaces the pixels contained within a manually defined rectangular aperture enclosing a star, with pixels having the same mean and standard deviation as those in the immediate surrounding area. The stars are identified by eye, combining UV, optical, and NIR data. Depending on the galactic latitude of the target and the band considered, the number of stars removed varies from a few per image to several dozens. The \textit{Spitzer} 3.6~$\mu$m and 4.5~$\mu$m images are generally the most affected by the presence of foreground stars.

Once the images have been cleaned, we degrade their angular resolution to that of the SPIRE 350~$\mu$m band using the convolution kernels of \cite{aniano2011a}.

All images are then registered onto a common grid with a pixel size of 24\arcsec, close to the resolution of the SPIRE 350~$\mu$m band, corresponding to physical scales from 0.5~kpc (NGC~7793) to 1.7~kpc (NGC~1097). The pixel size is similar to the angular resolution so that individual pixels are physically independent from one another. This registration is carried out using \textsc{iraf}'s \textsc{wregister} module with the \textsc{drizzle} interpolant.

Finally, the background is subtracted from the registered images by averaging the mean values from typically several tens of $3\times3$ pixel boxes randomly located in empty regions around the galaxy.

\subsection{Uncertainties}

The uncertainties on the fluxes are computed by summing in quadrature the standard deviation of the background level measurements with the mean of the standard deviation of the pixel values in each of the $3\times3$ pixel boxes. A systematic uncertainty of 5\% is added to reflect calibration errors.

For the SED modelling, we only select pixels detected at least at a 1--$\sigma$ level in all bands. The lowest signal--to--noise images are the PACS 70~$\mu$m and 100~$\mu$m bands. While this necessarily increases the uncertainties on the total infrared (TIR) luminosity, these uncertainties are fully taken into account in the modelling and therefore propagated to the estimation of the physical parameters. This yields a final sample of 1364 pixels distributed over eight galaxies.

\section{Estimation of the physical parameters\label{sec:physical-parameters}}

\subsection{\textsc{cigale} modelling\label{ssec:modelling}}

To compute $k_i$, it is essential that we are in a good position to determine either the intrinsic UV emission (Eq.~\ref{eqn:k-IR-v1}) or the SFR (Eq.~\ref{eqn:k-IR-v2}). This means that we need to obtain reliable estimates of either the UV attenuation or the SFR. In principle for deriving new hybrid SFR estimators, using Eq.~\ref{eqn:k-IR-v2} would appear to be the most direct option as there multiple conversion factors ($c_{UV}$) available in the literature. However in practice, these conversion factors strongly rely on multiple hypotheses, and in particular on the assumed shape of the SFH, which 1. is difficult to know a priori, 2. is not always adequate \citep{boquien2014a}, 3. varies from region to region and from galaxy to galaxy, and 4. makes the estimated SFR somewhat model dependent. To avoid such issues it appears safer to rely on Eq.~\ref{eqn:k-IR-v1} and therefore on the UV attenuation, which has a weaker dependence on the exact SFH. This means that while we do not explicitly provide hybrid SFR estimators, we provide relations to correct the UV for the attenuation and we leave the reader the choice of the most adequate UV--to--SFR conversion coefficient for their case, limiting model dependence and ensuring enhanced flexibility compared to hybrid estimators provided in the literature.

At the same time, we also want to understand whether and how hybrid estimators depend on local physical properties such as the SFR or the stellar mass, so we also need to be in a position to compute those. To reach these objectives, we model the FUV to FIR emission pixel--by--pixel  using the python CIGALE\footnote{\url{http://cigale.lam.fr/}} SED modelling code (Boquien et al.) version 0.9.

Briefly, the model is based on an energy balance principle: the energy absorbed by dust in the UV, optical, and NIR domains is self--consistently re--emitted in the mid-- and far--infrared. Each SED is computed from the following high--level procedure:

\begin{itemize}
 \item The star formation history (SFH) is modelled with two decaying exponentials. The first one with a long $e$--folding time of a few billion years models the long term star formation of the galaxy that has created the bulk of the total stellar mass. The second one with a much shorter $e$--folding time under 100 Myr models the latest episode of star formation. This modelling of the SFH represents a compromise. Because we are using only broad band data, it is not possible to constrain the shape of the SFH in more detail. At the same time we are studying regions of a typical size of 1~kpc whose SFH is expected to vary with more amplitude and more rapidly than in the case of entire galaxies. Modelling the SFH with two decaying exponentials allows us to take such variations into account without unnecessary complexity that would not be constrained with broad band observations. We will see later that this flexibility is important to include the impact of a variable SFH on the estimation of other physical properties, and as a consequence on the computation of $k_i$ (see Sect.~\ref{ssec:variations-kIR}).
 \item The stellar spectrum is calculated by convolving the \cite{bruzual2003a} single stellar populations with the SFH, assuming a \cite{chabrier2003a} initial mass function and a metallicity of $Z=0.02$, which corresponds to $12+\log O/H=8.90$ assuming the solar metallicity ($Z=0.0134$) and oxygen abundance ($12+\log O/H=8.69$) from \cite{asplund2009a}. One of the main uncertainties on the stellar models is the impact of thermally pulsating asymptotic giant branch stars. In total, systematics can induce mass inaccuracies of up to 0.3~dex \citep{conroy2013a}. In addition, due to the degeneracy between the metallicity, the age, and the attenuation, the choice of the metallicity can also have an impact on the estimates of the physical properties we present in Sect.~\ref{ssec:parameters-maps}. We evaluate these uncertainties in Appendix~\ref{sec:impact-metals}.
 \item The nebular emission, which includes both the recombination lines and the continuum (free--free, free--bound, and 2--photon processes) but does not include lines from photodissociation regions, is computed from dedicated CLOUDY \citep{ferland1998a} templates based on the models of \cite{inoue2011a}. The emission is directly proportional to the production rate of Lyman continuum photons, with the Lyman continuum escape fraction assumed to be negligible and the dust absorbing 10\% of Lyman photons. The ionisation parameter is assumed to be $10^{-2}$, which has an impact on line ratios but little impact on broadband fluxes.
 \item The stellar and nebular emissions are then attenuated using a power--law--modified starburst curve \citep{calzetti1994a,calzetti2000a}: $\mathrm{A(\lambda)=A(\lambda)_{SB}\times(\lambda/550~nm)^\delta}$, with $\delta$ the free parameter modifying the slope of the attenuation curve. We also include an ultraviolet bump with a variable strength ranging from no bump at all to a strong bump similar to that of the Milky Way. This modelling allows for a flexible shape for the attenuation law and also for a variable differential reddening between the emission due to stars younger and older than 10~Myr, following the model of \cite{charlot2000a}. This means that stellar populations of all ages, including old stars, contribute to dust heating and that stars older than 10~Myr are less attenuated than their younger counterparts.
 \item Finally, the energy absorbed by the dust is re--emitted in the IR using the latest \cite{dale2014a} templates. We assume a broad of range for $\alpha$, a ``parameter that represents the relative contributions of the different local SEDs'' \citep{dale2002a} and defined as $dM_d(U)\propto U^{-\alpha}dU$, with $M_d$ the dust mass and $U$ the radiation field intensity.
\end{itemize}
The full grid consists of a little over 80 million models. The list of the main parameters and their respective values is presented in Table~\ref{tab:cigale-parameters}.
\begin{table*}
 \centering
 \begin{tabularx}{\textwidth}{lX}
 \hline\hline
  Free parameters&Values\\\hline
  $e$--folding time of the old population&0.001, 1, 2, 3, 4, 5, 6, 7, 8 Gyr\\
  $e$--folding time of the latest episode of star formation&5, 10, 25, 50, 100 Myr\\
  Fraction of the stellar mass formed in the latest episode of star formation&0, 0.0001, 0.0005, 0.001, 0.005, 0.01, 0.05, 0.1, 0.25\\
  Age of the oldest stars&13 Gyr\\
  Age of the onset of the latest episode of star formation&5, 10, 25, 50, 100, 200, 350, 500 Myr\\\hline
  E(B$-$V) of the young population&0.01, 0.05, 0.10, 0.15, 0.20, 0.25, 0.30, 0.35, 0.40, 0.45, 0.50, 0.55, 0.60, 0.65, 0.70 mag\\
  Differential reddening&0.25, 0.50, 0.75\\
  $\delta$&$-0.5$, $-0.4$, $-0.3$, $-0.2$, $-0.1$, 0.0\\
  Ultraviolet bump strength&0 (no bump), 1, 2, 3 (Milky Way--like)\\\hline
  $\alpha$&1.5000, 1.5625, 1.6250, 1.6875, 1.7500, 1.8125,\\
          &1.8750, 1.9375, 2.0000, 2.2500, 2.5000, 2.7500,\\
          &3.0000, 3.2500, 3.5000, 3.7500, 4.0000\\\hline\hline
  Fixed parameters&Value\\\hline
  Metallicity&0.02\\\hline
  Ionisation parameter&$10^{-2}$\\
  Dust absorbed fraction&10\%\\
  Escape fraction&0\%\\\hline
 \end{tabularx}
 \caption{Main CIGALE parameters\label{tab:cigale-parameters}}
\end{table*}

\begin{figure}[!htbp]
 \includegraphics[width=\columnwidth]{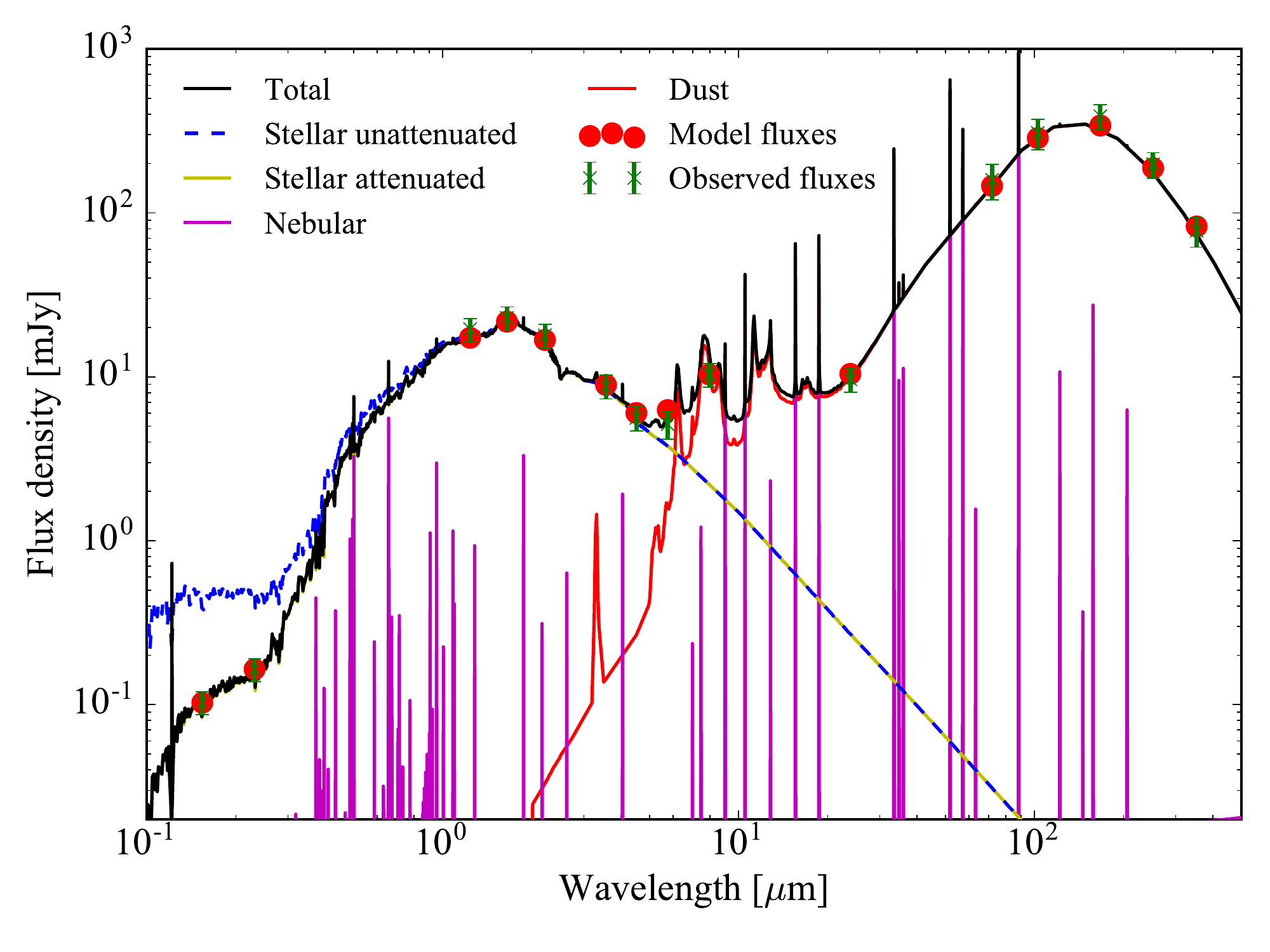}
\caption{Example of a typical best--fit (median reduced $\chi^2$ of the best--fits for the entire sample). This fit corresponds to a region to the East of NGC~4736. Different components are shown: unattenuated stellar emission (dashed blue line), attenuated stellar emission (yellow line), nebular emission (magenta line), and dust emission (red line). The model fluxes in broad bands are shown as red dots and the observed fluxes as green crosses with their corresponding uncertainties.\label{fig:typical-fit}}
\end{figure}

In the final step, we fit these models to the observed SED. For each of the output parameters, we compute the marginalised probability distribution functions (PDF) from the $\chi^2$ distribution. The estimated values of the parameters and their uncertainties are taken as the likelihood--weighted means and standard deviations. The precision and uncertainty of the evaluation of the physical properties is described in Appendix \ref{sec:model-uncertainties}. Note that as the different physical components contribute to part or all of the bands considered and as we fit all of them simultaneously, all components potentially have an effect on the determination of the physical properties. For instance, a change in the SFH will also have on effect on the stellar mass and on the attenuation, and therefore indirectly on the relative amount of dust heating by younger and older stars. An example of a best--fit with the stellar, nebular, and dust components is shown in Fig.~\ref{fig:typical-fit}.

\subsection{Parameters maps\label{ssec:parameters-maps}}

\begin{figure*}[!htbp]
 \includegraphics[width=.495\textwidth]{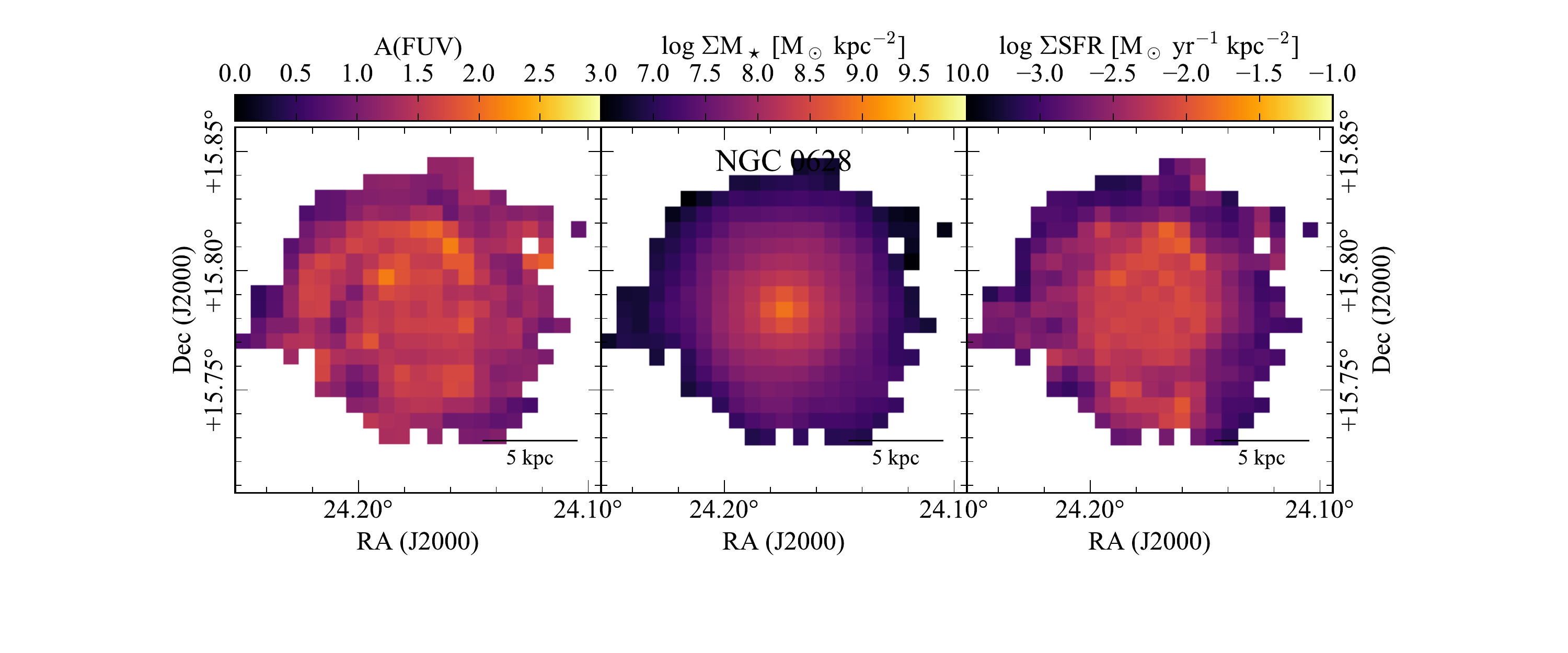}
 \includegraphics[width=.495\textwidth]{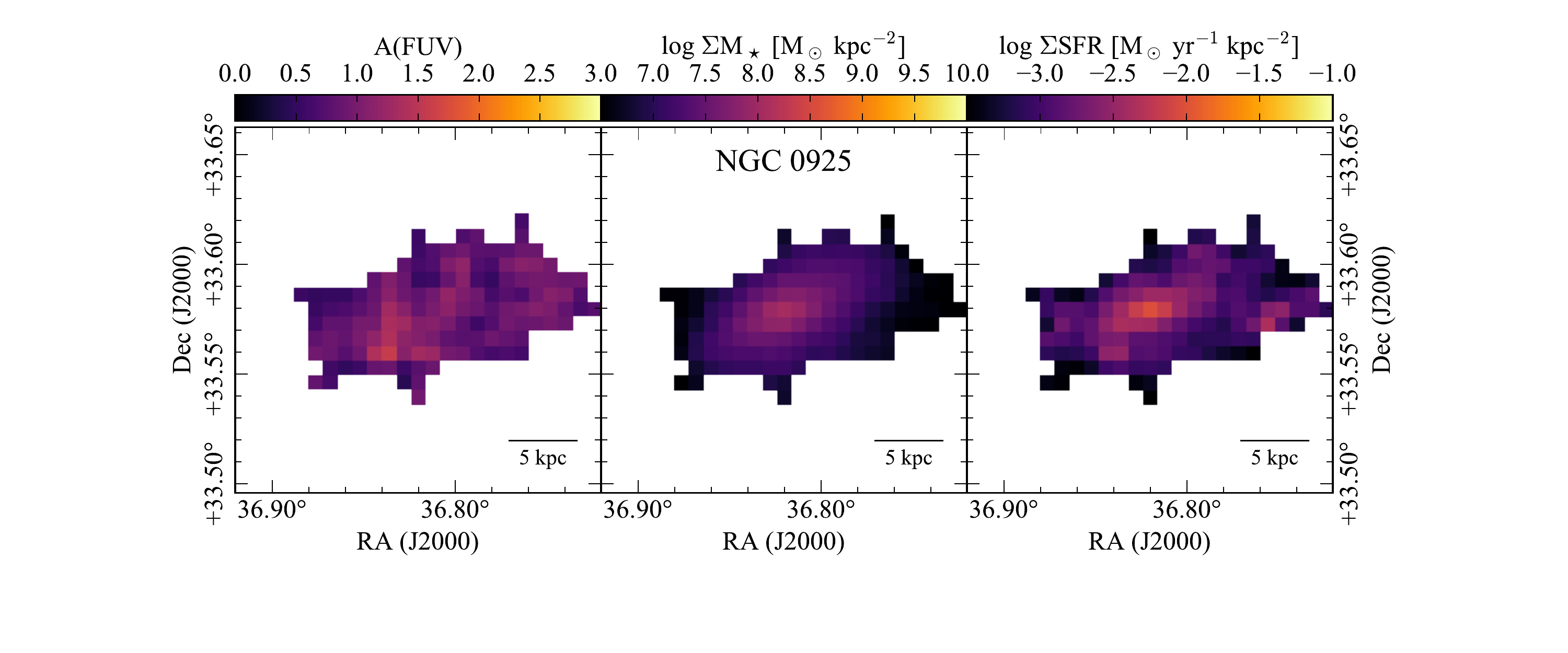}
 \includegraphics[width=.495\textwidth]{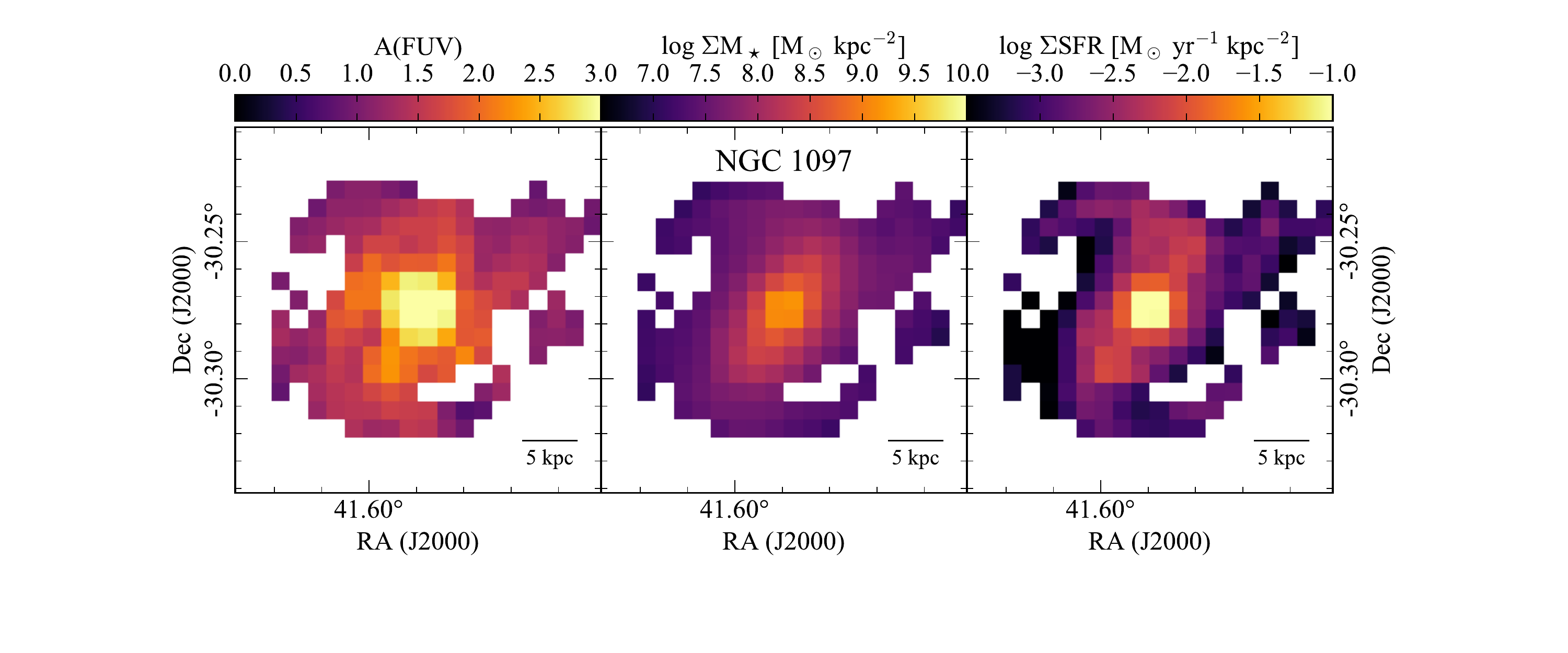}
 \includegraphics[width=.495\textwidth]{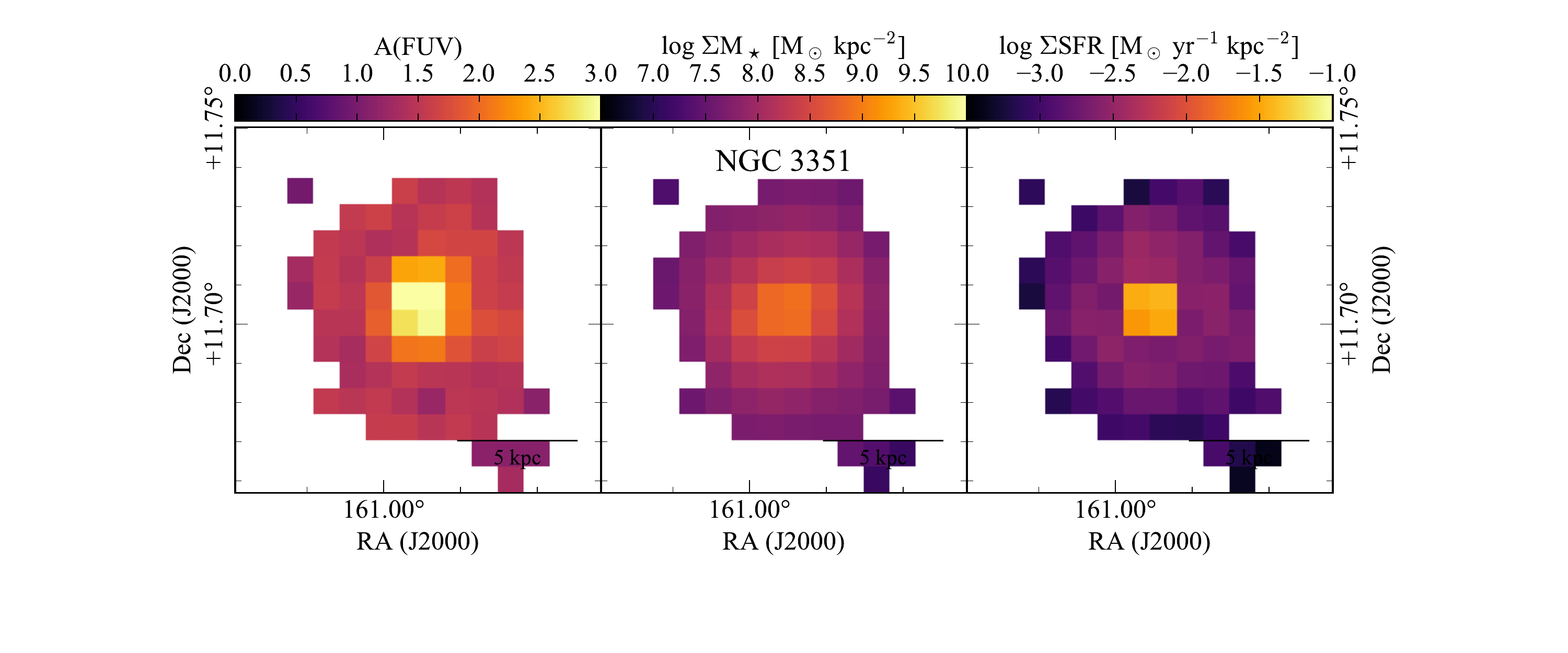}
 \includegraphics[width=.495\textwidth]{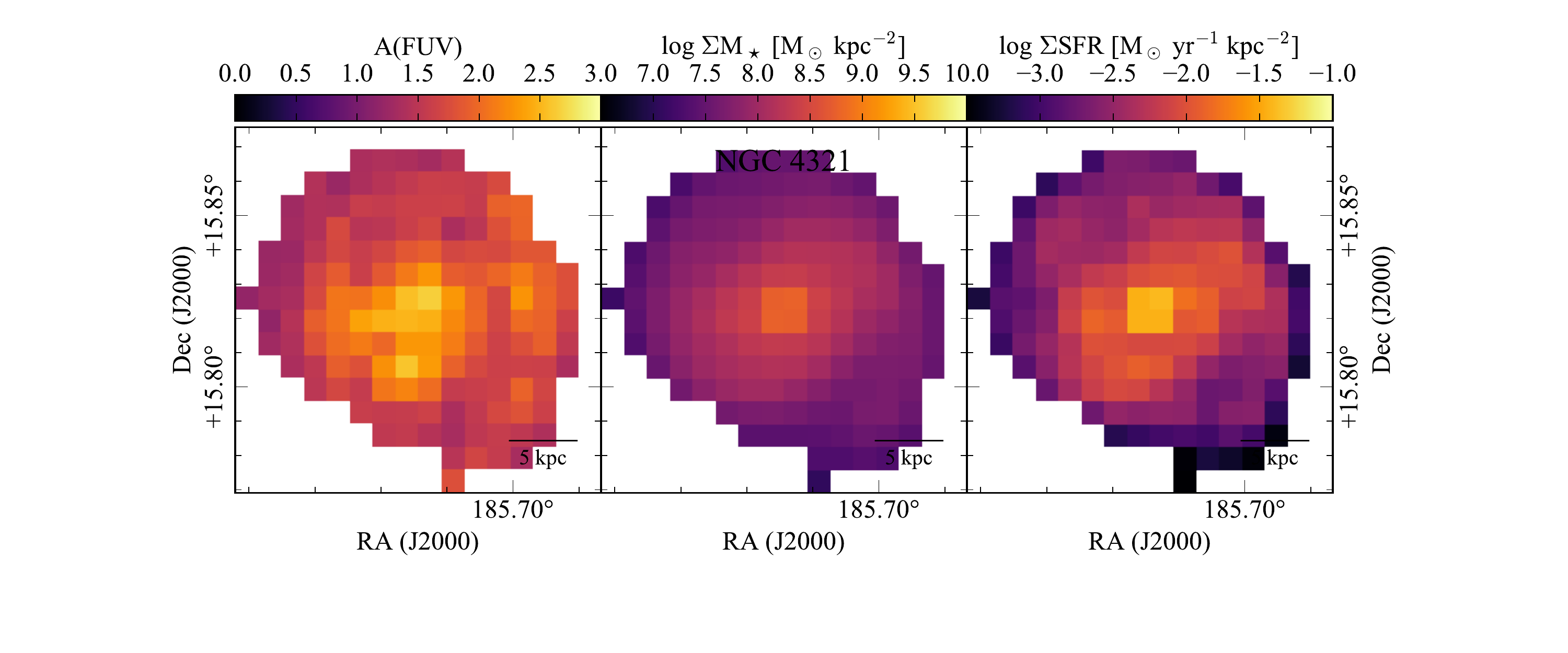}
 \includegraphics[width=.495\textwidth]{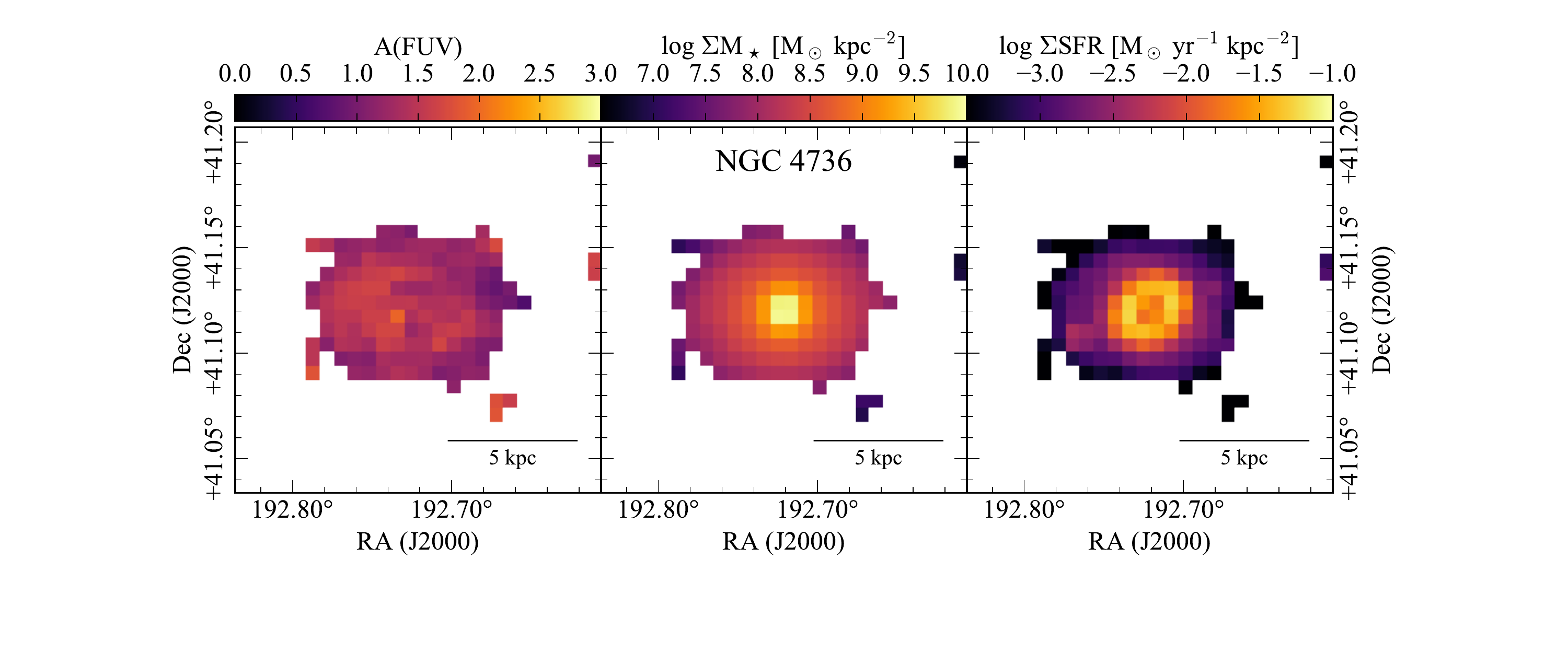}
 \includegraphics[width=.495\textwidth]{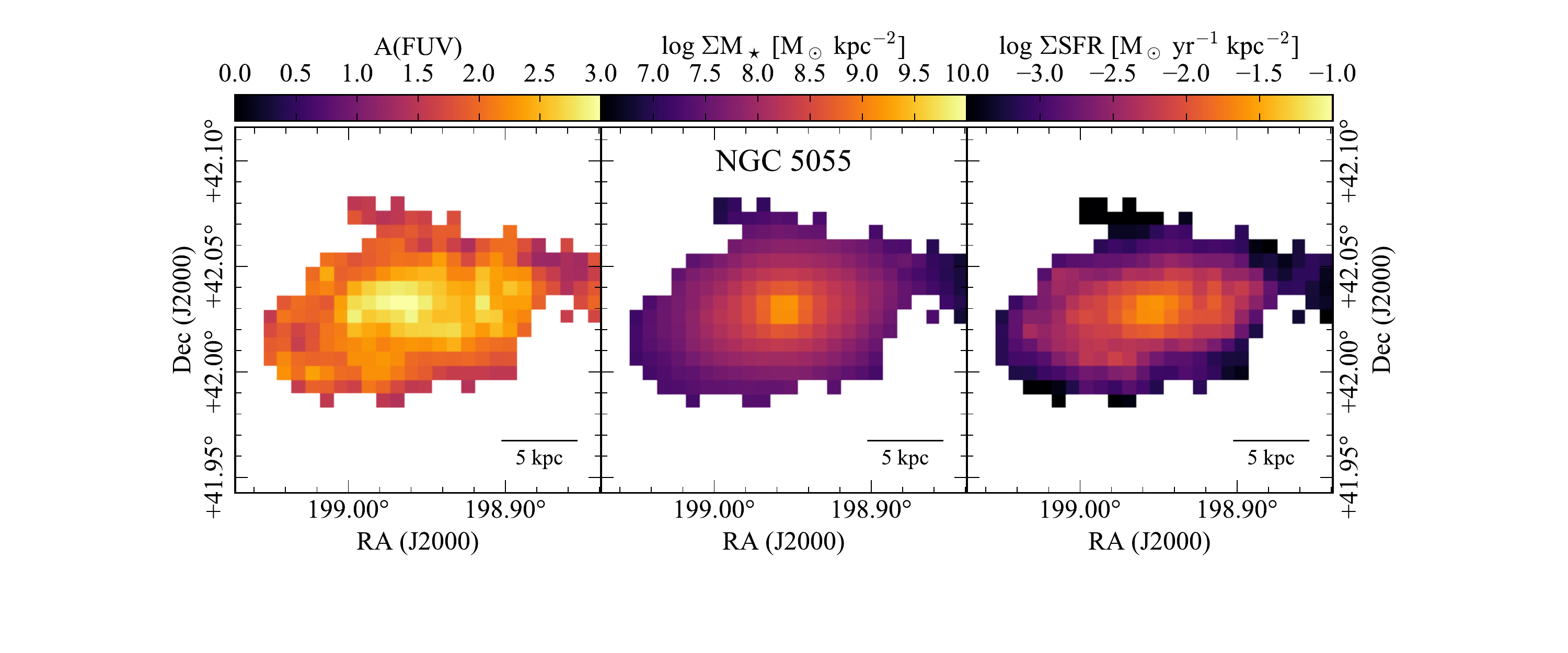}
 \includegraphics[width=.495\textwidth]{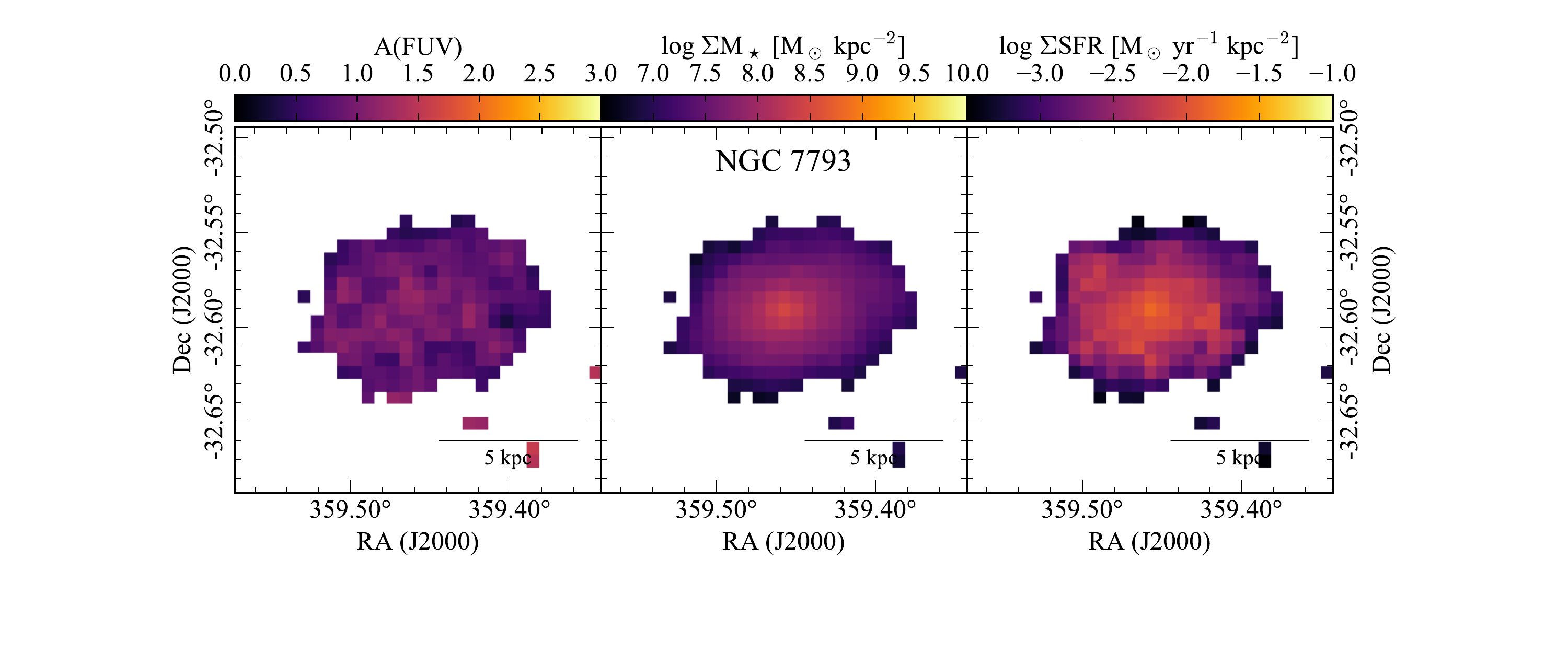}\\
\caption{FUV attenuation (left), stellar mass surface density (centre), and the current SFR surface density (right) estimated with CIGALE for all the galaxies in the sample. All surface densities are corrected for the inclination. For easier comparison between each galaxy, the colour scale is fixed across all objects.\label{fig:maps-parameters}}
\end{figure*}

For each galaxy in the sample, we show in Fig.~\ref{fig:maps-parameters} the maps of three of the main physical properties estimated by CIGALE: the FUV attenuation, the stellar mass surface density, and the current SFR surface density.

First of all, the FUV attenuation shows a broad range of behaviours. Galaxies such as NGC~925 and NGC~7793, which are the two lowest metallicity galaxies in the sample,
have consistently low attenuations across the disk. At the opposite end, NGC~5055 exhibits high values of the attenuation on a large fraction of the disk. Conversely, NGC~1097 and NGC~3351 have a very strong attenuation in the inner regions but more moderate values further out in the disk. The higher attenuation is probably due to the high concentration of molecular gas and dust in the central parts of these two well--known barred galaxies.

The stellar mass maps are markedly different from attenuation maps. We see that all galaxies have a very smooth, radially declining stellar mass profile. This shows the ability of the model to recover older stellar populations even in regions of intense star formation where they could be outshined by younger stellar populations at some wavelengths.

Finally, the SFR maps also present interesting features. There are broadly decreasing radial gradients in the SFR density. The nuclear starbursts in NGC~1097 and NGC~3351 present very strong peaks compared to the typical star formation across the rest of the disk. Such nuclear starbursts are expected due to the central gas concentration in barred galaxies \citep[e.g.,][]{tabatabaei2013a}. A strong peak is also present in NGC~4321 and NGC~5055. Conversely NGC~628, NGC~925, and NGC~7793 have much shallower radial gradients. Of particular interest, NGC~4736 has a strong star--forming ring that shows prominently on the map.

\section{Estimation of $k_i$ and dependence on the local physical parameters\label{sec:kIR-dependence-parameters}}

The estimation of the FUV attenuation with CIGALE allows us to compute $k_i$ for each pixel and for each IR band. This easily comes from Eq.~\ref{eqn:k-IR-v1}:
\begin{equation}
 k_i=\frac{L\left(UV\right)_{obs}}{L\left(IR\right)_i}\times\left[10^{A\left(UV\right)/2.5}-1\right].\label{eqn:ki-auv}
\end{equation}
This means that for individual IR bands, the derivation of $k_i$ relies on the modelling only for the determination of the UV attenuation (see Appendix \ref{sec:comp-FUV} why SED modelling is required to estimate the UV attenuation). The UV and the IR luminosities are directly obtained from the observations. In the case of the TIR only, $k_i$ also depends on the estimation of the TIR luminosity by the model, a very securely determined quantity as shown in Appendix \ref{sec:model-uncertainties}.

\subsection{Distribution of $k_i$ in galaxies\label{ssec:distrib-ki}}

\begin{figure*}[!htbp]
 \includegraphics[width=.49\textwidth]{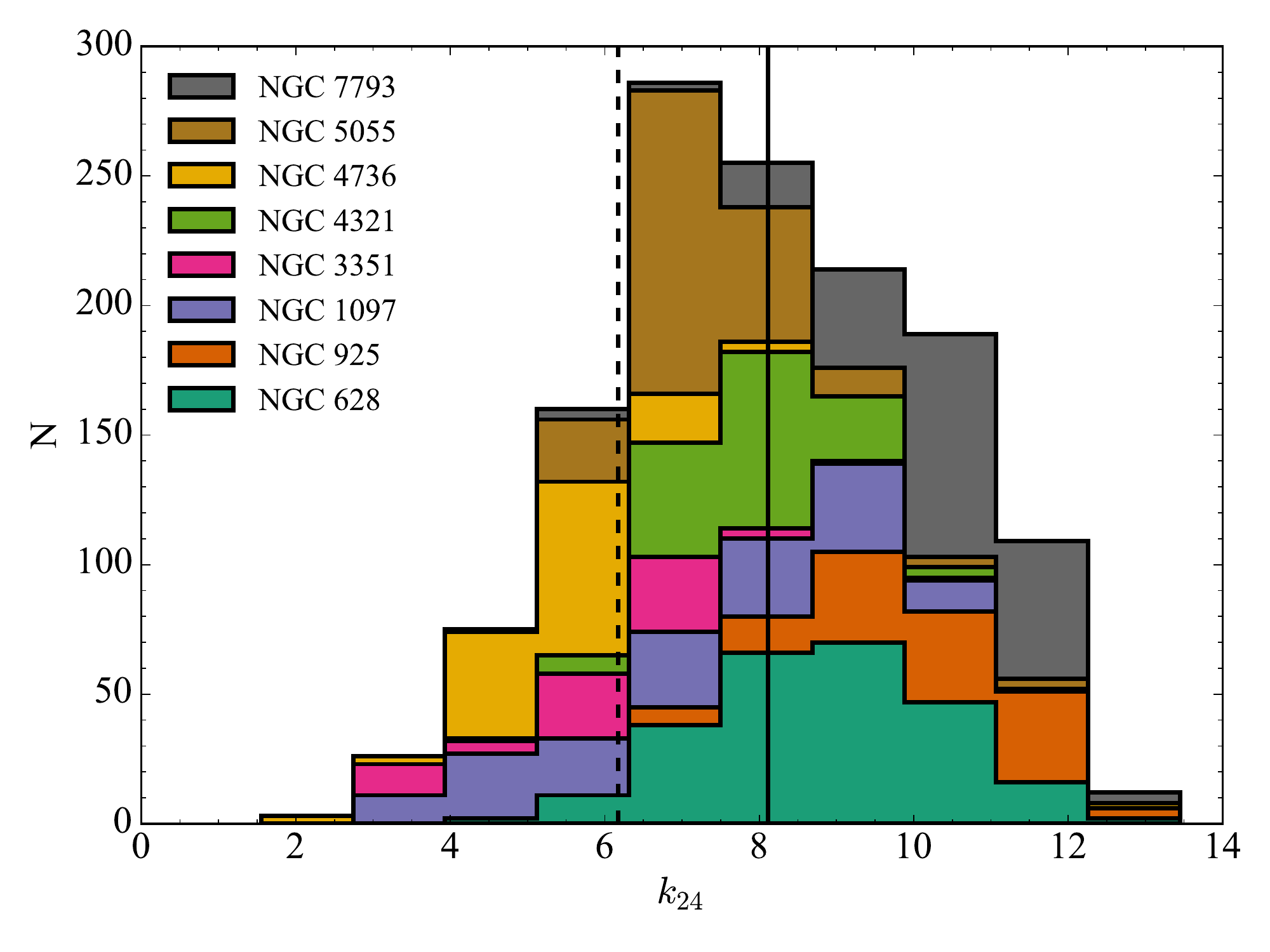}
 \includegraphics[width=.49\textwidth]{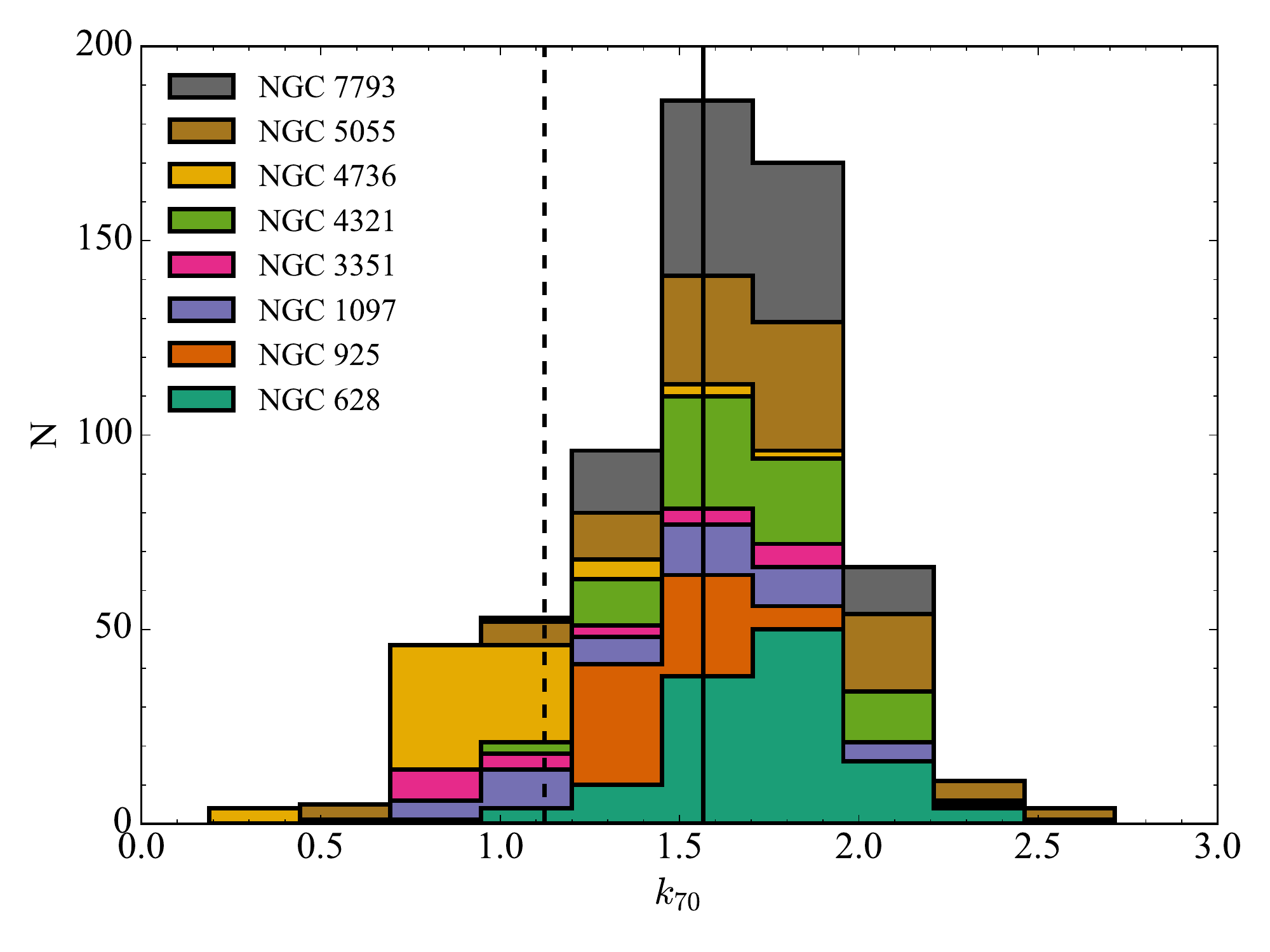}\\
 \includegraphics[width=.49\textwidth]{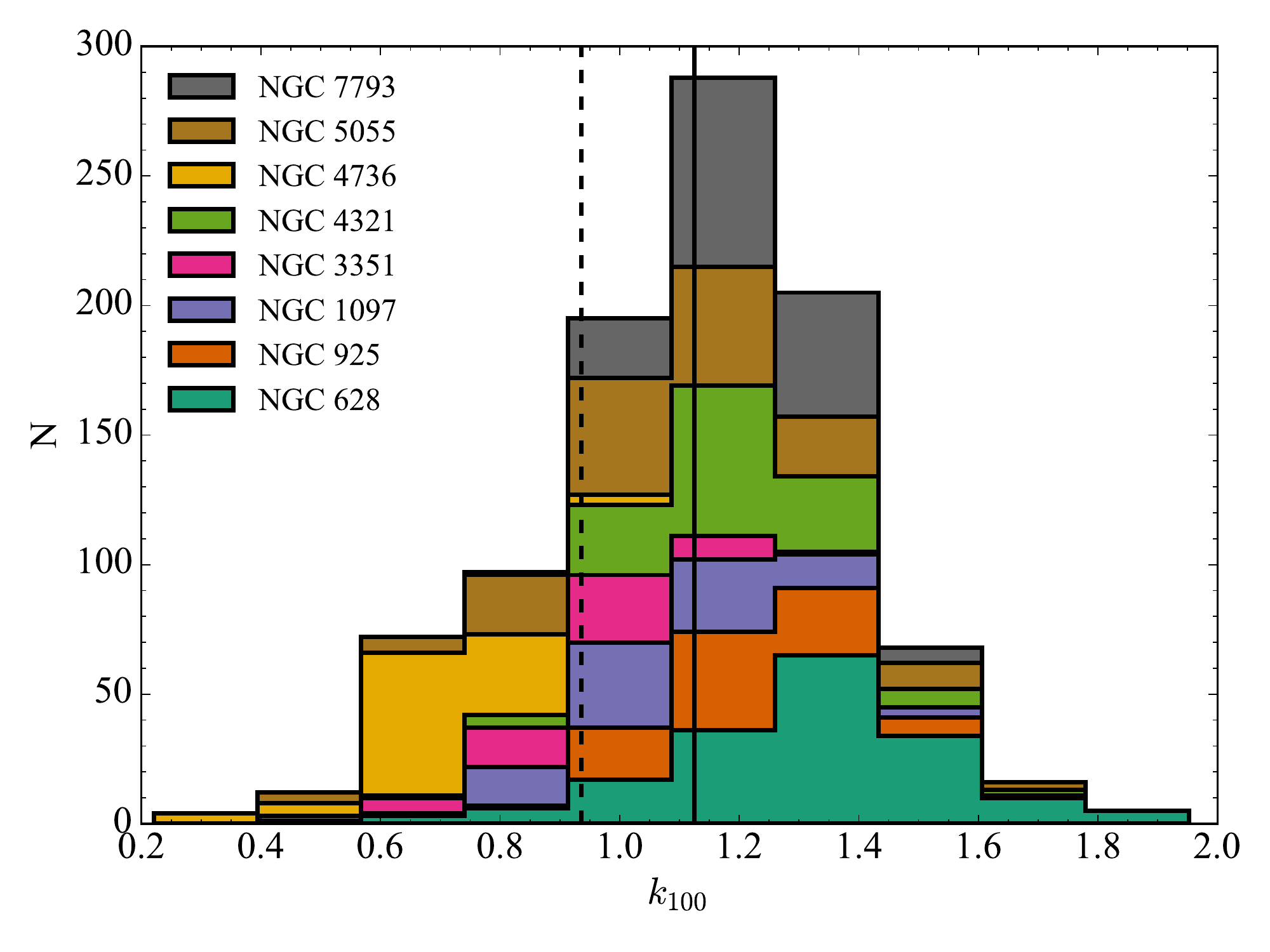}
 \includegraphics[width=.49\textwidth]{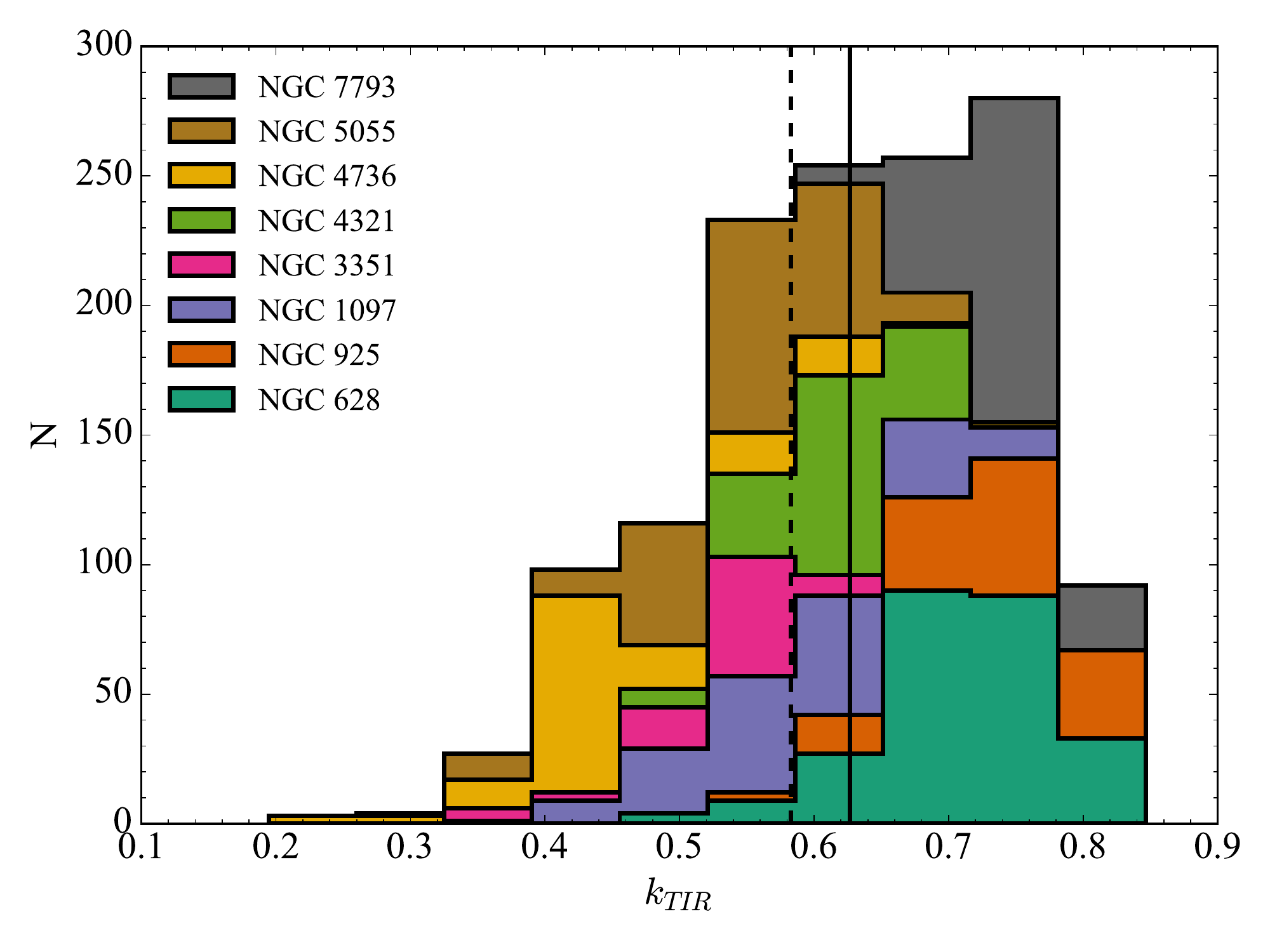}
 \caption{Stacked distributions of $k_i$ from 24~$\mu$m (top left) to 100~$\mu$m and the TIR (bottom right) for NGC~628, NGC~925, NGC~1097, NGC~3351 NGC~4321, NGC~4736, NGC~5055, and NGC~7793 from blue (bottom distribution) to red (top distribution). The solid black line indicates the mean value for the sample whereas the dashed black line indicates the luminosity--weighted mean value. We have selected only regions detected at least at a 5--$\sigma$ level in the relevant band. This affects mostly the 70~$\mu$m and 100~$\mu$m bands.\label{fig:hist-k}}
\end{figure*}

In Fig.~\ref{fig:hist-k} we present the distributions of $k_i$ for all the galaxies in the sample at 24~$\mu$m, 70~$\mu$m, and 100~$\mu$m, in addition to the TIR. We drop bands at longer wavelengths as it has been shown that their emission is mainly driven by dust heated by the old stellar populations \citep{bendo2010a,bendo2012a,bendo2015a,boquien2011a,lu2014a}.
The first important aspect to note is the broad width of the distributions. For instance, as we can see in Table~\ref{tab:kIR}, $k_{24}$ ranges from 1.55 to 13.45 ($\left<k_{24}\right>=8.11\pm 2.10$), so a variation of over a factor 8, and $k_{TIR}$ ranges from 0.19 to 0.85 ($\left<k_{TIR}\right>=0.63\pm0.12$) so a variation of over a factor 4.
\begin{table*}
\centering
\begin{tabular}{cccccccc}
\hline\hline
Band&min($k_i$)&max($k_i$)&$\left<k_i\right>$&$\left<k_i\right>_{L_i}$&$k_i$ \citep{hao2011a}&$k_i$ \citep{liu2011a}\\\hline
24  &1.55      &13.45     &8.11$\pm$2.10     &6.17$\pm$2.17           &3.89$\pm$0.15         &6.0\\
70  &0.19      &2.71      &1.57$\pm$0.39     &1.12$\pm$0.50           &                      &\\
100 &0.22      &1.95      &1.12$\pm$0.25     &0.94$\pm$0.29           &                      &\\
TIR &0.19      &0.85      &0.63$\pm$0.12     &0.58$\pm$0.12           &0.46$\pm$0.12         &\\\hline
\end{tabular}
\caption{From left to right columns: minimum, maximum, mean, and luminosity--weighted mean of $k_i$ at 24~$\mu$m, 70~$\mu$m, 100~$\mu$m, and for the TIR.\label{tab:kIR}}
\end{table*}

In order to test whether the $k_i$ values from different galaxies are drawn from the same underlying distribution or not, we have carried out Kolmogorov--Smirnov tests. We find that except for a few galaxy pairs at each wavelength that have $p \ge 0.01$ (with $p$ the probability that the values have been drawn from the same underlying distribution), the samples appear to have been drawn from different distributions. This alone already shows that there must be important physical differences driving the variation of $k_i$ between each of those galaxies.

A first indication of the physical origin of these variations can be obtained if we consider carefully the $k_{24}$ and $k_{TIR}$ distributions in relation to the morphological type of the galaxies. The earlier type a galaxy is, the lower $k_i$ should be because the fraction of diffuse emission should be higher in such galaxies \citep{sauvage1992a}. It is therefore not surprising that NGC~4736, the earliest type galaxy in the sample, shows some of the lowest $k_i$. At the other end of the Hubble sequence of spiral galaxies, NGC~925 and NGC~7793 are respectively of SABd and SAd types. We see in Fig.~\ref{fig:hist-k} that they indeed have among the highest values of $k_i$. This shows that the diversity in galaxy types likely contributes to the differences in the distribution of $k_i$. Our results here show the possible dangers of blindly applying estimators to individual galaxies and regions within galaxies if they have markedly different intrinsic physical properties compared to the calibration samples of the estimators.

Related to the first point, when we consider our entire sample we find, as mentioned earlier, that $\left<k_{24}\right>=8.11\pm 2.10$ and $\left<k_{TIR}\right>=0.58\pm0.12$. This is higher than the values determined by \cite{hao2011a} ($\left<k_{25}\right>=3.89\pm0.15$\footnote{While defined from the IRAS 25~$\mu$m band rather than from the Spitzer 24~$\mu$m band, \cite{kennicutt2009a} showed that $L(24~\mu m)/L(25~\mu m)=0.98\pm0.06$. We therefore make no distinction between those two bands.} and $\left<k_{TIR}\right>=0.46\pm0.12$ for the TIR) and \cite{liu2011a} ($k_{24}=6.0$). To compare more readily with values derived for entire galaxies, we have also computed the luminosity--weighted means: $\left<k_{24}\right>_{L(IR)}=6.17\pm2.17$ and $\left<k_{TIR}\right>_{L(IR)}=0.58\pm0.12$. These values are closer the results of \cite{hao2011a} and \cite{liu2011a}. This shows that at such scales, the global value for each galaxy is driven by the most luminous regions. Nevertheless, applying a variable $k_i$ has a clear impact on the integrated dust--corrected FUV luminosity. For instance, adopting $k_{24}=3.89$ ($k_{24}=6.0$) can lead to galaxy--integrated dust--corrected FUV luminosities lower (higher) by 38\% (40\%) when compared to luminosities computed with a variable $k_{24}$. Deviations can be much larger locally. There are also important variations from one galaxy to another, showing the necessity to have a $k_i$ adapted for each galaxy. The actual impact of using a potentially inadequate $k_i$ naturally depends on the fraction of UV photons absorbed by dust. The lower the attenuation, the less the impact an error on $k_i$ has. We will see in Sect.~\ref{sec:kir-unresolved} how we can relate locally derived $k_i$ to global values adapted to unresolved objects.

\begin{figure*}[!htbp]
 \includegraphics[width=.495\textwidth]{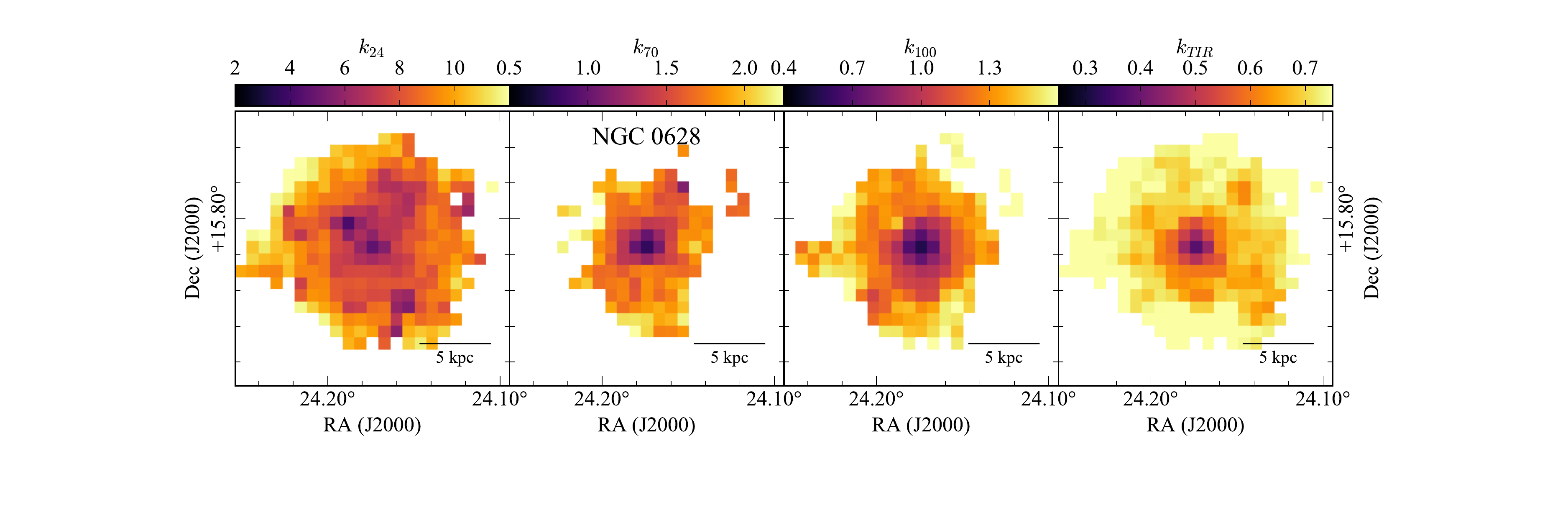}
 \includegraphics[width=.495\textwidth]{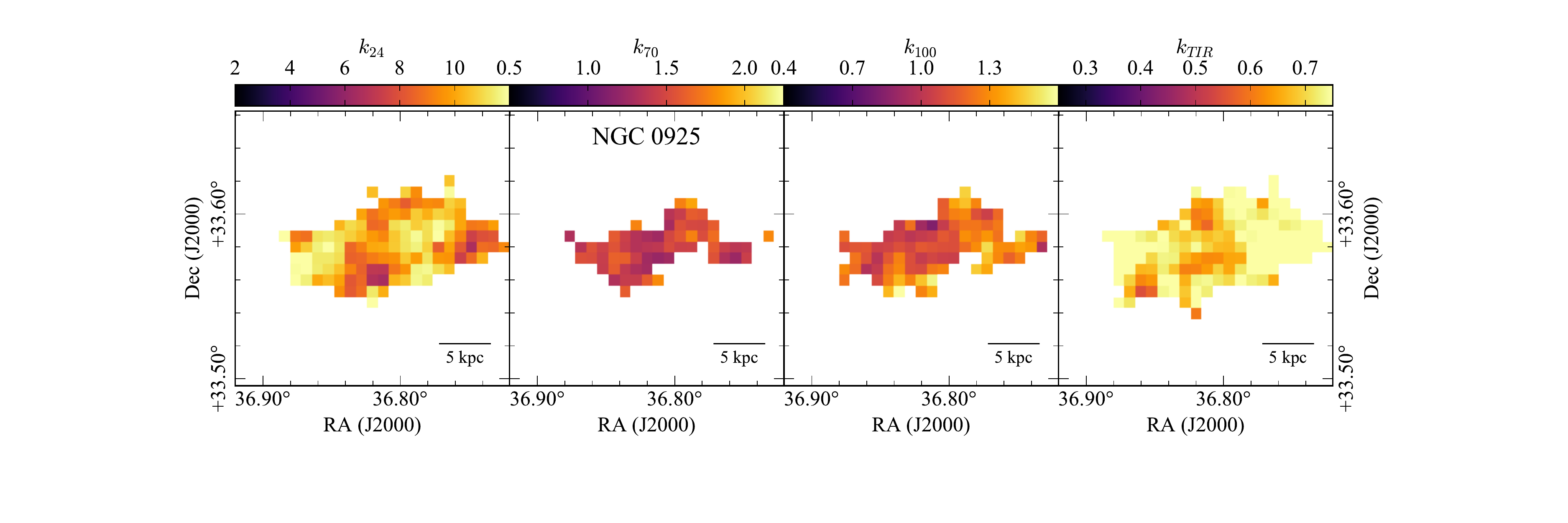}
 \includegraphics[width=.495\textwidth]{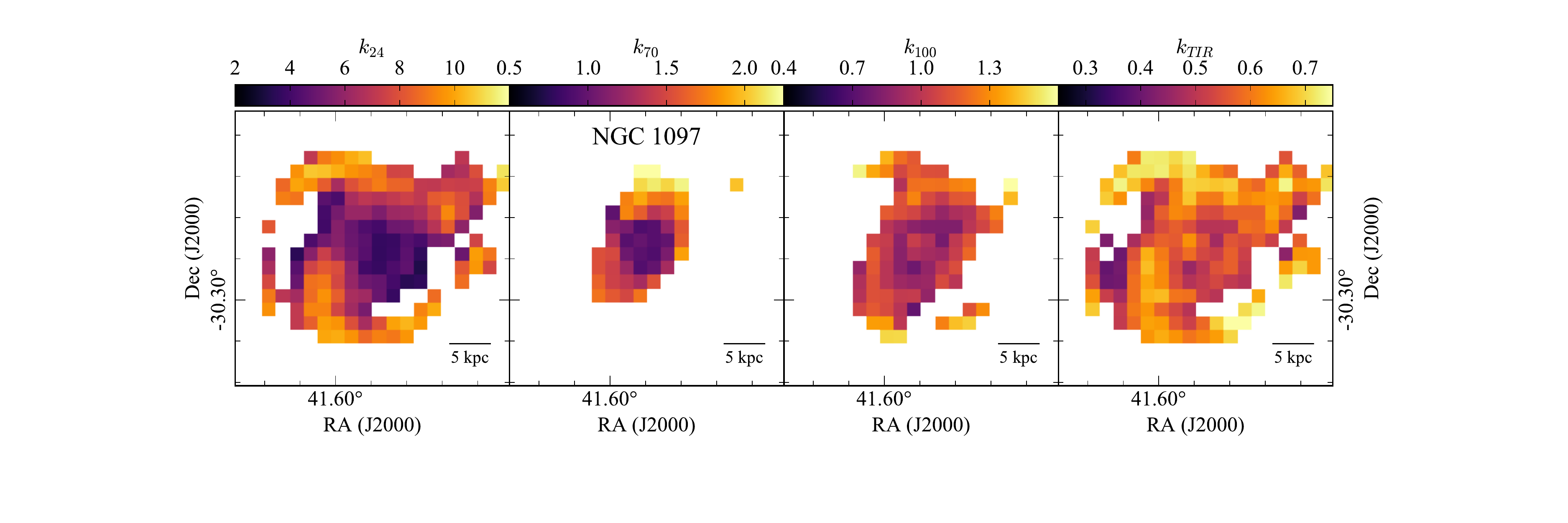}
 \includegraphics[width=.495\textwidth]{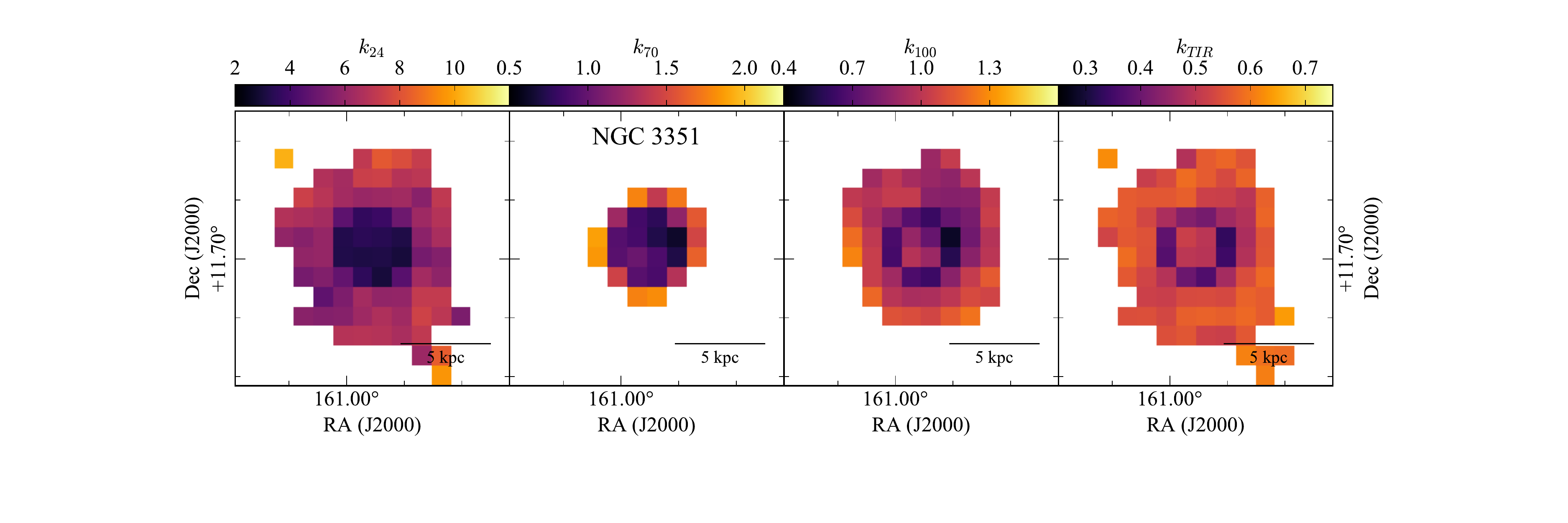}
 \includegraphics[width=.495\textwidth]{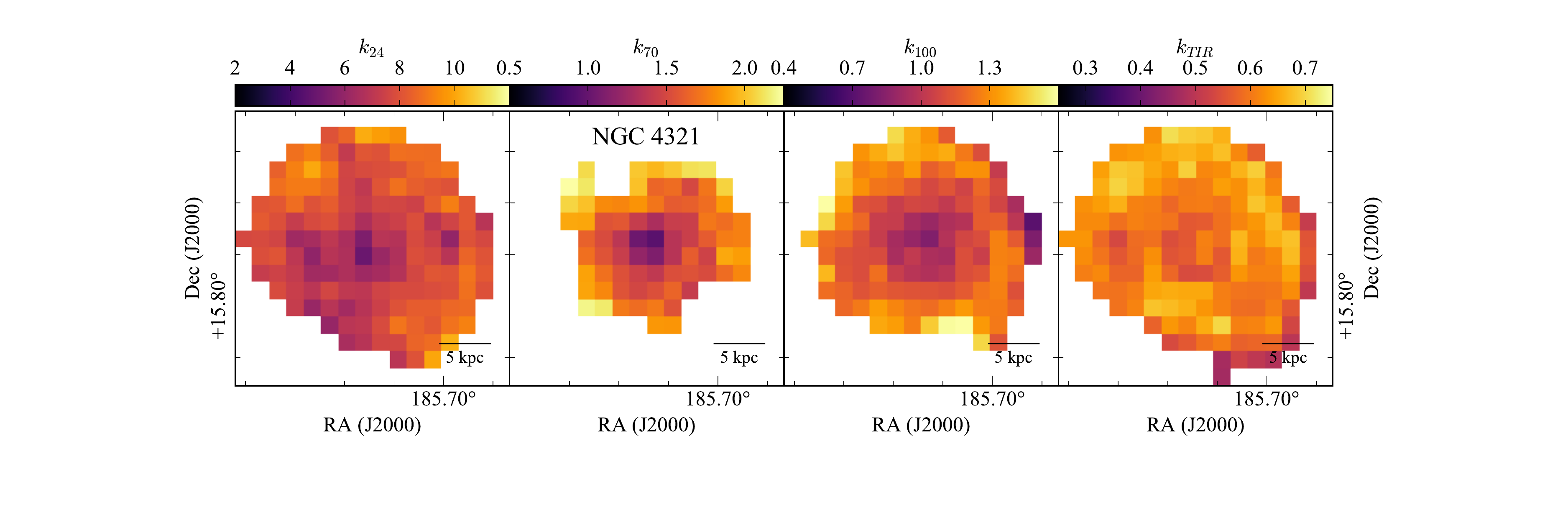}
 \includegraphics[width=.495\textwidth]{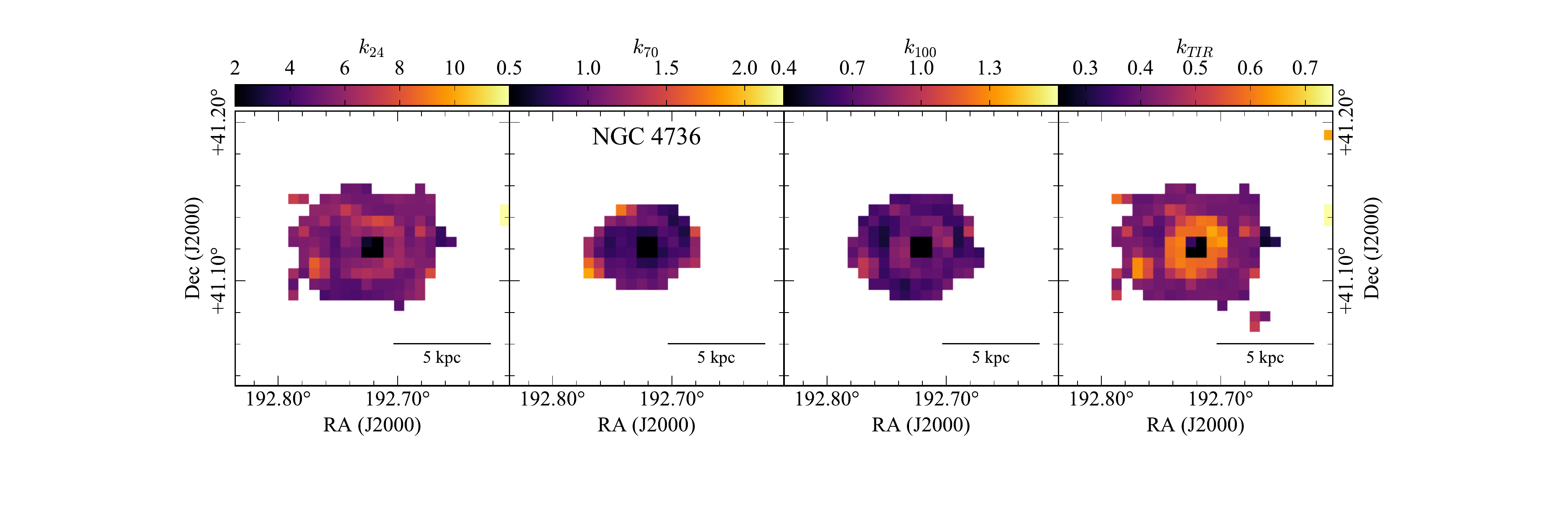}
 \includegraphics[width=.495\textwidth]{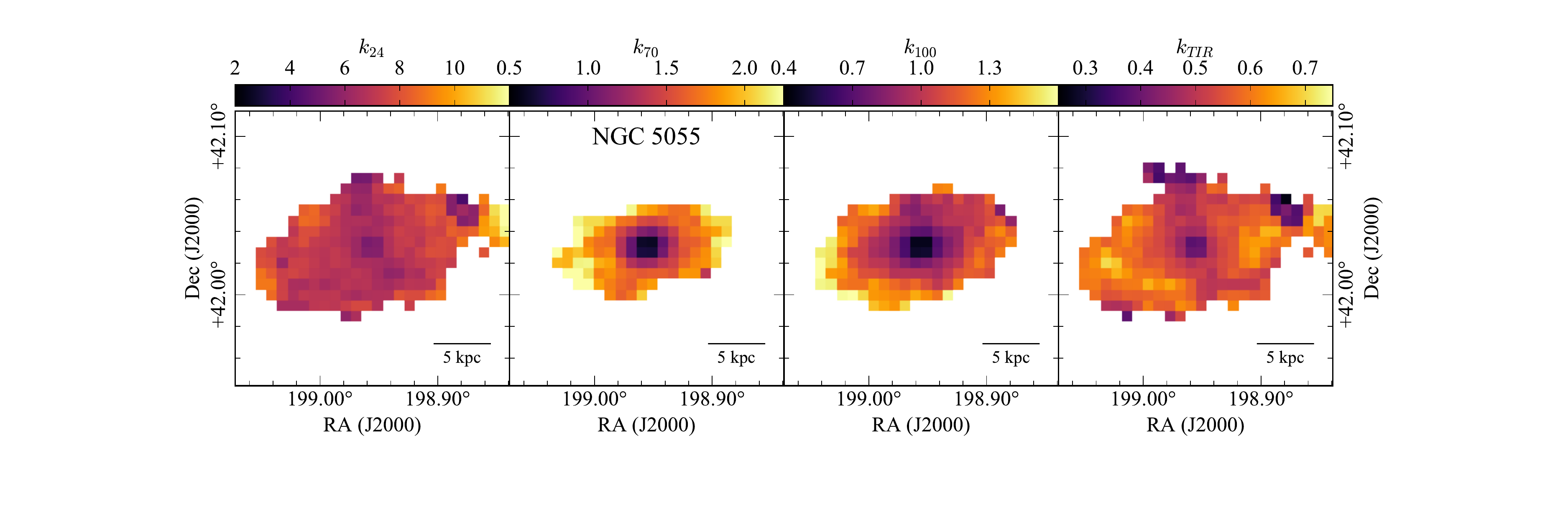}
 \includegraphics[width=.495\textwidth]{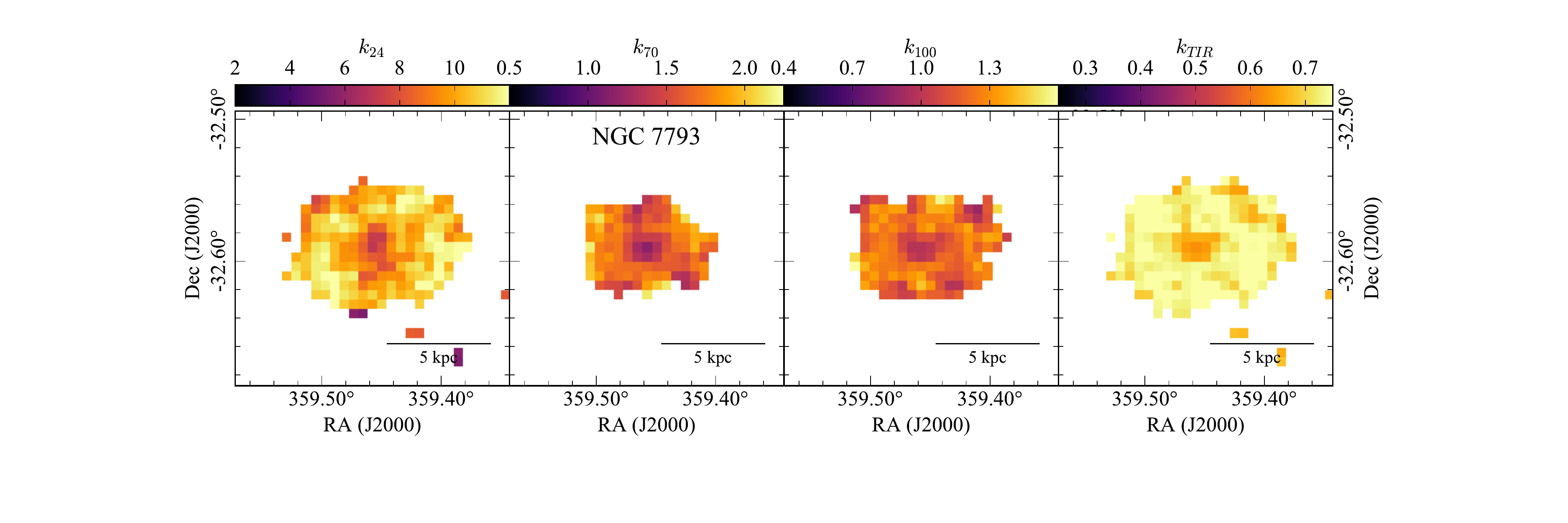}\\
\caption{Maps of $k_i$ \label{fig:maps-k} at 24~$\mu$m, 70~$\mu$m, 100~$\mu$m, and for the TIR, from left to right, for each of the galaxies in the sample. Only pixels detected at least at a 5--$\sigma$ level in the corresponding band are shown here.}
\end{figure*}

We can complement this first insight by examining the maps of $k_i$ at 24~$\mu$m, 70~$\mu$m, 100~$\mu$m, and for the TIR that we present in Fig.~\ref{fig:maps-k}.
The distribution of $k_i$ shows in general a broad gradient, with larger $k_i$ in the outer parts of the galaxies. Some complex structures can also be seen with variations from one band to the other, probably translating variations of the relative influence of the different the dust heating sources. The structures also appear to vary with wavelength translating the different modes of dust heating for dust emitting at different wavelengths. The variations in the SNR with wavelength do not allow however for a comparison over the full extent of each galaxy. While the lower SNR in some bands makes difficult the study of $k_i$ in outer regions, this has a limited impact on the computation of the SFR in such regions. Indeed, most of star formation will be visible in the UV and only a small fraction in the IR. Therefore even a large uncertainty for $k_i$ will translate into a lower uncertainty on the attenuation--corrected UV and thus on the SFR. The pixel--by--pixel study of inner regions still allows us to gain considerable information on $k_i$ in the most interesting range for hybrid estimators.

\subsection{Variation of $k_i$ with physical properties\label{ssec:variations-kIR}}

\begin{figure*}[!htbp]
 \includegraphics[width=\textwidth]{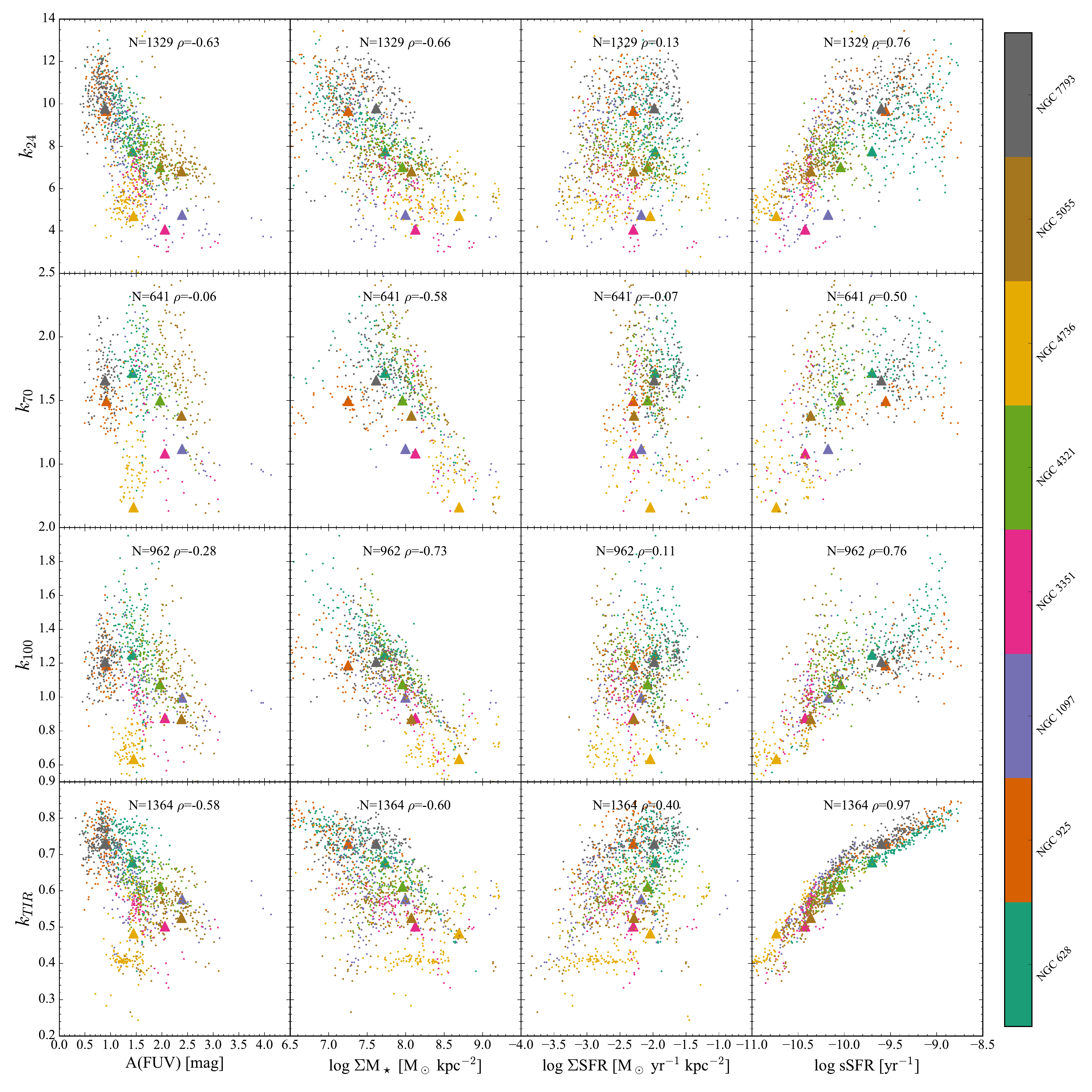}
 \caption{$k_i$ versus the FUV attenuation (left), the stellar mass surface density (centre--left), the SFR surface density averaged over 100~Myr (centre--right), and the sSFR averaged over 100~Myr (right) for all the galaxies in the sample. Each dot represents one pixel and the colour indicates the galaxy following the colour bar on the right. We have selected only regions detected at least at a 5--$\sigma$ level in the relevant band. This affects mostly the 70~$\mu$m and 100~$\mu$m bands as they are the shallowest. The number of regions and the Spearman's rank correlation coefficient are shown at the top of each plot. Finally, triangles represent the global value for each galaxy in the sample, taking into account all pixels detected at least at a 1--$\sigma$ level in all bands. \label{fig:k-vs-params}}
\end{figure*}

As we have just seen, there are strong variations of $k_i$ within and between galaxies. To identify the physical origin of these variations, in Fig.~\ref{fig:k-vs-params} we plot $k_i$ versus the FUV attenuation, the stellar mass surface density, the SFR surface density, and the sSFR, the latter two averaged over the past 100~Myr.
A first inspection already provides us with interesting information beyond the systematic galaxy--to--galaxy differences in $k_i$ we found in Sect.~\ref{ssec:distrib-ki}:

\begin{enumerate}
 \item there is very little to no variation of $k_i$ as a function of either the FUV attenuation and the SFR surface density,
 \item there is a decrease of $k_i$ with an increase of the stellar mass surface density,
 \item there is a clear increase of $k_i$ with the sSFR.
\end{enumerate}

Separating the respective impact of the different physical properties considered here is difficult as they are not independent from one another. For instance, as we can see in Fig.~\ref{fig:maps-parameters}, inner regions in our sample will tend to have higher attenuations, higher stellar mass surface densities, and they will also tend to have more star formation than outer regions. This means that even if the variation of $k_i$ is exclusively due to one of these physical properties, in some cases some correlations may also appear with other physical properties. We concentrate here on the relations between $k_i$, the stellar mass surface density, and the sSFR.

Let us first examine in detail the variations of $k_i$ with the stellar mass surface density. There is a clearly defined decreasing trend of $k_i$ with increasing stellar mass surface densities ($-0.73\le\rho\le-0.58$, with $\rho$ Spearman's rank correlation coefficient). This trend suggests that evolved stellar populations, which contribute the bulk of the total stellar mass, may play an important role in governing $k_i$. This role is probably not exclusive though. If we consider that these populations are responsible at all wavelengths for the diffuse emission, then for a fixed SFR we can expect that at low (high) stellar mass surface density a larger (smaller) fraction of the IR emission will be due to UV--emitting stellar populations. If this is indeed the case, there should be an excellent correlation between $k_i$ and the sSFR. Even though there are important uncertainties on the average SFR over 100~Myr, we see a positive trend between the average sSFR over 100~Myr and $k_i$ ($0.50\le\rho\le0.97$).

To further understand the origin of the relations of $k_i$ with the different physical properties, we need to distinguish IR bands based on whether their emission is proportional or not to the incident radiation field intensity. This is the case for the TIR and the 24~$\mu$m band. The reason is obvious for the TIR, which by definition measures the total dust--absorbed luminosity. The 24~$\mu$m also behaves similarly because it is dominated by stochastic heating of very small grains \citep[see in particular Fig.~15 from][note however that the contamination by the emission from grains at the equilibrium can become significant for large radiation field intensities, which would render this approximation invalid]{draine2007a}. This linear relation between the absorbed and emitted luminosities means that if, for instance, 50\% of the absorbed luminosity originates from old stellar populations, correspondingly 50\% of the emission can also be attributed to old stellar populations. In consequence, the fraction of IR emission due to star formation can be indirectly estimated from the sSFR. This explains the excellent correlation between $k_i$ and the sSFR at 24~$\mu$m and for the TIR (top--right and bottom--right panels of Fig.~\ref{fig:k-vs-params}). The relation between $k_i$ and the stellar mass surface density appears however much weaker. This is expected as this only probes the radiation field from old stars and therefore only gives a weak and indirect information on the fraction of the emission due to young stars.

Conversely, the blackbody emission that dominates the 70~$\mu$m and 100~$\mu$m bands is in general not linearly proportional to the incident radiation field intensity. As the radiation field intensity increases (decreases), the emission peaks at shorter (longer) wavelengths following Wien's displacement law. This induces a super--linear or a sub--linear variation of the emission with respect to the radiation field intensity, depending on whether the band probes the Wien or the Rayleigh--Jeans tail of the blackbody. This means that unlike for the 24~$\mu$m and for TIR, the sSFR alone is not sufficient to estimate the fraction of the emission that is related to star formation. This is probably because, regions with the same sSFR can have different radiation field intensities, and therefore different dust temperatures and different $k_i$. However as suggested by the higher Spearman's rank correlation coefficient between $k_{100}$ and the sSFR, the issue is not as severe at 100~$\mu$m as it is at 70~$\mu$m. This is likely due to its closer location to the peak of the blackbody emission in terms of luminosity. As the absolute value of the derivative of the blackbody emission with respect to wavelength is smaller near the luminosity peak, it somewhat weakens the dependence on the temperature and therefore it makes the sSFR a better estimator of $k_{100}$ than for $k_{70}$ as the emission is more directly linked to the radiation field intensity. These differences from one band to the other and the contrast with the 24~$\mu$m band and the TIR illustrate the complexity to disentangle the emission due to star formation and old stellar populations for bands that are sensitive to the emission of dust at thermal equilibrium.

\section{Recipes to correct the FUV for the attenuation\label{sec:hybrid-relations}}

Following our results, a powerful approach to correct the UV for the attenuation by dust with hybrid estimators would be to parametrise the variation of $k_i$ against the sSFR or the stellar mass surface density. However, such a parametrisation would actually be difficult to put in practice. Parametrisations involving the sSFR would imply that we somehow already know the intrinsic FUV emission, defeating the purpose of attenuation correction. As for the stellar mass, its determination is somewhat dependent on the assumed model of stellar populations. With improvements on models, a parametrisation on the stellar mass would become invalid if current models prove to have systematic biases. Therefore, rather than relying on derived physical properties, we attempt to parametrise $k_i$ directly on observed quantities that are good tracers of the sSFR (FUV$-$NIR colours, Sect.~\ref{ssec:param-colours}) and of the stellar mass surface density (luminosity densities per unit area, Sect.~\ref{ssec:param-NIR}). This approach has the advantage of being purely observational, not relying on the sometimes strong assumptions behind empirical sSFR estimators, and not requiring a full--fledged SED modelling to compute the SFR, which may not be possible for lack of appropriate data or experience.

\subsection{Observed colours as a parametrisation for $k_i$\label{ssec:param-colours}}

One approach to parametrise $k_i$ is to estimate the sSFR of a galaxy is through its colour. For instance \cite{salim2005a} showed that the $\mathrm{NUV}-r$ colour of a galaxy is tightly linked to its birthrate parameter (the ratio of the average SFR over the last 100~Myr to the lifetime average SFR), a quantity closely related to the sSFR. This means that in principle we should be able to parametrise $k_i$ against $\mathrm{NUV}-r$. One obstacle in doing so in our case is that we do not have reliable optical data for the entire sample. Instead of relying on the $r$--band emission, we rather explore the effectiveness of a parametrisation against FUV$-$NIR colours. The FUV band is chosen not to require additional UV data and the NIR bands are good proxies for the stellar mass \citep{bell2001a}.

\begin{figure*}[!htbp]
 \includegraphics[width=\textwidth]{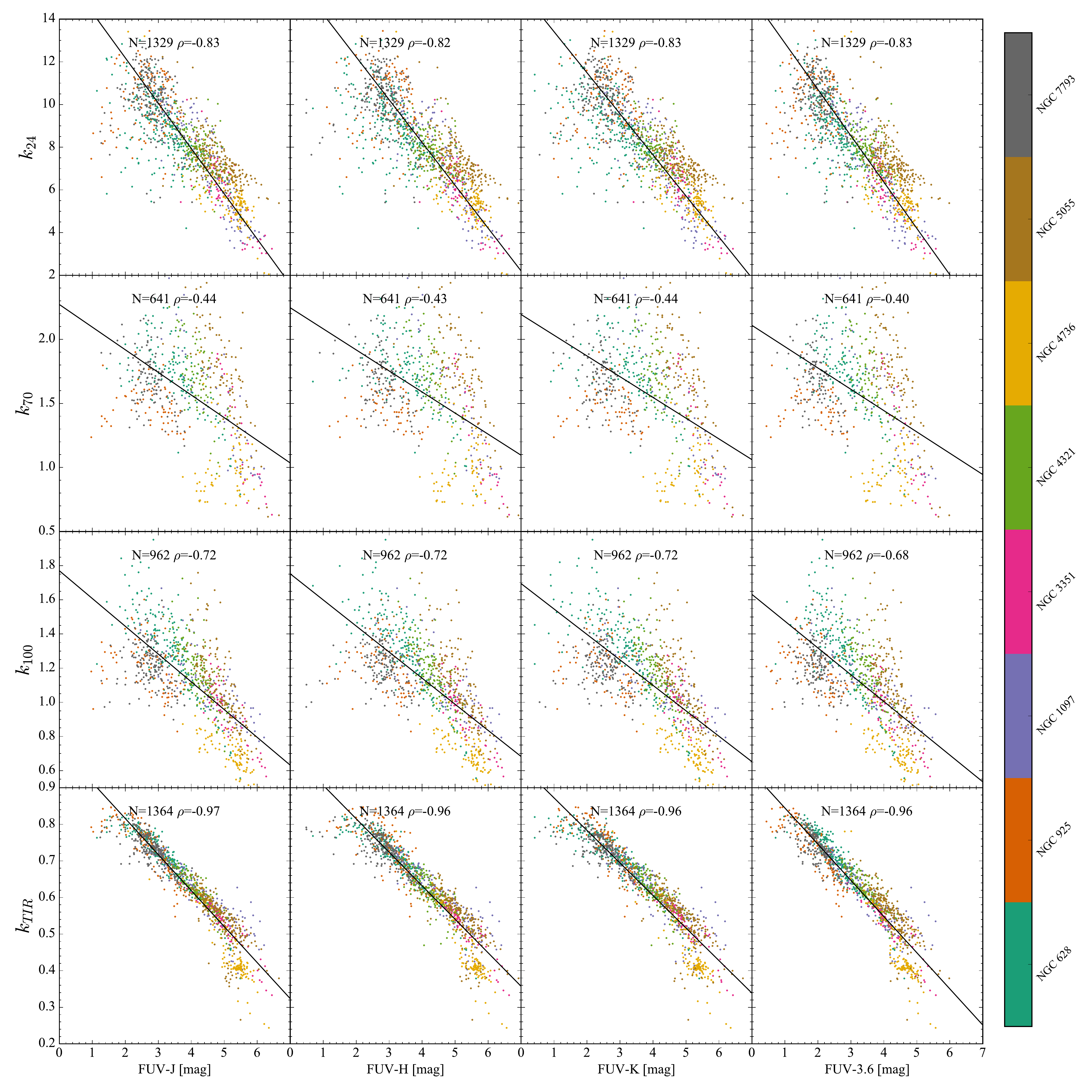}
 \caption{$k_i$ versus the FUV$-$J, FUV$-$H, FUV$-$Ks, and FUV$-$3.6 colours (AB magnitudes) for each galaxy in the sample at 24~$\mu$m, 70~$\mu$m, 100~$\mu$m, and for the TIR. The colour of each dot indicates the galaxy following the colour bar to the right. We have only selected regions detected at least at a 5--$\sigma$ level in the relevant band. This affects mostly the 70~$\mu$m and 100~$\mu$m bands. The number of regions and Spearman's rank correlation coefficient are shown at the top of each plot. The black lines represent the best linear fits using an Orthogonal Distance Regression algorithm (see the caption of Table~\ref{tab:kIR-colours} for a more detailed description). The best fit parameters along with the corresponding uncertainties are provided in Table~\ref{tab:kIR-colours}.\label{fig:kIR-colours}}
\end{figure*}

In Fig.~\ref{fig:kIR-colours} we show the variation of $k_i$ versus the FUV$-$J, FUV$-$H, FUV$-$Ks, and FUV$-$3.6~$\mu$m colours.
We see that there is a very clear relation between $k_i$ and the FUV$-$NIR colours at 24~$\mu$m ($-0.83\le\rho\le-0.82$) and for the TIR ($-0.96\le\rho\le-0.97$). The relation is however less clear at 100~$\mu$m ($-0.72\le\rho\le-0.68$) and especially at 70~$\mu$m ($-0.44\le\rho\le-0.40$). For the latter two bands, it appears however that individual galaxies tend to follow their own relation. One explanation is that as we have seen previously, the sSFR alone is not sufficient to estimate reliably the fraction of the TIR emission emerging in a given IR band. This problem is naturally non existent for the TIR which by definition includes all of the dust emission and not just a fraction as for individual bands. The 24~$\mu$m, being dominated by non--equilibrium emission appears to suffer much less from this issue.

To provide an easy--to--use recipe to estimate $k_i$ from FUV$-$NIR colours, we compute the best linear fit of the form:
\begin{equation}
k_i=a_\mathrm{FUV-NIR}+b_\mathrm{FUV-NIR}\times\mathrm{(FUV-NIR)}, 
\end{equation}
taking the uncertainties on both $k_i$ and FUV$-$NIR into account. Selecting all pixels detected at least at 1--$\sigma$ in all bands and at 5--$\sigma$ in band $i$, the coefficients $a_\mathrm{FUV-NIR}$ and $b_\mathrm{FUV-NIR}$ along with the corresponding uncertainties are presented in Table~\ref{tab:kIR-colours}.
\begin{table*}
 \centering
 \begin{tabular}{ccccccc}
  \hline\hline
  IR band&colour&$a$&$b$&$\sigma_{ab}$&$\Delta k_i$\\\hline
24&FUV$-$J&$  16.429\pm   0.265$&$  -2.122\pm   0.066$&$  -0.017$&$   0.000\pm   1.393$\\
24&FUV$-$H&$  16.236\pm   0.248$&$  -2.005\pm   0.059$&$  -0.014$&$   0.000\pm   1.446$\\
24&FUV$-$K&$  15.332\pm   0.215$&$  -1.923\pm   0.055$&$  -0.011$&$   0.000\pm   1.404$\\
24&FUV$-$3.6&$  15.044\pm   0.227$&$  -2.169\pm   0.068$&$  -0.015$&$ 0.000\pm   1.405$\\\hline
70&FUV$-$J&$   2.272\pm   0.145$&$  -0.177\pm   0.035$&$  -0.005$&$   0.000\pm   0.339$\\
70&FUV$-$H&$   2.247\pm   0.143$&$  -0.164\pm   0.033$&$  -0.005$&$   0.000\pm   0.340$\\
70&FUV$-$K&$   2.194\pm   0.132$&$  -0.162\pm   0.033$&$  -0.004$&$   0.000\pm   0.340$\\
70&FUV$-$3.6&$   2.109\pm   0.122$&$  -0.166\pm   0.035$&$  -0.004$&$ 0.000\pm   0.347$\\\hline
100&FUV$-$J&$   1.770\pm   0.120$&$  -0.163\pm   0.029$&$  -0.003$&$  0.000\pm   0.180$\\
100&FUV$-$H&$   1.752\pm   0.118$&$  -0.153\pm   0.027$&$  -0.003$&$  0.000\pm   0.180$\\
100&FUV$-$K&$   1.695\pm   0.107$&$  -0.149\pm   0.027$&$  -0.003$&$  0.000\pm   0.179$\\
100&FUV$-$3.6&$   1.632\pm   0.101$&$  -0.157\pm   0.030$&$  -0.003$&$0.000\pm   0.190$\\\hline
TIR&FUV$-$J&$   1.012\pm   0.099$&$  -0.098\pm   0.024$&$  -0.002$&$  0.000\pm   0.039$\\
TIR&FUV$-$H&$   0.998\pm   0.097$&$  -0.092\pm   0.023$&$  -0.002$&$  0.000\pm   0.043$\\
TIR&FUV$-$K&$   0.961\pm   0.088$&$  -0.089\pm   0.022$&$  -0.002$&$  0.000\pm   0.043$\\
TIR&FUV$-$3.6&$   0.943\pm   0.083$&$  -0.099\pm   0.025$&$  -0.002$&$0.000\pm   0.041$\\\hline
\end{tabular}
 \caption{Coefficients to estimate $k_i$ at 24~$\mu$m, 70~$\mu$m, 100~$\mu$m and for TIR from luminosities in the NIR: $k_i=a+b\times\mathrm{colour}$, with the colour in AB magnitudes. The uncertainties on each coefficient are indicated along with the covariance $\sigma_{ab}$. The last column indicates the mean offset and the standard deviation between the measured $k_i$ and the estimation from the fit. The fitting procedure is carried out using the \textsc{odr} module of the \textsc{python scipy} package.\label{tab:kIR-colours}}
\end{table*}
This means that we are now in a position to parametrise $k_i$ simply from the FUV$-$NIR colours.

We should note that this approach is somewhat similar to that of \cite{arnouts2013a}. While they do not explore the variations of $k_i$, they have derived relations to estimate observationally the IR luminosity based solely on the 24~$\mu$m and on the NUV$-$\textit{r} and \textit{r}$-$K colours, with results that are better than 0.3~dex. In our case, the absence of $r$-band data does not allow us to adopt the same strategy for the estimation of $k_i$.

For reference, we have also carried out a similar derivation for the NUV band. Results are presented in Appendix~\ref{sec:kir-NUV}.

\subsection{The NIR as a parametrisation for $k_i$\label{ssec:param-NIR}}

A complementary approach to parametrise $k_i$ is to estimate the stellar mass surface density of a galaxy through its NIR emission. Indeed, as old stellar populations dominate the NIR emission, the latter acts as a proxy for the total stellar mass even though in principle some colour terms can be required \citep{bell2001a}.

\begin{figure*}[!htbp]
 \includegraphics[width=\textwidth]{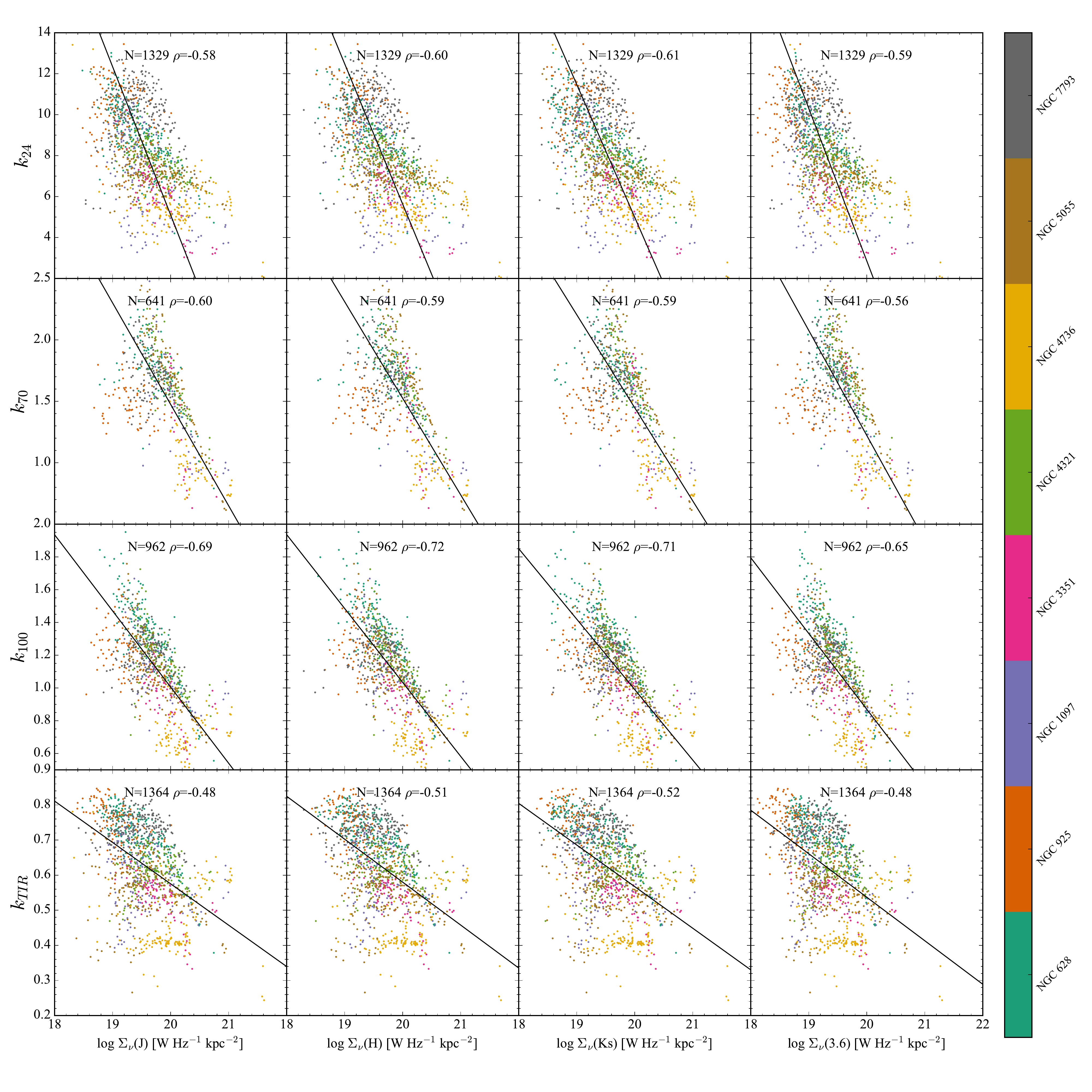}
 \caption{$k_i$ versus the J, H, Ks, and 3.6~$\mu$m band luminosity densities per unit area in W~Hz$^{-1}$~kpc$^{-2}$ for each galaxy in the sample at 24~$\mu$m, 70~$\mu$m, 100~$\mu$m, and for the TIR. The colour of each dot indicates the galaxy following the colour bar on the right. We have selected only regions detected at least at a 5--$\sigma$ level in the relevant band. This affects mostly the 70~$\mu$m and 100~$\mu$m bands. The number of regions and the Spearman's rank correlation coefficient are shown at the top of each plot. The black lines represent the best linear fits using an Orthogonal Distance Regresion algorithm (see the caption of Table~\ref{tab:kIR-colours} for a more detailed description) taking into account the uncertainties on both axes. The best fit parameters along with the corresponding uncertainties are provided in Table~\ref{tab:kIR-NIR}.\label{fig:kIR-NIR}}
\end{figure*}

In Fig.~\ref{fig:kIR-NIR} we show the variation of $k_i$ versus the J, H, Ks, and 3.6~$\mu$m luminosity densities per unit area.
As expected, this approach provides us with a stronger correlation at 70~$\mu$m even though the scatter remains large ($-0.60\le\rho\le-0.56$ versus $-0.44\le\rho\le-0.40$). At 100~$\mu$m the results are similar as with the FUV$-$NIR colour ($-0.72\le\rho\le-0.65$ versus $-0.72\le\rho\le-0.68$). Finally, for the 24~$\mu$m band and the TIR, the correlation appear much weaker ($-0.61\le\rho\le-0.58$ versus $-0.83\le\rho\le-0.82$ at 24~$\mu$m and $-0.52\le\rho\le-0.48$ versus $-0.97\le\rho\le-0.96$ for the TIR). This shows that taking into account the NIR luminosity density per unit area is especially important for bands sensitive to the blackbody emission of the dust.

To provide an easy--to--use recipe to estimate $k_i$ from the NIR luminosity densities per unit area, we compute the best linear fit of the form:
\begin{equation}
k_i=a_{NIR}+b_{NIR}\times \log\Sigma_\nu(NIR),
\end{equation}
taking the uncertainties on both $k_i$ and $\Sigma_\nu(NIR)$ into account. Selecting all pixels detected at least at 1--$\sigma$ in all bands and at 5--$\sigma$ in band $i$, the coefficients $a_{NIR}$ and $b_{NIR}$ along with the corresponding uncertainties are presented in Table~\ref{tab:kIR-NIR}.
\begin{table*}
 \centering
 \begin{tabular}{ccccccc}
  \hline\hline
  IR band&NIR band&$a$&$b$&$\sigma_{ab}$&$\Delta k_i$\\\hline
24&J&$   5.103\pm   0.350$&$  -7.238\pm   0.691$&$   0.199$&$   0.000\pm   2.801$\\
24&H&$   5.629\pm   0.294$&$  -6.857\pm   0.621$&$   0.140$&$   0.000\pm   2.693$\\
24&K&$   4.987\pm   0.323$&$  -6.492\pm   0.557$&$   0.150$&$   0.000\pm   2.645$\\
24&3.6&$   2.858\pm   0.556$&$  -7.449\pm   0.732$&$   0.378$&$ 0.000\pm   2.784$\\\hline
70&J&$   1.475\pm   0.053$&$  -0.826\pm   0.125$&$   0.002$&$   0.000\pm   0.312$\\
70&H&$   1.525\pm   0.051$&$  -0.786\pm   0.118$&$   0.001$&$   0.000\pm   0.311$\\
70&K&$   1.448\pm   0.053$&$  -0.758\pm   0.113$&$   0.002$&$   0.000\pm   0.311$\\
70&3.6&$   1.222\pm   0.074$&$  -0.855\pm   0.132$&$   0.007$&$ 0.000\pm   0.327$\\\hline
100&J&$   1.005\pm   0.041$&$  -0.465\pm   0.081$&$   0.002$&$  0.000\pm   0.183$\\
100&H&$   1.034\pm   0.039$&$  -0.451\pm   0.078$&$   0.001$&$  0.000\pm   0.179$\\
100&K&$   0.990\pm   0.042$&$  -0.430\pm   0.074$&$   0.002$&$  0.000\pm   0.179$\\
100&3.6&$   0.870\pm   0.058$&$  -0.463\pm   0.083$&$   0.004$&$0.000\pm   0.193$\\\hline
TIR&J&$   0.575\pm   0.037$&$  -0.118\pm   0.056$&$   0.001$&$  0.000\pm   0.103$\\
TIR&H&$   0.580\pm   0.034$&$  -0.122\pm   0.055$&$   0.001$&$  0.000\pm   0.101$\\
TIR&K&$   0.567\pm   0.038$&$  -0.119\pm   0.052$&$   0.001$&$  0.000\pm   0.101$\\
TIR&3.6&$   0.537\pm   0.050$&$  -0.124\pm   0.058$&$   0.002$&$0.000\pm   0.103$\\\hline
 \end{tabular}
 \caption{Coefficients to estimate $k_i$ at 24~$\mu$m, 70~$\mu$m, 100~$\mu$m and for TIR from luminosities in the NIR: $k_i=a+b\times\left(\log\Sigma_\nu(NIR)-20\right)$, with $\Sigma_\nu$ the luminosity density per unit area in terms of W~Hz$^{-1}$~kpc$^{-2}$. The fitting procedure and the computation of the uncertainties are done in the same way as for Table~\ref{tab:kIR-colours}.\label{tab:kIR-NIR}}
\end{table*}
This means that we are now in a position to parametrise $k_i$ simply from observations in the NIR.

For reference, we have also carried out a similar derivation for the NUV band. Results are presented in Appendix~\ref{sec:kir-NUV}.

\subsection{Comparison of attenuation corrected FUV with different methods\label{ssec:comp-methods}}

\begin{figure*}[!htbp]
 \includegraphics[width=\columnwidth]{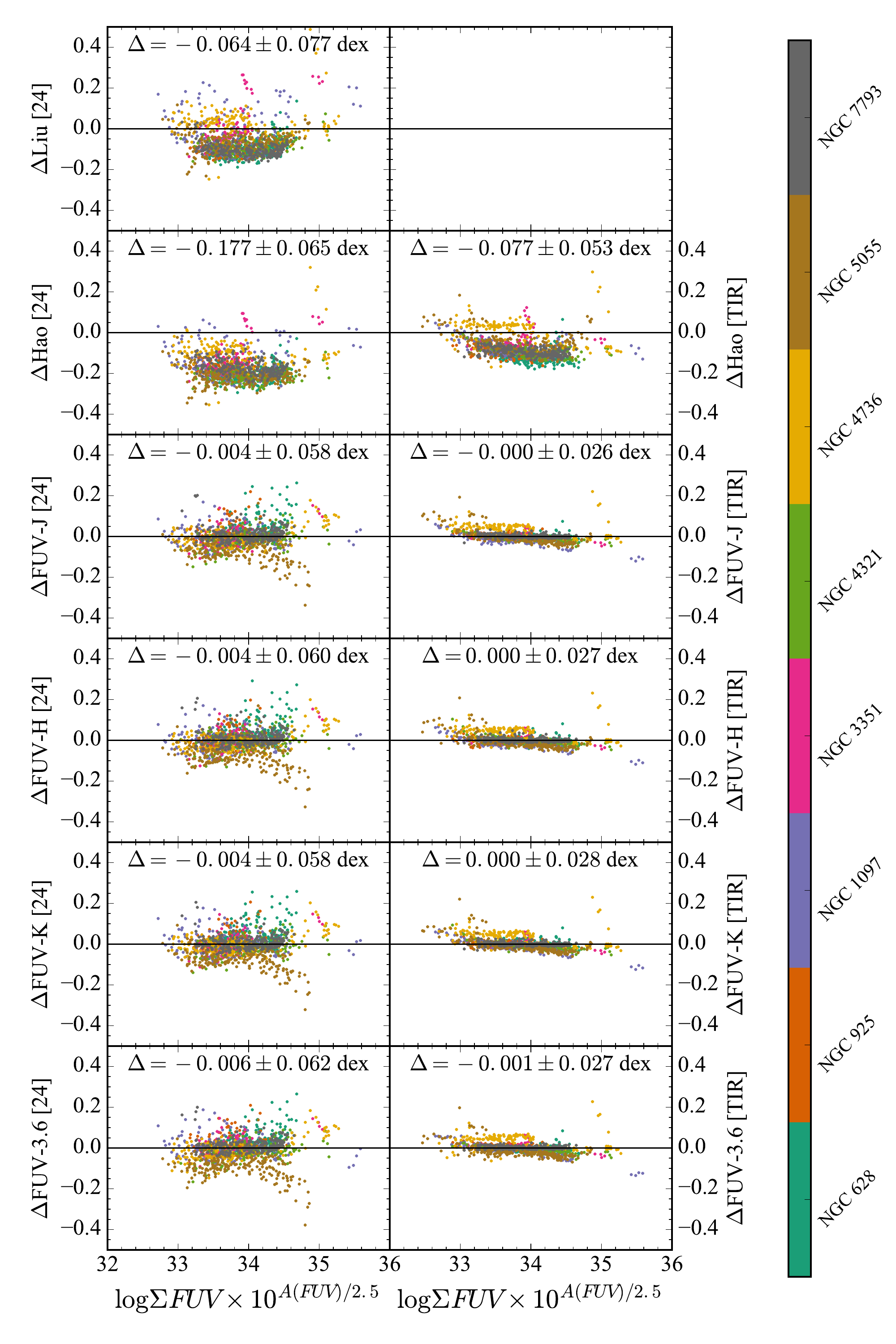}
 \includegraphics[width=\columnwidth]{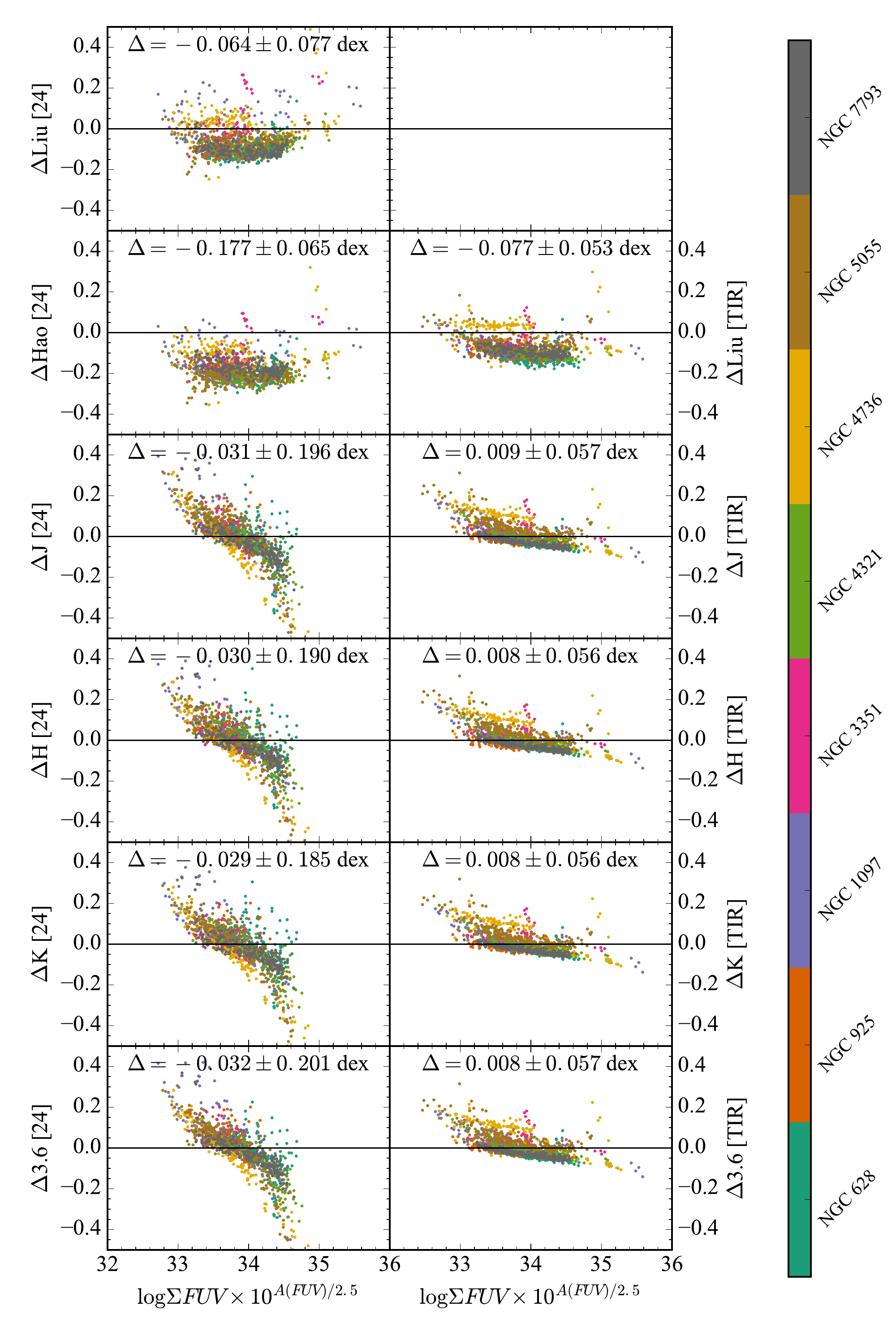}
 \caption{Left: Difference between the CIGALE attenuation--corrected FUV luminosities per unit area and the ones estimated using a constant $k_i$ from \cite{liu2011a} and \cite{hao2011a} ($y$--axis, first two rows), a variable $k_i$ estimated from one of the linear relations with the FUV$-$NIR colours given in Table \ref{tab:kIR-colours} ($y$--axis, other rows), and the estimated FUV attenuation directly obtained from the CIGALE SED modelling ($x$--axis). The first column is based on the 24~$\mu$m band and the second column on the TIR. Right: same but estimating $k_i$ from one of the linear relations with the NIR luminosity density per unit area given in Table~\ref{tab:kIR-NIR}.\label{fig:comp-sigma-FUV}}
\end{figure*}

To examine the impact of a variable $k_i$, we compare in Fig.~\ref{fig:comp-sigma-FUV} the estimated attenuation--corrected FUV luminosities per unit area using 1. a constant $k_i$ from \cite{liu2011a} and \cite{hao2011a} ($y$--axis, first two rows), 2. a variable $k_i$ estimated from one of the linear relations with the FUV$-$NIR colours given in Table~\ref{tab:kIR-colours} and with the NIR luminosity density per unit area given in Table~\ref{tab:kIR-NIR} ($y$--axis, other rows), and 3. the estimated FUV attenuation directly obtained from the CIGALE SED modelling ($x$--axis). We carry out this comparison at 24~$\mu$m (left column) and for the TIR (right column).

The relation of \cite{liu2011a} leads to values of the attenuation--corrected FUV that are on average compatible with that from our modelling by $-0.064\pm0.077$~dex at 24~$\mu$m whereas \cite{hao2011a} leads to values that are lower by $-0.177\pm0.062$~dex at 24~$\mu$m and $-0.077\pm0.053$~dex for the TIR. Conversely our relations give a mean offset smaller than 0.032~dex at 24~$\mu$m and smaller than 0.009~dex for the TIR.

If estimates of the attenuation--corrected FUV luminosity are very good on average, some regions are nevertheless clearly discrepant, especially when estimating $k_i$ from the NIR luminosity density per unit area. At 24~$\mu$m this is particularly the case of regions with $\log\Sigma FUV\times10^{A(FUV)/2.5}>34$~W~kpc$^{-2}$, with a strong underestimate of the attenuation--corrected FUV luminosity: $\Delta=-0.510\pm0.405$~dex, using the estimator based on the 3.6~$\mu$m band. This offset is due to the strong deviation from the $k_i$--NIR best fit shown in Fig.~\ref{fig:kIR-NIR}. Where the fit led to $k_i<0$ for the most extreme regions, we assumed $k_i=0$. This illustrates one of the main limits of relying on the NIR: it does not trace locally the presence of intense star--forming episodes. Regions with a particularly high sSFR will then see their attenuation-corrected FUV emission underestimated. This effect can be seen more strongly at 24~$\mu$m than for the TIR.

The clear luminosity--dependent trends with the attenuation--corrected FUV we see when estimating $k_i$ from the NIR luminosity density per unit area are nearly non--existent when estimating $k_i$ from the FUV$-$NIR colours. In the latter case, there is a slightly higher but still very small systematic offset and a smaller scatter. This means that the FUV$-$NIR colours based estimators will provide us with results that are much less affected by possible FUV luminosity dependent biases.

Another interesting result is that the scatter is systematically lower for the TIR than at 24~$\mu$m. We also retrieve this effect when considering the relations of \cite{hao2011a}. This likely comes from the much smaller dynamical range of estimated $k_i$ for the TIR compared to 24~$\mu$m: from 0.35 to 0.90 for the former and from 2.07 to 14.09 for the latter, when considering the 3.6~$\mu$m band. In other words, the absorbed FUV emission from young stars contributes a much more variable fraction of the 24~$\mu$m than for the TIR. This is unsurprising as for a fixed absorbed FUV luminosity, the dust SED is very dependent on the local physical conditions, as we have seen earlier in Sect.~\ref{ssec:variations-kIR}. This is not the case for the TIR, however. Considering energy conservation, at the equilibrium the total luminosity emitted by the dust in the IR is exactly the luminosity it absorbs from the UV to the near--IR. This means that the TIR depends only on the absorbed luminosity by dust, and not on its temperature. This emphasises the importance of the TIR as an SFR estimator even if it can also be contaminated by dust emission unrelated to recent star formation. The induced variations appear small enough that the uncertainties on FUV dust correction remain small, or at least smaller than for individual IR passbands.

Finally, we need to keep in mind a caveat of this comparison. We are limited here to the case of the 24~$\mu$m band and the TIR, which are the most favourable cases for the FUV$-$NIR estimators. Given the visibly poorer results at 70~$\mu$m and 100~$\mu$m, which are particularly clear when comparing Fig.~\ref{fig:kIR-colours} with Fig.~\ref{fig:kIR-NIR}, the NIR luminosity density per unit area would yield better (albeit not perfect) estimates of $k_i$ in such cases.

\section{Estimators for unresolved galaxies\label{sec:kir-unresolved}}

We have derived new hybrid estimators at local scales by parametrising $k_i$ on FUV$-$NIR colours and on NIR luminosity densities per unit area. These estimators are directly applicable when resolved observations are available. However, this is generally not the case beyond the local universe and the application to unresolved observations is non--trivial. This is because in that case the effective $\left<k_i\right>$ for an entire galaxy does necessarily not correspond to the one derived for a given FUV$-$NIR colour or NIR luminosity density per unit area. The reason is that at the scale of a galaxy, the effective $\left<k_i\right>$ is weighted towards the more luminous regions\footnote{In our sample, typically the 20\% to 30\% most luminous pixels dominate the integrated luminosity. The global $\left<k_i\right>$ appear to be driven by those pixels.}, which in star--forming spiral galaxies are often located in the inner regions, which in turn have a redder FUV$-$NIR colour and higher NIR luminosity density per unit area. If we simply considered the average FUV$-$NIR colour or the average NIR luminosity density per unit area of the galaxy, this would lead to an overestimate of $\left<k_i\right>$. For example, when considering the 3.6~$\mu$m luminosity density, the estimated total attenuated FUV luminosity from the 24~$\mu$m band (in effect $k_{24}\times L(24)$) is up to 70\% higher when applying the relations in Table~\ref{tab:kIR-NIR} on integrated galaxies rather than when applying these relations at local scales. This is especially the case for galaxies where star formation is very concentrated, such as NGC~1097 for instance. For galaxies such as NGC~628 and NGC~925 for which star formation is more spread out, the difference is within 10\%. Interestingly, the relations based on the FUV$-$NIR provide very similar results when applied on local or global fluxes. We therefore concentrate here on extending to unresolved galaxies the relations based on the NIR luminosity densities per unit area.

By making some assumptions on the distributions of the old and young stellar populations across the disk, it is possible to derive a relation to extend our local $k_i$ estimates to global ones. To explore this possibility, we have assumed that the SFR surface density and the NIR luminosity density per unit area both follow decaying exponential profiles. The derivation, which is detailed in Appendix~\ref{sec:derivation-entire-gal}, leads to the following effective $\left<k_i\right>$ when using the NIR luminosity density per unit area:
\begin{equation}\label{eqn:effective-kir}
 \left<k_i\right>=a_{NIR}+b_{NIR}\times\left[\log\Sigma_{\nu\mathrm{,~NIR}}\left(0\right)-\frac{2}{\ln 10}\times\frac{r_{SFR}}{r_{NIR}}-20\right],
\end{equation}
with $\Sigma_{\nu\mathrm{,~NIR}}\left(0\right)$ the luminosity density per unit area in the NIR at the centre of the galaxy in W~Hz$^{-1}$~kpc$^{-2}$, $r_{SFR}$ the scale length of the SFR exponential profile, and $r_{NIR}$ the scale length of the NIR exponential profile. Such relations therefore require a priori knowledge of the young and old stellar populations scale lengths. They are not always available or easy to estimate. In such cases, it may be valuable to use average values. Using the GALEX Nearby Galaxies Survey sample \citep{gildepaz2007a}, \cite{munoz2007a} measured the SFR and the stellar mass surface density scale lengths of a large number of galaxies. Using a sub--sample of 131 galaxies for which both quantities have been computed reliably and assuming that the NIR luminosity density per unit area and stellar mass surface density scale lengths are similar, we find that $\left<r_{SFR}/r_{NIR}\right>\simeq1.18\pm0.46$. Injecting this into Eq.~\ref{eqn:effective-kir}, we find:
\begin{equation}
 \left<k_i\right>=a_{NIR}+b_{NIR}\times\left[\log\Sigma_{\nu\mathrm{,~NIR}}\left(0\right)-21.03\right],\label{eqn:kIR-resolved}
\end{equation}
or equivalently:
\begin{equation}
 \left<k_i\right>=a_{NIR}+b_{NIR}\times\left[\log \frac{L_{\nu\mathrm{,~NIR}}}{2 \times \pi \times r_{NIR}^2}-21.03\right].
\end{equation}
The dispersion found from the \cite{munoz2007a} sample induces a typical uncertainty on $\left<k_i\right>$ of $\pm1.7$ at 24~$\mu$m, $\pm0.3$ at 70~$\mu$m, $\pm0.2$ at 100~$\mu$m, and $\pm0.05$ for the TIR. This can be considered as a lower limit as deviations from exponential profiles for either the SFR or the NIR will also increase the uncertainty on $\left<k_i\right>$.

To test whether the previous relations provide us with reasonable estimates of $\left<k_i\right>$, we have applied Eq.~\ref{eqn:kIR-resolved} to all the galaxies in our sample from 24~$\mu$m to the TIR, and for each of the four NIR bands considered. In Table~\ref{tab:comp-kIR} we compare these estimates with an effective $\left<k_i\right>$ derived from SED modelling: we compute the global attenuation--corrected FUV luminosity by correcting each individual pixel for the attenuation and then we apply Eq.~\ref{eqn:k-IR-v1} to compute $\left<k_i\right>$ from the integrated luminosities.

\begin{table*}[!htbp]
 \centering
 \begin{tabular}{ccccccc}\hline\hline
  Galaxy&Band&$\left<k_i(\textrm{CIGALE})\right>$&$\left<k_i(J)\right>$&$\left<k_i(H)\right>$&$\left<k_i(Ks)\right>$&$\left<k_i(3.6)\right>$\\\hline
NGC  628&24&$7.76$&$7.17$&$7.18$&$7.06$&$7.57$\\
NGC  925&24&$9.67$&$12.49$&$12.17$&$11.86$&$12.73$\\
NGC 1097&24&$4.77$&$5.36$&$5.17$&$4.97$&$4.66$\\
NGC 3351&24&$4.07$&$6.83$&$6.66$&$6.49$&$6.83$\\
NGC 4321&24&$7.02$&$7.57$&$7.40$&$7.11$&$7.31$\\
NGC 4736&24&$4.71$&$0.84$&$1.04$&$1.15$&$0.86$\\
NGC 5055&24&$6.82$&$5.65$&$5.43$&$5.31$&$5.42$\\
NGC 7793&24&$9.80$&$9.57$&$9.56$&$9.44$&$10.10$\\\hline
NGC  628&70&$1.72$&$1.71$&$1.70$&$1.69$&$1.76$\\
NGC  925&70&$1.50$&$2.32$&$2.27$&$2.25$&$2.36$\\
NGC 1097&70&$1.12$&$1.50$&$1.47$&$1.45$&$1.43$\\
NGC 3351&70&$1.08$&$1.67$&$1.64$&$1.62$&$1.68$\\
NGC 4321&70&$1.50$&$1.76$&$1.73$&$1.70$&$1.73$\\
NGC 4736&70&$0.66$&$0.99$&$1.00$&$1.00$&$0.99$\\
NGC 5055&70&$1.38$&$1.54$&$1.50$&$1.48$&$1.52$\\
NGC 7793&70&$1.66$&$1.98$&$1.98$&$1.97$&$2.05$\\\hline
NGC  628&100&$1.25$&$1.14$&$1.14$&$1.13$&$1.16$\\
NGC  925&100&$1.19$&$1.48$&$1.46$&$1.45$&$1.48$\\
NGC 1097&100&$1.00$&$1.02$&$1.00$&$0.99$&$0.98$\\
NGC 3351&100&$0.88$&$1.12$&$1.10$&$1.09$&$1.12$\\
NGC 4321&100&$1.08$&$1.16$&$1.15$&$1.13$&$1.15$\\
NGC 4736&100&$0.64$&$0.73$&$0.73$&$0.74$&$0.75$\\
NGC 5055&100&$0.87$&$1.04$&$1.02$&$1.01$&$1.03$\\
NGC 7793&100&$1.21$&$1.29$&$1.29$&$1.29$&$1.32$\\\hline
NGC  628&TIR&$0.68$&$0.61$&$0.61$&$0.61$&$0.62$\\
NGC  925&TIR&$0.73$&$0.70$&$0.70$&$0.69$&$0.70$\\
NGC 1097&TIR&$0.58$&$0.58$&$0.57$&$0.57$&$0.57$\\
NGC 3351&TIR&$0.50$&$0.60$&$0.60$&$0.59$&$0.60$\\
NGC 4321&TIR&$0.61$&$0.62$&$0.61$&$0.61$&$0.61$\\
NGC 4736&TIR&$0.48$&$0.51$&$0.50$&$0.50$&$0.50$\\
NGC 5055&TIR&$0.53$&$0.58$&$0.58$&$0.57$&$0.58$\\
NGC 7793&TIR&$0.73$&$0.65$&$0.65$&$0.65$&$0.66$\\\hline

 \end{tabular}
 \caption{Comparison of the effective $\left<k_i\right>$ coefficient from 24~$\mu$m to the TIR for each galaxy derived either from the CIGALE model or estimated from Eq.~\ref{eqn:kIR-resolved} using the central NIR luminosity density per unit area in the J, H, Ks, or 3.6~$\mu$m band.\label{tab:comp-kIR}}
\end{table*}

Taking the estimates based on the CIGALE modelling as a reference, we find that Eq.~\ref{eqn:kIR-resolved} provides us with excellent estimates. In the vast majority of cases Eq.~\ref{eqn:kIR-resolved} gives closer estimates of $\left<k_i\right>$ than adopting the fixed values of \cite{hao2011a}. There are some cases where the estimates are particularly discrepant though. This concerns the three earlier--type galaxies in our sample: NGC~1097, NGC~3351, and NGC~4736. This is especially clear at 24~$\mu$m. If we inspect the parameters maps of NGC~1097 and NGC~3351 shown in Fig.~\ref{fig:maps-parameters} we see that star formation in the galaxy is strongly dominated by the nuclear starburst. As these two galaxies clearly deviate from the assumption of a decaying exponential profile, it is therefore expected that $\left<k_i\right>$ estimated from CIGALE will be heavily weighted towards the local $k_i$ in the most luminous regions. For NGC~4736, the issue may be different. The four central pixels, which all overlap with the nucleus, present a suspiciously large estimate of the stellar mass. After close inspection, it turns out that the fit of these pixels is particularly poor. Depending on the actual AGN type, it is possible that a very significant part of the flux from these regions actually originates from the active nucleus, with hot dust strongly increasing the IR emission, which leads to an overestimate of the stellar mass and of the SFR, leading to an uncertain impact on the sSFR, and to an overall misestimate of the physical properties we are interested in. In which case this means that for these pixels Eq.~\ref{eqn:kIR-resolved} does not provide us with reliable results. More generally, this also means that the estimators we provide in this article should not be used in regions where physical processes other than star formation contribute in a non--negligible way to the emission.

In conclusion, the new relations we provide can be reasonably extended to entire galaxies as long as there is no strong deviation from the assumption of exponential profiles. For other profiles, similar relations can also be derived following the method outlined in Appendix~\ref{sec:derivation-entire-gal}.

\section{Limits and recommendations\label{sec:limits}}

\subsection{Validity range of the new estimators}

One of the main dangers in using the new parametrisations we have derived is extrapolating the relations provided in Tables \ref{tab:kIR-colours} and \ref{tab:kIR-NIR} beyond the domain upon which they have been derived. In that case $k_i$ could be assigned unrealistically low or high values. This is in part because we assume a simple linear relation between $k_i$ and the FUV$-$NIR colours or NIR luminosity densities per unit area. In reality $k_i$ is physically bounded. For instance $k_{TIR}\simeq0$ if the TIR is exclusively due to old stellar populations and $k_{TIR}\lesssim1$ if the TIR is exclusively due to the youngest stellar populations, with the actual maximum value depending on the fraction of the dust luminosity coming from absorbed FUV photons. The range of $k_i$ values probed in this study is probably representative of the typical boundaries one may encounter as we have a wide dynamical range in terms of physical conditions. Nevertheless, it is important to apply the new estimators exclusively in the range in which they have been derived. In terms of dust luminosity per unit area, the estimators we have defined are valid over: 
\begin{itemize}
 \item $4.93\le\log\Sigma(24)\le8.36$~L$_\odot$~kpc$^{-2}$,
 \item $5.83\le\log\Sigma(70)\le8.99$~L$_\odot$~kpc$^{-2}$,
 \item $5.84\le\log\Sigma(100)\le8.97$~L$_\odot$~kpc$^{-2}$,
 \item $6.12\le\log\Sigma(TIR)\le9.19$~L$_\odot$~kpc$^{-2}$.
\end{itemize}
In terms of NIR luminosity densities per unit area, they have been defined over:
\begin{itemize}
 \item $18.31\le\log\Sigma_\nu(J)\le21.62$~W~Hz$^{-1}$~kpc$^{-2}$,
 \item $18.30\le\log\Sigma_\nu(H)\le21.70$~W~Hz$^{-1}$~kpc$^{-2}$,
 \item $18.31\le\log\Sigma_\nu(Ks)\le21.62$~W~Hz$^{-1}$~kpc$^{-2}$,
 \item $18.12\le\log\Sigma_\nu(3.6)\le21.30$~W~Hz$^{-1}$~kpc$^{-2}$.
\end{itemize}
And finally in terms of FUV$-$NIR colours (AB magnitudes):
\begin{itemize}
 \item $\mathrm{0.97<FUV-J<6.66}$~mag,
 \item $\mathrm{0.49<FUV-H<6.92}$~mag,
 \item $\mathrm{0.41<FUV-Ks<6.72}$~mag,
 \item $\mathrm{0.44<FUV-3.6<5.98}$~mag.
\end{itemize}

Another possible limit is that the new estimators have been derived on a sample of eight star--forming galaxies. While we have ensured to cover early and late--type spiral galaxies, such derivations are by nature dependent on the sample used. As such they should not be used on galaxies earlier than Sa. Their use on irregular galaxies is equally uncertain. We also caution against using them when there is any indication of a strong active nucleus. A type 1 nucleus can strongly contaminate the FUV and the NIR bands \citep{ciesla2015a}. More generally, active nuclei provide an additional heating mechanism independent from stellar populations.

\subsection{Choice of the estimators}

The new hybrid estimators that we have derived in this work represent a generalisation and an extension to new wavelengths of hybrid estimators previously published in the literature. We have shown that if these estimators physically depend on the sSFR and on the stellar mass surface density, they can be efficiently parametrised based on FUV$-$NIR colours or on the NIR luminosity densities per unit area. We have seen in Sect.~\ref{ssec:comp-methods} that they can bring important improvements on FUV attenuation estimates, and therefore on the SFR. This however comes at the cost of a slightly increased complexity with the additional requirement of NIR observations. If we take the estimators of \cite{hao2011a} as our baseline comparisons, there are several main scenarios. We suggest here a few possibilities to guide the choice of the estimator.

\subsubsection{Dependence on wavelength}

We have shown that each of the two methods we have presented have symmetric strengths and weaknesses. For the 24~$\mu$m and for the TIR, we suggest to estimate $k_i$ from the FUV$-$NIR colour. For the 70~$\mu$m and 100~$\mu$m bands, we suggest to estimate $k_i$ from the NIR luminosity density per unit area.

\subsubsection{Resolved observations of low redshift galaxies}

In case the study benefits from resolved observations of nearby galaxies, given the all--sky coverage in the NIR brought by 2MASS and WISE, we recommend to apply the estimators provided in this paper. This is especially important at 24~$\mu$m as this band appears to exhibit large variations of $k_i$. Even though we cannot compare our results with the \cite{hao2011a} estimators at 70~$\mu$m and 100~$\mu$m, this is also likely to be true. For the TIR, given the small dynamical range of $k_i$, the difference between our estimators and that of \cite{hao2011a} will be much smaller on average. If data are available, given the excellent relation with the FUV$-$NIR colours, we recommend to apply the $k_i$ derived in this paper.

Finally, the new estimators have been computed on spatial scales ranging from 0.5~kpc (NGC~7793) to 1.7~kpc (NGC~1097). Their application on much smaller scales is not advised, if only because a number of assumptions for the modelling (fully sampled initial mass function, continuous SFH, etc.) will necessarily break down at some point at small spatial scales \citep{boquien2015a}.

\subsubsection{Unresolved observations of low redshift galaxies}

To deal with the case of unresolved galaxies, we have extended our new locally derived estimators to entire galaxies in Sect.~\ref{sec:kir-unresolved}. We showed in Table~\ref{tab:comp-kIR} that they appear to provide closer estimates of the effective $k_i$ than when adopting a constant value from \cite{hao2011a}. The main difficulty is that it requires at least the central NIR luminosity density per unit area of the galaxy, if we take the SFR and NIR scale lengths provided by \cite{munoz2007a}. If the galaxies are nevertheless resolved in the NIR only, Eq.~\ref{eqn:kIR-resolved} can be directly applied. This still requires the assumptions behind it to be valid. That is, the young and old stellar populations follow exponentially decaying profiles. This particular point is even more important when there is no resolved observation available in the NIR. In this case it may be safer to apply the average relations of \cite{hao2011a}, which have been derived on a larger sample of unresolved galaxies.

\subsubsection{Note on the conversion from FUV luminosity to SFR}

To compute the SFR from the relations given in this paper one needs to adopt a conversion factor, for instance one of those given in \cite{boquien2014a} for a \cite{chabrier2003a} IMF. However in this study, we have shown that the sSFR is an important parameter to understand how to combine the FUV and the IR to correct the former for the presence of dust. In other words, it is important to take the SFH into account to correct for the attenuation. But at the same time, the SFH also has an impact on the conversion factor from the FUV luminosity to the SFR, due to the contamination of the FUV emission by long--lived stars \citep{johnson2013a,boquien2014a}, whereas a constant SFR over 100~Myr is generally assumed \citep[e.g.,][]{kennicutt1998a,kennicutt2011a}. In other words, as $k_i$ depends on the exact SFH of the galaxy, the FUV--to--SFR conversion factor also depends on the SFH, albeit in a different way. This emphasises the necessity to take the impact of the SFH into account to reliably measure the SFR of galaxies from hybrid estimators.

\section{Summary\label{sec:conclusion}}

In this article, we have investigated why and how hybrid SFR estimators based on the combination of UV and IR emission ($L(UV)_{int}=L(UV)_{obs}+k_i\times L(IR)_i$) depend on the local physical conditions. To do so, using CIGALE (Boquien et al., Burgarella et al. in prep.), we have modelled the FUV--to--FIR SED of eight nearby spatially--resolved star--forming spiral galaxies drawn from the KINGFISH sample. This has allowed us to characterise their local physical properties such as the stellar mass, the SFR, the sSFR, the TIR luminosity, and the UV attenuation at a typical scale of 1~kpc. Our main findings are the following:

\begin{enumerate}
 \item There are important region--to--region variations within galaxies and galaxy--to--galaxy variations of $k_i$, from 1.55 to 13.45 at 24~$\mu$m for instance (for comparison, \cite{hao2011a} and \cite{liu2011a} found a constant factor of $3.89\pm0.15$ and $6.0$ respectively). This shows that hybrid estimators using a fixed value for $k_i$ cannot be appropriate for the full diversity of galaxies and may provide systematically biased estimates when applied on galaxies whose physical properties differ from that of the original calibration sample.
 \item When considering the combination of the FUV with the IR luminosity, $k_i$ varies with the average sSFR over 100~Myr: increasing values of sSFR also yield increasing values of $k_i$. The reason is that as the sSFR increases, the IR emission is increasingly linked to recent star formation and the relative importance of dust heating by older stellar populations diminishes. This is more particularly the case for $k_{24}$ and $k_{TIR}$. However, being sensitive to the blackbody emission of the dust, $k_{70}$ and $k_{100}$ show a more complex behaviour with a particular sensitivity to the stellar mass surface density.
 \item Exploiting the physical insights provided by these correlations, we have parametrised $k_i$ against a. the FUV$-$NIR colours, and b. the NIR luminosity densities per unit area. As a result, these new parametrisations bring strong improvements compared to a constant $k_i$ to correct the FUV for the attenuation. This shows that when using individual passbands in the IR, it is crucial to take into account the variability of $k_i$. The TIR emission is not as sensitive to a variation of the diffuse emission and the difference with the $k_{TIR}$ of \cite{hao2011a} remains minor.
 \item Building on the success of the parametrisation of $k_i$ for the FUV, we have expanded those to the NUV (see Appendix \ref{sec:kir-NUV}). We also found clear correlations of $k_i$ with the sSFR and the stellar mass surface density, showing that we can use such estimators efficiently to correct the NUV emission for the presence of dust and retrieve the SFR in the absence of FUV data.
 \item Assuming exponentially decaying radial profiles for the stellar populations, we have expanded the parametrisation on the NIR luminosity densities per unit area to the case of unresolved galaxies. We have found that it provides better estimates of the effective $\left<k_i\right>$ than adopting a constant factor as for classical hybrid estimators. This shows that these new estimators can work well both on resolved and unresolved data.
\end{enumerate}

These new estimators provide better estimates than classical hybrid estimators published in the literature. By statistically taking into account the impact of dust heated by old stellar populations they constitute an important step towards universal estimators.

\begin{acknowledgements}
We would like to thank the referee whose useful comments have helped making the paper clearer.
MB would like to thank Andrew Connolly for enlightening discussions about fitting techniques and the handling of uncertainties. This publication was financed by the FIC-R Fund, allocated to the
project 30321072. FST acknowledges financial support from the Spanish Ministry of Economy and Competitiveness (MINECO) under grant number AYA2013-41243-P.

This research made use of the NASA/IPAC Infrared Science Archive, which is operated by the Jet Propulsion Laboratory, California Institute of Technology, under contract with the National Aeronautics and Space Administration; APLpy, an open-source plotting package for Python hosted at \url{https://aplpy.github.com}; Astropy, a community-developed core Python package for Astronomy \citep{astropy2013a}; the IPython package \citep{ipython2007a}; matplotlib, a Python library for publication quality graphics \citep{matplotlib}; SciPy \citep{scipy}. This publication makes use of data products from the Two Micron All Sky Survey, which is a joint project of the University of Massachusetts and the Infrared Processing and Analysis Center/California Institute of Technology, funded by the National Aeronautics and Space Administration and the National Science Foundation. Based on observations made with the NASA Galaxy Evolution Explorer. GALEX is operated for NASA by the California Institute of Technology under NASA contract NAS5-98034. This work is based in part on observations made with the Spitzer Space Telescope, which is operated by the Jet Propulsion Laboratory, California Institute of Technology under a contract with NASA. Support for this work was provided by NASA through an award issued by JPL/Caltech. {\it Herschel} is an ESA space observatory with science instruments provided by European-led Principal Investigator consortia and with important participation from NASA.

\end{acknowledgements}
\bibliographystyle{aa}
\bibliography{article}

\appendix

\section{Impact of the assumption on the metallicity\label{sec:impact-metals}}

To model the local FUV--to--FIR SED of galaxies, we have assumed a constant metallicity $Z=0.02$ throughout the disk. The reason for doing so is two--fold. First, the selected sample is metal--rich. Measurements using the \cite{kobulnicky2004a} estimator yield high average oxygen abundances $\left<12+\log O/H\right>=8.79$ to $9.19$ \citep{kennicutt2011a}. $Z=0.02$ corresponds to an oxygen abundance $12+\log O/H=8.90$ assuming the solar metallicity ($Z=0.0134$) and oxygen abundance ($12+\log O/H=8.69$) from \cite{asplund2009a}. This assumption therefore appears reasonable as the sample is made of galaxies with similar metallicities. The second reason is that there are age--metallicity and age--attenuation degeneracies, which make disentangling their respective effects an especially difficult problem. To limit the impact of these degeneracies, we have decided to set a constant metallicity $Z=0.02$.

Nevertheless, to assess the actual impact of the assumption of the metallicity on our results, we have also run the models assuming $Z=0.008$. We find that the average SFR over 100~Myr is then lower by $0.17\pm0.12$~dex. The stellar mass, however, only increases by $0.02\pm0.03$~dex. At the same time, the FUV attenuation is also systematically lower by $0.07\pm0.05$~mag. This has a direct impact on $k_i$, yielding values lower by $\sim8$\% but preserving the relations with the physical parameters examined in Sect.~\ref{ssec:variations-kIR}. This difference is probably an upper limit because it is likely that only few regions in our sample have a metallicity as low as $Z=0.008$. In conclusion, the impact of the assumption of the metallicity appears to remain minor compared to the total dynamic range spanned by the physical properties and by $k_i$.

\section{Uncertainties on estimated physical parameters\label{sec:model-uncertainties}}

If we want to understand what impacts hybrid estimators, it is important to determine whether the physical parameters we derive are measured reliably. To ascertain this point, we need to ensure that the models we consider can reasonably reproduce the observations (Sect.~\ref{ssec:reproducing-observations}) and that we can retrieve the intrinsic physical parameters precisely and accurately (Sect.~\ref{ssec:precision-accuracy}).

\subsection{Ability of the models to reproduce observations\label{ssec:reproducing-observations}}

\begin{figure}[!htbp]
 \includegraphics[width=\columnwidth]{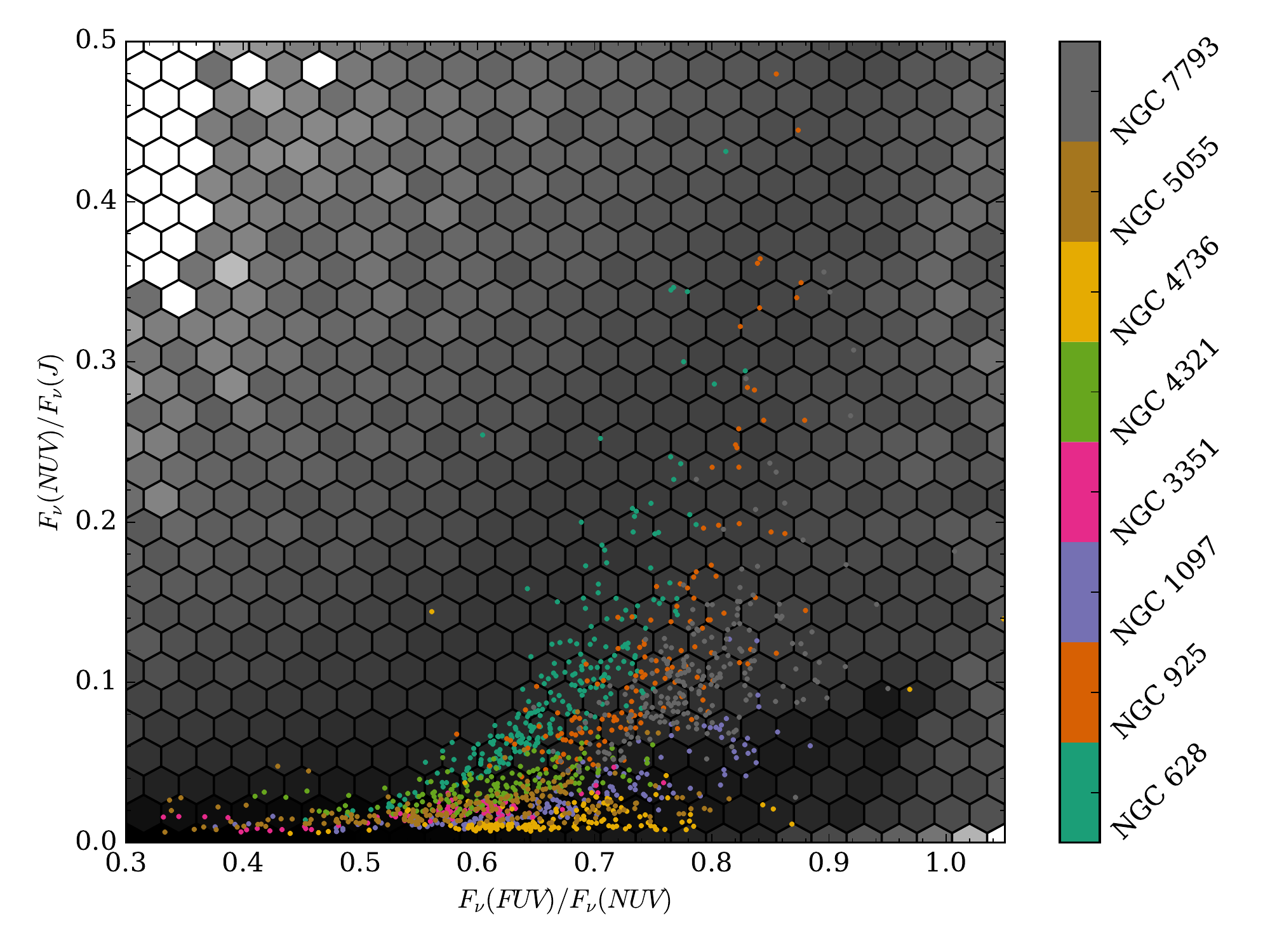}
 \includegraphics[width=\columnwidth]{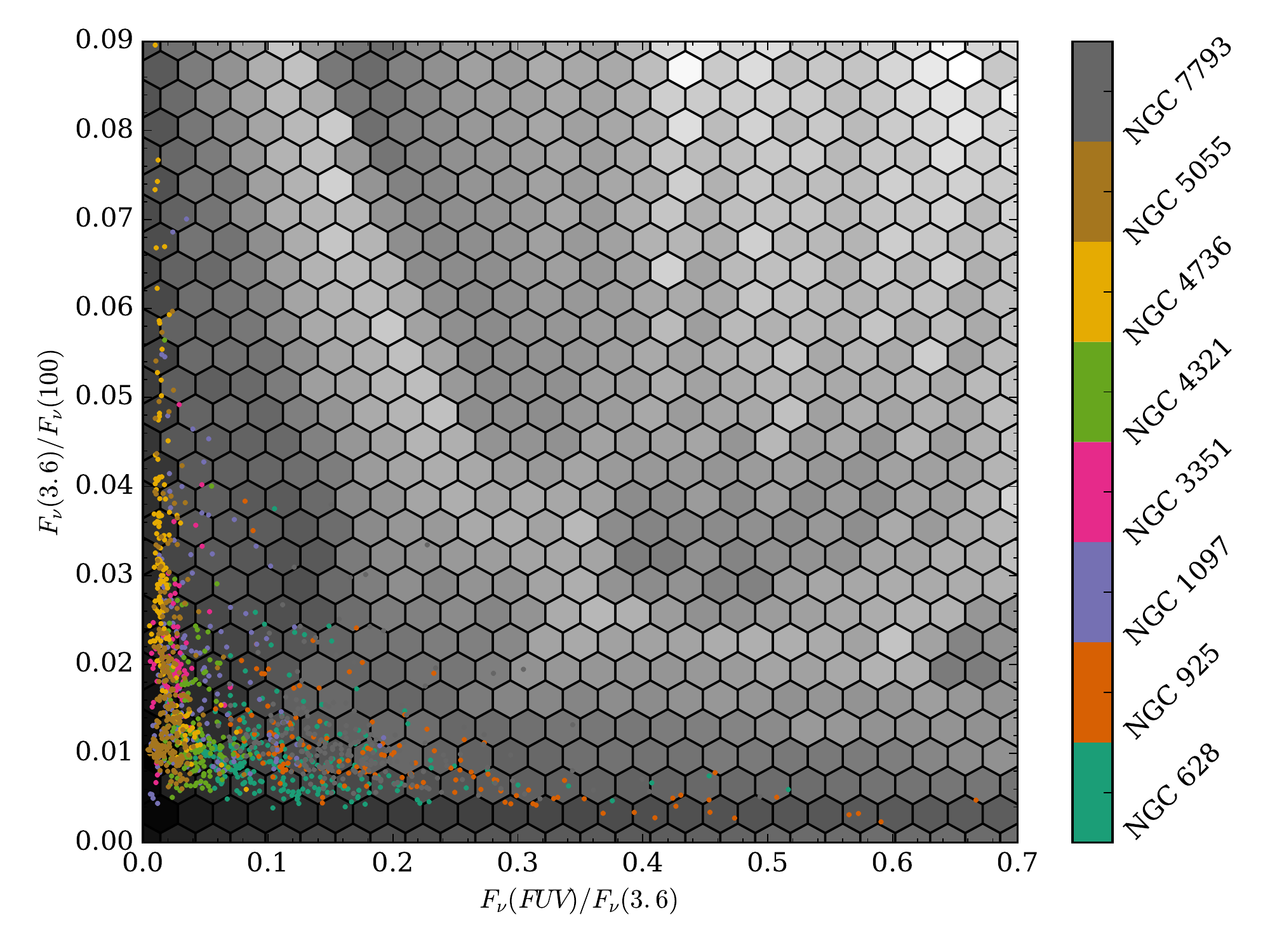}
 \includegraphics[width=\columnwidth]{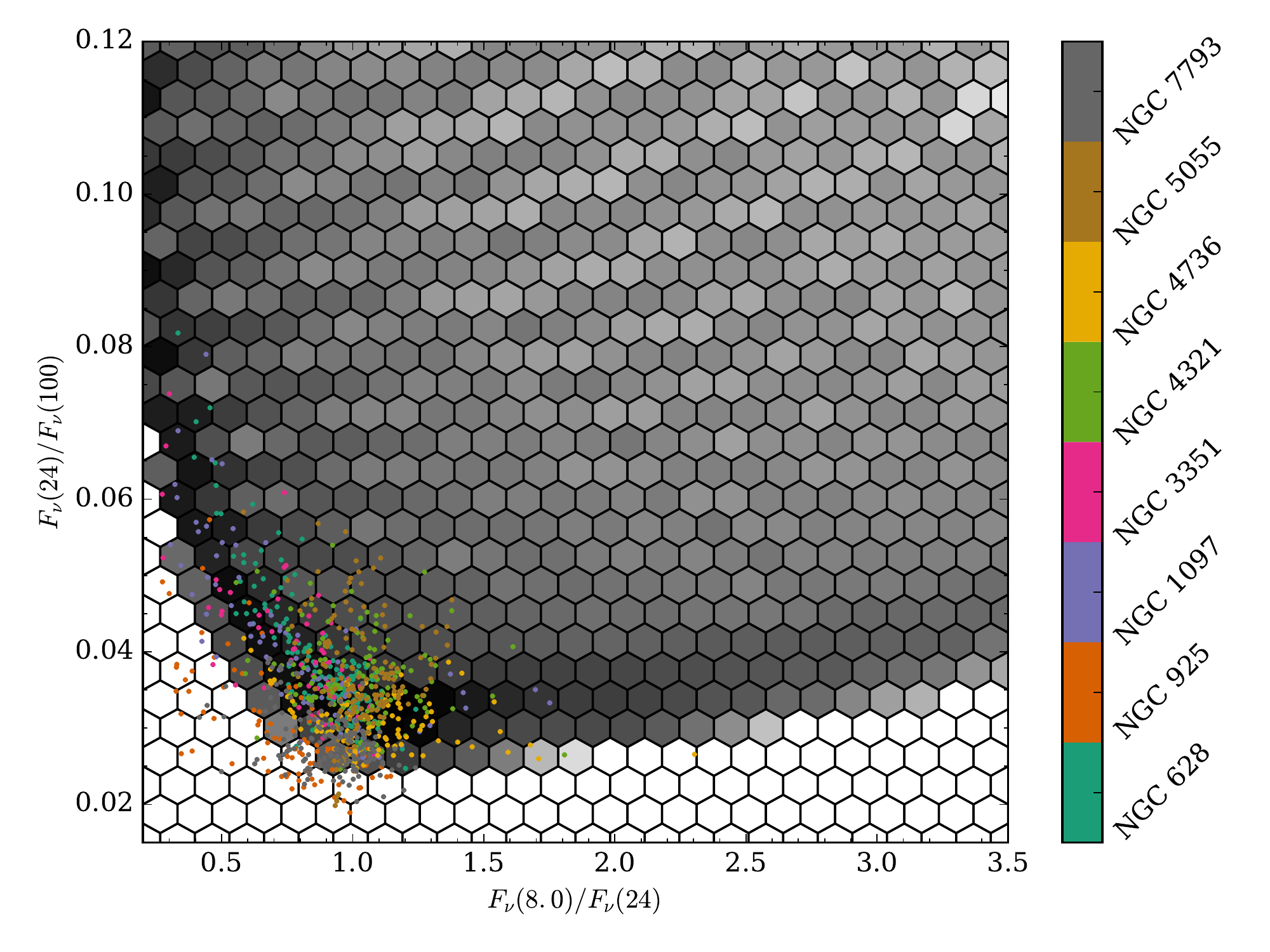}
\caption{Comparison of some of the observed colours for all the regions in the sample (coloured points) with the colour space covered by the models (grey hexagons). The selected colours are representative of the young and old stellar populations as well as of the dust. The colour of each dot indicates the galaxy following the colour bar on the right. The density of models is indicated by the shade of the hexagonal bins, following a logarithmic scale. Black indicates the highest density of models whereas white indicates the absence of models in the particular bin.\label{fig:comp-models}}
\end{figure}

The most basic requirement for properly modelling galaxies or regions within galaxies is that the adopted set of models is able to reproduce the observations. To verify this, we compare in Fig.~\ref{fig:comp-models} some of the observed colours of our sample with the range of theoretical colours covered by the models.

We clearly see that the models well reproduce the observations for the selected colours. However, when examining the mid-- and far--infrared colours we find that there is a number of regions with low 8.0--to--24 and 24--to--100 ratios that cannot be reproduced with the \cite{dale2014a} dust templates. They belong to different galaxies but in particular to NGC~925 and NGC~7793. It appears that the overwhelming majority corresponds to faint regions with a rather low signal--to--noise ratio. Here we include all regions detected down to a 1--$\sigma$ level in all bands even though the derivation of $k_i$ is then limited to regions detected at least at the 5--$\sigma$ level in the considered band. The effect is mostly prevalent at 70~$\mu$m and to a lesser extent at 100~$\mu$m. Only considering the regions detected at the 5--$\sigma$ level, we find fewer discrepant regions and most of them appear to be compatible with this noise level. This is a lower limit as we implicitly assume here that the 24~$\mu$m does not contribute significantly to the uncertainty. However, as we will see in Sect.~\ref{ssec:precision-accuracy}, even with the low signal--to--noise ratio of the 70~$\mu$m and 100~$\mu$m bands we can retrieve the TIR luminosities with excellent precision and accuracy as dust emission is constrained simultaneously using 7 bands from 8~$\mu$m to 350~$\mu$m.

To assess whether a different method would provide us with different estimates, we have compared the dust luminosities computed from the modelling to the ones estimated from the empirical recipes of \cite{galametz2013a}, which have been calibrated on local scales using the more complex \cite{draine2007a} dust models from 24~$\mu$m to 250~$\mu$m, we find TIR luminosities within 0.03~dex on average for the most discrepant galaxy (NGC~925). This suggests that the low signal--to--noise ratio bands do not have a dramatic influence on the estimation of the TIR luminosity.

\subsection{Precision and accuracy\label{ssec:precision-accuracy}}
Another important aspect to model galaxies or regions within galaxies is a reliable quantification of the precision and accuracy of the derived physical properties.

The precision can be estimated by examining the uncertainties computed from the marginalised PDF. We find that on average the FUV attenuation and the stellar mass have small mean relative uncertainties of $9.8\pm2.7$\% and $21.2\pm10.2$\% respectively. The uncertainties on the SFR are much larger: $52.6\pm23.2$\% for the current SFR and $97.6\pm23.3$\% for the average SFR over 100~Myr ($\mathrm{\left<SFR\right>_{100}}$). This reflects the difficulty there is to constrain the SFH of galaxies from broadband data. Two different SFHs can yield very similar SEDs because of various degeneracies, such as the degeneracy between the strength of an episode of star formation and its age. At the same time, it shows it is important to include diverse SFHs in the modelling to estimate properly the uncertainties. Failing this, the PDF of some physical properties may be narrower than they should be, which in turn would induce an underestimate of the uncertainties of these physical properties.

To estimate the accuracy of the estimates, we need to quantify whether the modelling allows us to retrieve the true value of the physical properties on average. To do so, a standard approach is to model the SED of galaxies or regions within galaxies whose physical properties are perfectly known. We can then compare these exact values to the values we estimate from the modelling. The results are dependent on the physical characteristics of the sample, on the set of models adopted, and on the set of bands used. Therefore they cannot be established in general. Such a test has to be carried out independently for each study.

In the first step for this test, we build a dedicated simulated catalogue of regions for which we know the physical properties with infinite accuracy. For this, we follow the procedure outlined in \cite{giovannoli2011a} and \cite{boquien2012a}. We model the SED of each region in our sample using the set of models described in Table~\ref{tab:cigale-parameters} and we compute the SED of the best fit model along with the associated physical properties. We then simulate an observation of the SED of the best model by adding a random Gaussian noise to the flux in each band based on the 1--$\sigma$ uncertainty of the observed fluxes. This yields a simulated catalogue of observed regions in galaxies for which we know the true physical properties. The main advantage of this method is that it yields a set of SEDs that are very similar to the observed sample. Conversely the limitation is the models used for the simulated catalogue are not independent from the models used to fit this catalogue. This means that while this method allows us to test whether we can retrieve the physical properties in an internally consistent way, it will not detect systematic differences between models and observations such as systematic biases in the stellar evolution tracks for instance.

\begin{figure*}[!htbp]
 \includegraphics[width=\columnwidth]{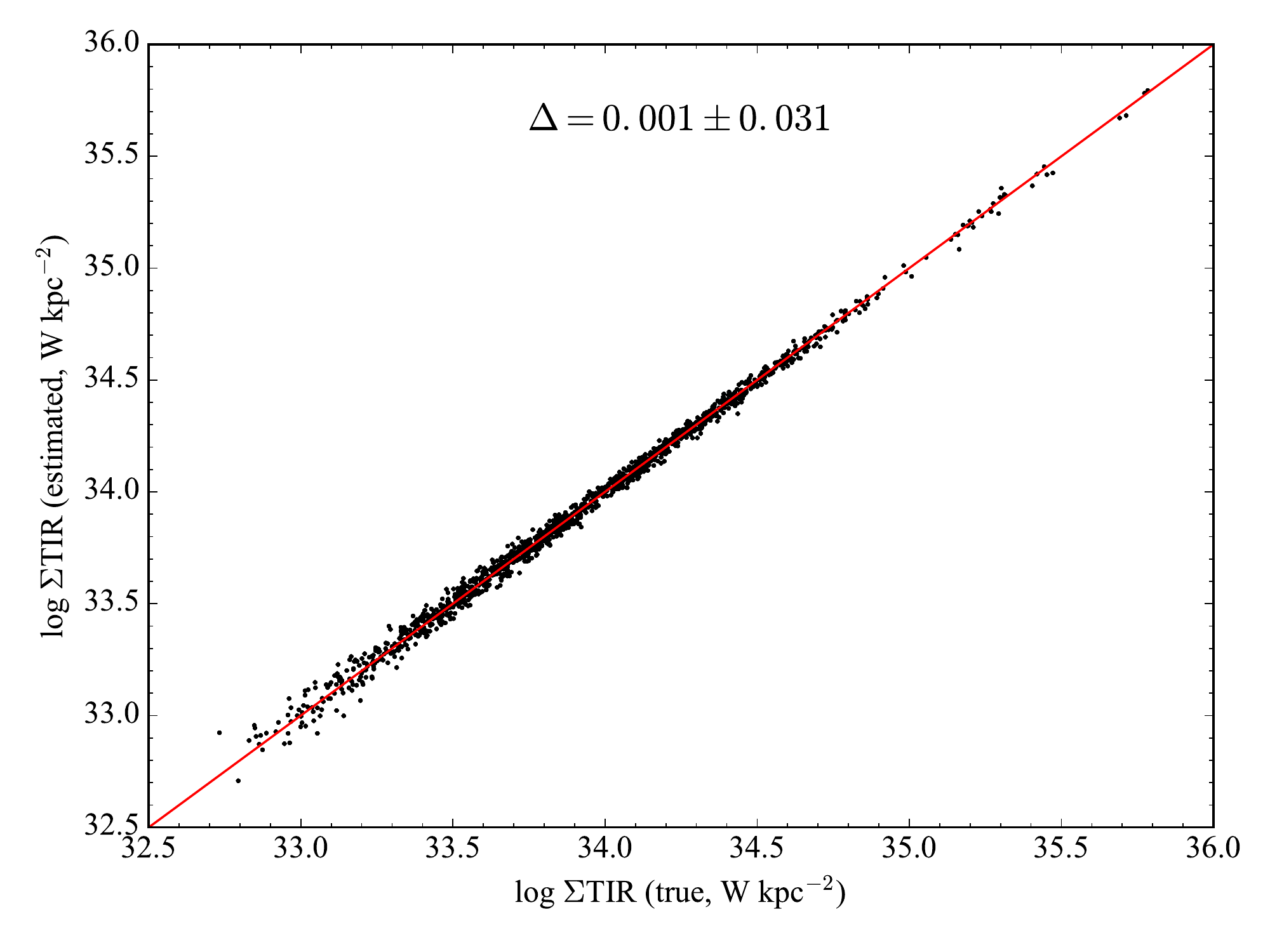}
 \includegraphics[width=\columnwidth]{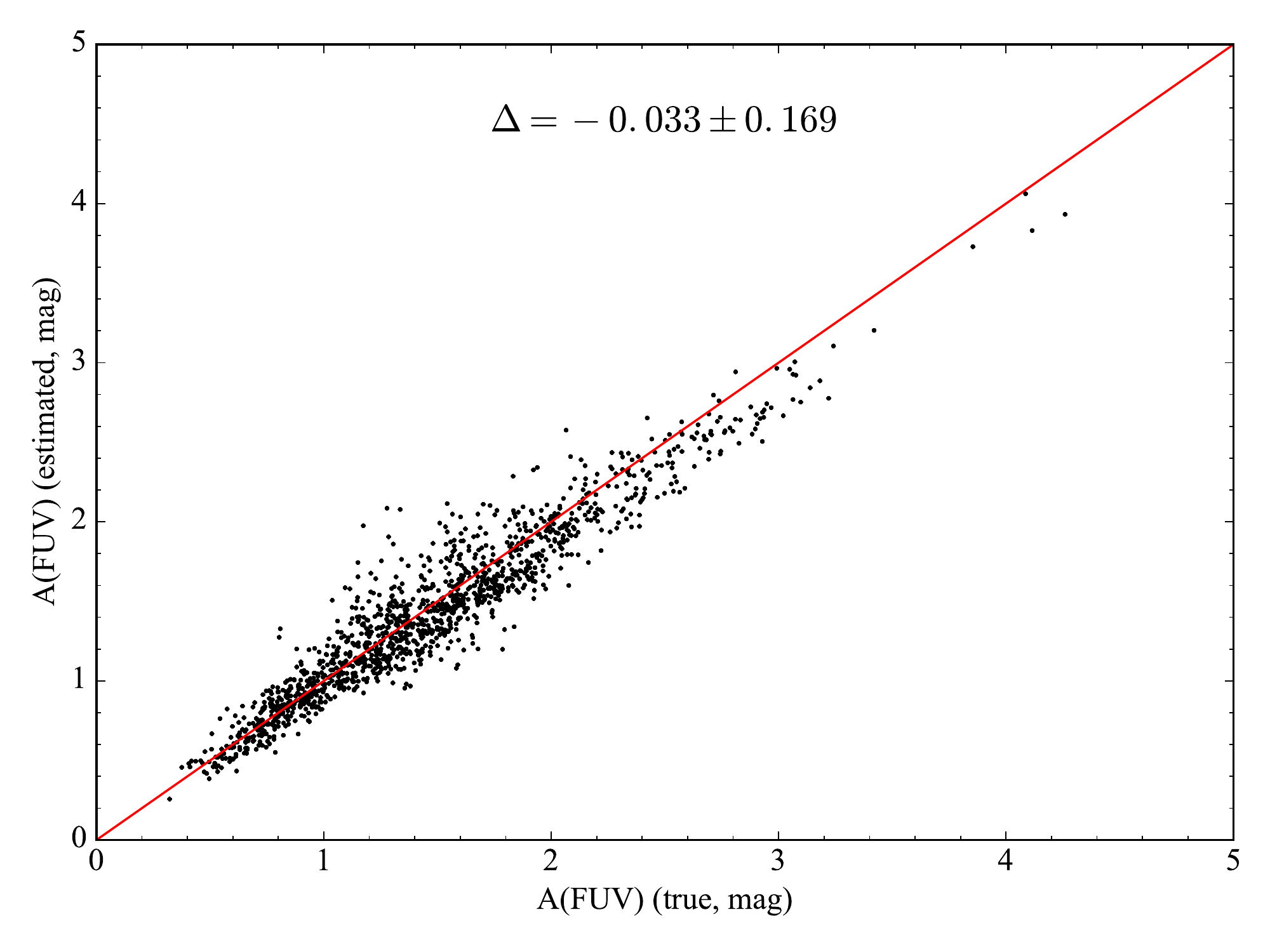}\\
 \includegraphics[width=\columnwidth]{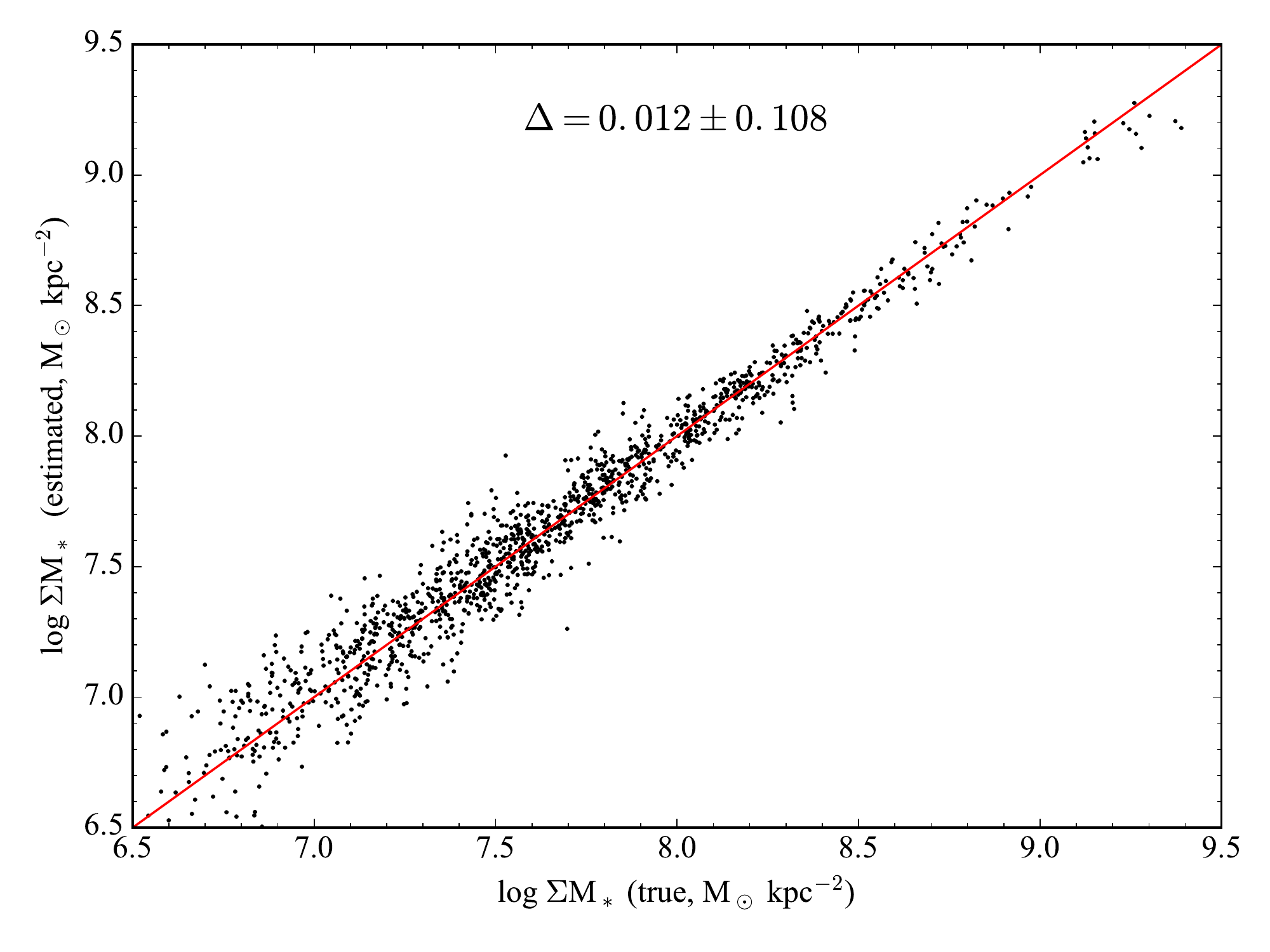}
 \includegraphics[width=\columnwidth]{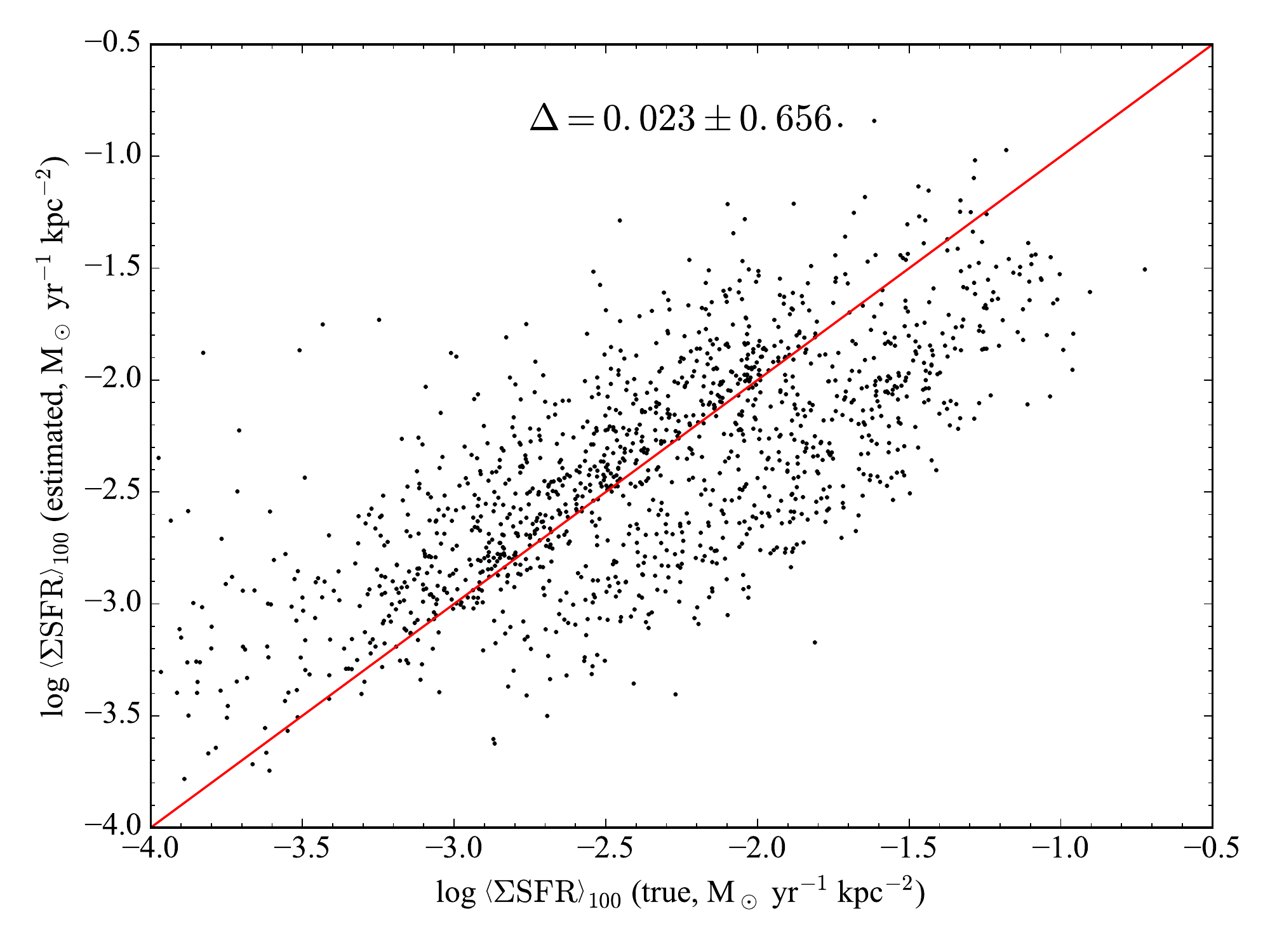}
\caption{Comparison between the true ($x$--axis) and estimated ($y$--axis) properties for the TIR luminosity (top--left), the FUV attenuation (top--right), the stellar mass (bottom--left), and the average SFR over 100~Myr (bottom--right). The red line indicates the one--to--one relation.\label{fig:mock}}
\end{figure*}

In the second step, we model and analyse this new catalogue as if it was made of real observations. We compare the true and estimated values in Fig.~\ref{fig:mock} for the TIR luminosity, the FUV attenuation, the stellar mass, and the average SFR over 100~Myr.
First of all, we find that the TIR luminosity, is very well constrained, with a logarithmic mean offet between the estimated and true values of $\Delta=0.001\pm0.031$~dex. It is therefore reliable. The FUV attenuation is also very well determined with only a slight systematic underestimate: $\Delta=-0.033\pm0.169$~mag. This is important as the FUV attenuation is the critical parameter in this study: it is directly involved in the computation of $k_i$ (Eq.~\ref{eqn:ki-auv}). Another very well constrained physical property is the stellar mass: $\Delta=0.012\pm0.108$~dex. The case of the SFR is however more complex. If we consider the current SFR (not shown here), the estimates are both very imprecise and very inaccurate: $\Delta=0.382\pm1.050$~dex. A large part of this discrepancy is due to artificial regions with extremely low SFR. Using the full PDF will result in large overestimates of the current SFR for these regions. However, we do not see similarly catastrophic overestimates of the average SFR. While the estimates for many regions are close to the true values, it shows that the SFR from the modelling should be interpreted with particular caution. This is a consequence of the difficulties to reconstruct the SFH of galaxies from broadband photometry, and therefore retrieve the current SFR. That being said, the average SFR over 100~Myr shows much more promising results: $\Delta=0.023\pm0.656$~dex, suggesting it is a considerably more accurate parameter than the current SFR. If the scatter is large, it remains much smaller than the dynamical range probed by the sample, approximately 3~dex. One drawback is that in the present case we only assume decreasing SFH, which means the average SFR will always be larger than the current one. For this study, to enjoy much more reliable estimates of the SFR we only consider them averaged over a timescale of 100~Myr. Such timescales also have the advantage that they are more comparable to the typical age of dust heating populations \citep{boquien2014a}, which is important to understand the physical origin of the variation of hybrid estimators.

\begin{figure}[!htbp]
 \includegraphics[width=\columnwidth]{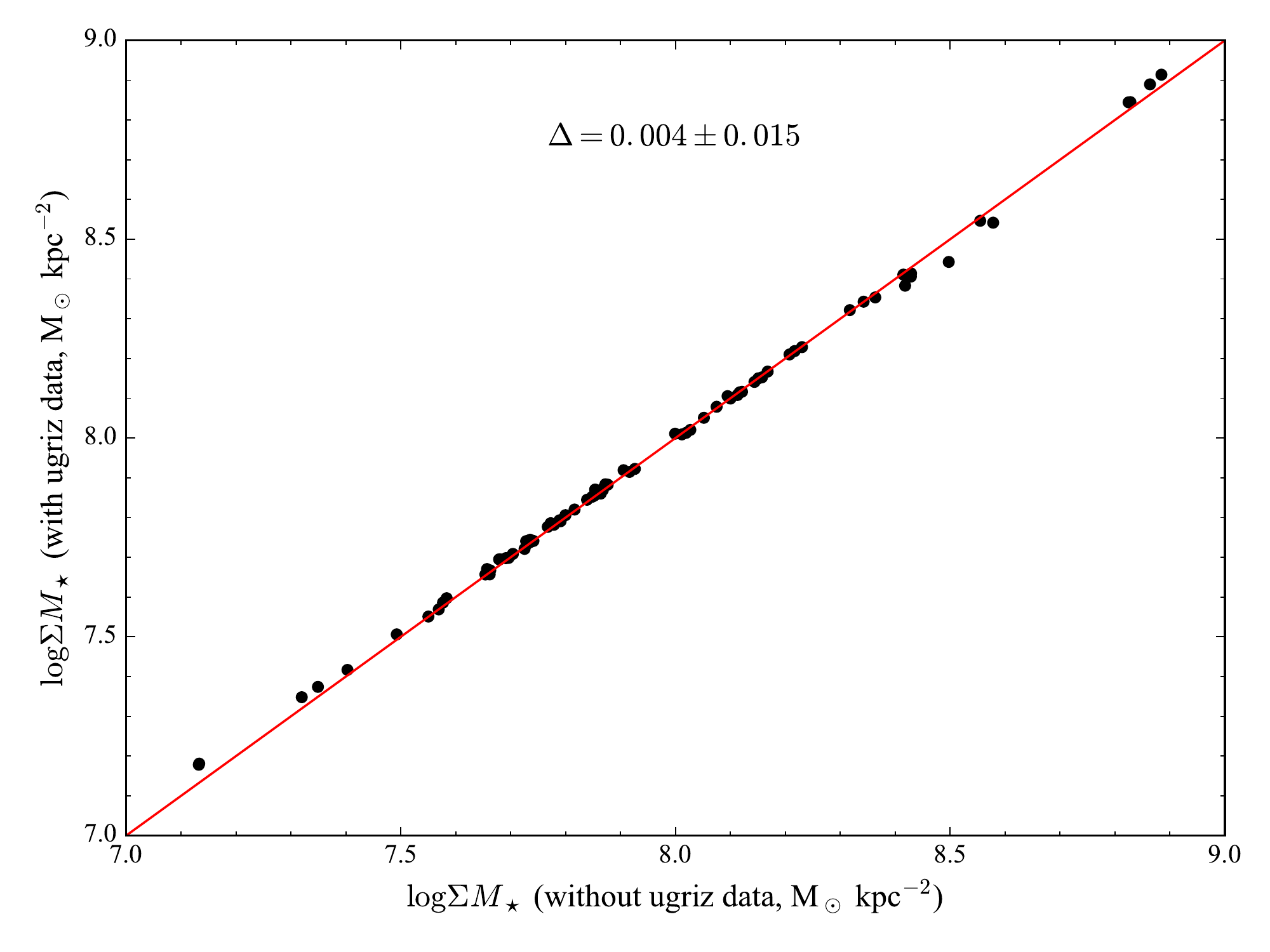}
\caption{Comparison between the estimate of the stellar mass of the regions of NGC~3351 without optical SDSS with data ($x$--axis) and with SDSS optical data ($y$--axis). The red line indicates the one--to--one relation.\label{fig:comp-optical}}
\end{figure}

Finally we examine the question of the consequence of waiving the optical data on the determination of the stellar mass. Indeed, some of the optical images available for KINGFISH galaxies have unreliable calibrations and in any case they are not homogeneous as they come from different telescopes and surveys. This means that in this study the stellar mass is essentially determined from the 2MASS J, H, and Ks bands along with the Spitzer 3.6~$\mu$m and 4.5~$\mu$m bands. To estimate the effect of the presence of optical data on the estimate the stellar mass we have run our model on NGC~3351 as, being in the SDSS footprint, it has well--calibrated ugriz data and therefore represents the most favourable case. We show the estimation of the stellar mass for this galaxy with and without ugriz data in Fig.~\ref{fig:comp-optical}.
The inclusion of SDSS optical data makes little difference with an increase of the stellar mass of $\Delta=0.004\pm0.015$~dex. Of more importance however is the question of the UV attenuation because it is critical to compute $k_i$ as we can see in Eq.~\ref{eqn:ki-auv}. Comparing the two estimates, we find that including SDSS optical data decreases the FUV attenuation by $\Delta=0.009\pm 0.062$~mag. We conclude that waiving the optical data does not have averse consequences on the reliability of our results.

Overall we conclude that the model estimates of the physical parameters we are particularly interested in, such as the FUV attenuation, the stellar mass, and the average SFR over 100~Myr are sufficiently accurate to study their impact on hybrid estimators.

\section{Comparison with popular recipes to compute the FUV attenuation\label{sec:comp-FUV}}

\begin{figure}[!htbp]
 \includegraphics[width=\columnwidth]{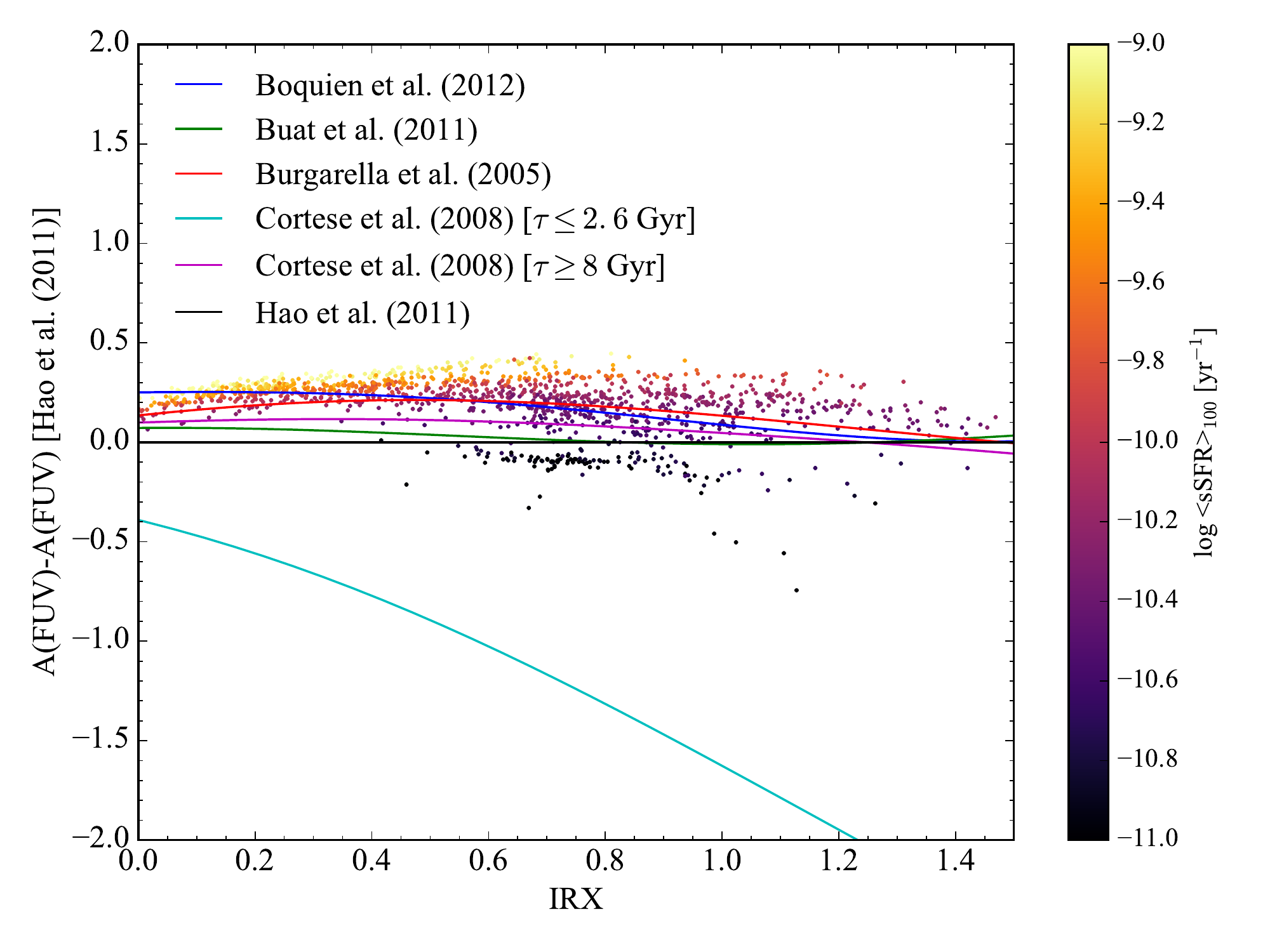}
\caption{Difference between the FUV attenuation estimated from various methods and from the recipe of \cite{hao2011a} versus IRX. Each line represents a different recipe that depends on IRX: \citet[blue]{boquien2012a}, \citet[green]{buat2011a}, \citet[red]{burgarella2005a}, and \citet[cyan ($\tau\le2.6$~Gyr) and purple ($\tau\ge8$~Gyr) ]{cortese2008a}. The points represent the difference between the FUV attenuation estimated from CIGALE and from \cite{hao2011a}. Their colour is indicative of the average sSFR over 100~Myr ($\left<sSFR\right>_{100}$), following the colour bar on the right.\label{fig:comp-AFUV-recipes}}
\end{figure}

Beyond the SED modelling approach adopted in this study, a popular method to compute the FUV attenuation is to relate it directly to IRX, the ratio of the TIR to the observed FUV luminosities. Several such recipes are available in the literature, for instance: \cite{cortese2008a} which is theoretically-motivated using an ``\`a la Sandage'' SFH ($\mathrm{SFR}(t)\propto t\times\exp(-t^2/2\tau^2)$), \cite{burgarella2005a}, \cite{buat2011a}, and \cite{boquien2012a} which are based on polynomial fits between A(FUV) and IRX with these values computed from SED modelling, and finally \cite{hao2011a} which is based on a purely empirical derivation. Ultimately as we can see in Fig.~\ref{fig:comp-AFUV-recipes}, they provide us with attenuation estimates that are typically within $\sim0.2$~mag of one another.

If these methods provide us with reliable estimates of A(FUV), one limitation that is specific to the context of this study is that for a given IRX, they each associate a unique A(FUV). This can be understood as different galaxies with the same IRX having the same fraction of dust heating by evolved stellar populations or in other words that the fraction of heating by evolved stellar populations can be exactly determined from IRX only. However in reality a given IRX can give rise to different A(FUV) depending on the dust heating sources. To account for this possibility, we have modelled each region in our sample with a large variety of SFH and attenuation curves. As a consequence we can have different amounts of dust heating by these older stellar populations. This explains why we see in Fig.~\ref{fig:comp-AFUV-recipes} that the difference between our A(FUV) estimates and that of individual recipes depends on the sSFR. So while all these methods provide us with consistent estimates of A(FUV), in this study we need to rely on SED modelling to include the effect of a variable SFH on dust heating and ensure our results do not depend on the intrinsic link in IRX--based methods between IRX and the fraction of dust heating by evolved stellar populations.

\section{NUV attenuation correction\label{sec:kir-NUV}}

As we have seen, it is possible to parametrise hybrid estimators in the FUV. This raises the question of whether the method we have developed is also applicable to the NUV, another frequently used star formation tracer in the absence of FUV observations. One of the main uncertainties in the NUV is the possible presence of a bump in the attenuation curve, which affects estimates of the NUV attenuation. Determining its amplitude from just the two GALEX bands is not possible as a change in the attenuation curve slope in the UV is degenerate with the presence of a bump in the NUV band. Answering this question would require additional data and ideally NUV spectra \citep[e.g.][]{buat2012a} that are not available for our sample. That being said, because we have included the presence of a bump of variable strength in the models, the uncertainty on its strength is taken into account when estimating the physical parameters. With this caveat in mind, we have computed for the NUV the relations between $k_i$ and the NUV$-$NIR colours and the NIR luminosity densities per unit area. For reference, we provide the best fit parameters in Table \ref{tab:kIR-colours-NUV} for the NUV$-$NIR colours and in Table~\ref{tab:kIR-NIR-NUV} for the NIR luminosity densities per unit area.

We find that we retrieve for the NUV the trends we found earlier for the FUV. The values for $k_i$ are systematically lower, probably due to a smaller attenuation in the NUV. For comparison, \cite{hao2011a} found $k_i=2.26\pm0.09$ at 25~$\mu$m and $k_{TIR}=0.27\pm0.02$ for the TIR. 

\begin{table*}
 \centering
 \begin{tabular}{ccccccc}
  \hline\hline
  IR band&colour&$a$&$b$&$\sigma_{ab}$&$\Delta k_i$\\\hline
24&NUV$-$J&$   9.636\pm   0.181$&$  -1.312\pm   0.050$&$  -0.009$&$  0.000\pm   0.893$\\
24&NUV$-$H&$   9.409\pm   0.165$&$  -1.200\pm   0.044$&$  -0.007$&$  0.000\pm   0.900$\\
24&NUV$-$K&$   8.868\pm   0.142$&$  -1.145\pm   0.041$&$  -0.006$&$   0.000\pm   0.874$\\
24&NUV$-$3.6&$   8.784\pm   0.152$&$  -1.347\pm   0.053$&$  -0.008$&$  0.000\pm   0.896$\\\hline
70&NUV$-$J&$   1.305\pm   0.145$&$  -0.088\pm   0.040$&$  -0.006$&$  0.000\pm   0.222$\\
70&NUV$-$H&$   1.293\pm   0.142$&$  -0.082\pm   0.038$&$  -0.005$&$  0.000\pm   0.222$\\
70&NUV$-$K&$   1.266\pm   0.131$&$  -0.080\pm   0.037$&$  -0.005$&$  0.000\pm   0.222$\\
70&NUV$-$3.6&$   1.216\pm   0.119$&$  -0.080\pm   0.041$&$  -0.005$&$  0.000\pm   0.225$\\\hline
100&NUV$-$J&$   1.039\pm   0.119$&$  -0.094\pm   0.033$&$  -0.004$&$  0.000\pm   0.119$\\
100&NUV$-$H&$   1.030\pm   0.116$&$  -0.088\pm   0.031$&$  -0.003$&$  0.000\pm   0.119$\\
100&NUV$-$K&$   0.996\pm   0.105$&$  -0.086\pm   0.030$&$  -0.003$&$   0.000\pm   0.118$\\
100&NUV$-$3.6&$   0.957\pm   0.099$&$  -0.090\pm   0.034$&$  -0.003$&$   0.000\pm   0.124$\\\hline
TIR&NUV$-$J&$   0.596\pm   0.099$&$  -0.058\pm   0.027$&$  -0.003$&$   0.000\pm   0.030$\\
TIR&NUV$-$H&$   0.588\pm   0.096$&$  -0.054\pm   0.025$&$  -0.002$&$   0.000\pm   0.030$\\
TIR&NUV$-$K&$   0.566\pm   0.086$&$  -0.052\pm   0.025$&$  -0.002$&$  0.000\pm   0.030$\\
TIR&NUV$-$3.6&$   0.555\pm   0.081$&$  -0.059\pm   0.028$&$  -0.002$&$  0.000\pm   0.030$\\\hline

 \end{tabular}
 \caption{Same as Table~\ref{tab:kIR-colours} for the NUV band.\label{tab:kIR-colours-NUV}}
\end{table*}

\begin{table*}
 \centering
 \begin{tabular}{ccccccc}
  \hline\hline
  IR band&NIR band&$a$&$b$&$\sigma_{ab}$&$\Delta k_i$\\\hline
24&J&$   3.259\pm   0.223$&$  -4.390\pm   0.445$&$   0.082$&$  0.000\pm   1.797$\\
24&H&$   3.601\pm   0.182$&$  -4.097\pm   0.389$&$   0.055$&$   0.000\pm   1.700$\\
24&K&$   3.238\pm   0.198$&$  -3.836\pm   0.343$&$   0.057$&$   0.000\pm   1.651$\\
24&3.6&$   1.887\pm   0.358$&$  -4.534\pm   0.474$&$   0.159$&$   0.000\pm   1.792$\\\hline
70&J&$   0.955\pm   0.043$&$  -0.358\pm   0.097$&$   0.001$&$  0.000\pm   0.202$\\
70&H&$   0.976\pm   0.042$&$  -0.342\pm   0.093$&$   0.000$&$   0.000\pm   0.202$\\
70&K&$   0.943\pm   0.044$&$  -0.328\pm   0.091$&$   0.001$&$   0.000\pm   0.203$\\
70&3.6&$   0.851\pm   0.058$&$  -0.356\pm   0.100$&$   0.004$&$   0.000\pm   0.207$\\\hline
100&J&$   0.653\pm   0.038$&$  -0.216\pm   0.075$&$   0.001$&$  0.000\pm   0.121$\\
100&H&$   0.665\pm   0.036$&$  -0.213\pm   0.072$&$   0.001$&$   0.000\pm   0.119$\\
100&K&$   0.645\pm   0.039$&$  -0.204\pm   0.069$&$   0.002$&$   0.000\pm   0.119$\\
100&3.6&$   0.593\pm   0.053$&$  -0.209\pm   0.076$&$   0.003$&$  0.000\pm   0.125$\\\hline
TIR&J&$   0.575\pm   0.037$&$  -0.118\pm   0.056$&$   0.001$&$  0.000\pm   0.103$\\
TIR&H&$   0.580\pm   0.034$&$  -0.122\pm   0.055$&$   0.001$&$   0.000\pm   0.101$\\
TIR&K&$   0.567\pm   0.038$&$  -0.119\pm   0.052$&$   0.001$&$  0.000\pm   0.101$\\
TIR&3.6&$   0.537\pm   0.050$&$  -0.124\pm   0.058$&$   0.002$&$  0.000\pm   0.103$\\\hline
 \end{tabular}
 \caption{Same as Table \ref{tab:kIR-NIR} for the NUV band.\label{tab:kIR-NIR-NUV}}
\end{table*}

\onecolumn
\section{Derivation of galaxy--wide $k_i$\label{sec:derivation-entire-gal}}

The new relations presented in Sect.~\ref{sec:hybrid-relations} to estimate $k_i$ from the FUV$-$NIR colours or the NIR bands are only applicable to resolved galaxies. As we have seen, individual regions in entire galaxies span a broad range in terms of $k_i$. However, the galaxy--wide $k_i$ is weighted towards more luminous regions that have red FUV$-$NIR colours and a high NIR luminosity density per unit area. If we assume that both the FUV and NIR luminosities per unit area and the SFR surface density follow exponential profiles of scale--lengths $r_{NIR}$ and $r_{SFR}$, we can compute an effective mean $k_i$ for the entire galaxy.
\begin{equation}
 \left<k_i\right>=\frac{\int_0^{2\pi}\int_0^{+\infty}k_i\left(r\right)\times\Sigma \mathrm{SFR}\left(r\right) \times r \times \mathrm{d}r \mathrm{d}\theta}{\int_0^{2\pi}\int_0^{+\infty}\Sigma \mathrm{SFR}\left(r\right) \times r \times \mathrm{d}r \mathrm{d}\theta}.
\end{equation}
In the case of the estimators based on the NIR luminosity density per unit area, if we assume exponential profiles:
\begin{eqnarray}
 k_i\left(r\right)&=&a_{NIR}+b_{NIR}\times\left[\log\Sigma_{\nu,~NIR}\left(r\right)-20\right],\\
 k_i\left(r\right)&=&a_{NIR}+b_{NIR}\times\left[\log\Sigma_{\nu,~NIR}\left(0\right)\times\exp\left(-r/r_{NIR}\right)-20\right],\\
 k_i\left(r\right)&=&a_{NIR}+b_{NIR}\times\left[\log\Sigma_{\nu,~NIR}\left(0\right)-\frac{r/r_{NIR}}{\ln 10}-20\right],\\
 k_i\left(r\right)&=&\underbrace{a_{NIR}+b_{NIR}\times\left[\log\Sigma_{\nu,~NIR}\left(0\right)-20\right]}_\alpha -\underbrace{\frac{b_{NIR}/r_{NIR}}{\ln 10}}_\beta\times r,
\end{eqnarray}
and
\begin{eqnarray}
 \Sigma\mathrm{SFR}\left(r\right)=\Sigma\mathrm{SFR}\left(0\right)\times\exp\left(-r/r_{SFR}\right).
\end{eqnarray}
Integrating over $2\pi$ and simplifying:
\begin{eqnarray}
 \left<k_i\right>&=&\frac{\int_0^{+\infty}\left[\alpha-\beta \times r\right] \times \exp\left(-r/r_{SFR}\right) \times r \times \mathrm{d}r}{\int_0^{+\infty}\exp\left(-r/r_{SFR}\right) \times r \times \mathrm{d}r},\\
 \left<k_i\right>&=&\alpha-\beta\times\frac{\int_0^{+\infty}\exp\left(-r/r_{SFR}\right) \times r^2 \times \mathrm{d}r}{\int_0^{+\infty}\exp\left(-r/r_{SFR}\right) \times r \times \mathrm{d}r},
\end{eqnarray}
and integrating over $r$:
\begin{eqnarray}
 \left<k_i\right>&=&\alpha-2\times\beta r_{SFR}.
\end{eqnarray}
We finally obtain:
\begin{eqnarray}
 \left<k_i\right>&=&a_{NIR}+b_{NIR}\times\left[\log\Sigma_{\nu\mathrm{,~NIR}}\left(0\right)-20\right]-2\times\frac{b_{NIR}}{\ln 10}\times\frac{r_{SFR}}{r_{NIR}},\\
 \left<k_i\right>&=&a_{NIR}+b_{NIR}\times\left[\log\Sigma_{\nu\mathrm{,~NIR}}\left(0\right)-\frac{2}{\ln 10}\times\frac{r_{SFR}}{r_{NIR}}-20\right].
\end{eqnarray}
This can also be expressed in terms of luminosity density rather than luminosity density per unit area:
\begin{equation}
 \left<k_i\right>=a_{NIR}+b_{NIR}\times\left[\log \frac{L_{\nu\mathrm{,~NIR}}}{2 \times \pi \times r_{NIR}^2}-\frac{2}{\ln 10}\times\frac{r_{SFR}}{r_{NIR}}-20\right].
\end{equation}

To estimate $k_i$ from the FUV$-$NIR colour, the computation is very similar and yields the following result:
\begin{equation}
 \left<k_i\right>=a_\mathrm{colour}+2.5\times b_\mathrm{colour}\times\left[\log\frac{\Sigma_{\nu\mathrm{,~NIR}}\left(0\right)}{\Sigma_{\nu\mathrm{,~FUV}}\left(0\right)}-\frac{2}{\ln 10}\times r_{SFR}\times\left[1/r_{NIR}-1/r_{FUV}\right]\right].
\end{equation}
As before, this can also be expressed in terms of luminosity densities rather than luminosity densities per unit area:
\begin{equation}
 \left<k_i\right>=a_\mathrm{colour}+2.5\times b_\mathrm{colour}\times\left[\log\left(\frac{L_{\nu\mathrm{,~NIR}}\left(0\right)}{L_{\nu\mathrm{,~FUV}}\left(0\right)}\times\frac{r_{FUV}^2}{r_{NIR}^2}\right)-\frac{2}{\ln 10}\times r_{SFR}\times\left[1/r_{NIR}-1/r_{FUV}\right]\right].
\end{equation}

\twocolumn

\end{document}